\documentclass[pdflatex,sn-mathphys-num]{sn-jnl}

\pdfoutput=1

\usepackage{graphicx}%
\usepackage{multirow}%
\usepackage{diagbox}
\usepackage{amsmath,amssymb,amsfonts}%
\usepackage{amsthm}%
\usepackage{mathrsfs}%
\usepackage[title]{appendix}%
\usepackage{xcolor}%
\usepackage{textcomp}%
\usepackage{manyfoot}%
\usepackage{booktabs}%
\usepackage{algorithm}%
\usepackage{algorithmicx}%
\usepackage{algpseudocode}%
\usepackage{listings}%
\usepackage{graphicx}
\usepackage{subcaption} 
\usepackage{enumitem}
\usepackage{comment}
\usepackage{titlesec}
\usepackage{graphicx}
\usepackage{subcaption}
\usepackage{booktabs}
\titleformat{\paragraph}[runin]{\normalfont\bfseries}{}{0pt}{}

\theoremstyle{thmstyleone}%
\theoremstyle{thmstyletwo}%
\theoremstyle{thmstylethree}%
\newtheorem{definition}{Definition}%
\begin{document}

\title{Ant swarm functional control via stigmergic Reinforcement Learning agents}

\author[1,3]{\fnm{Alessio} \sur{Pitteri}}\email{apitteri@fbk.eu}

\author[1]{\fnm{Andrea} \sur{Guizzo}}\email{aguizzo@fbk.eu}

\author*[2]{\fnm{Laura} \sur{Ferrarotti}}\email{lferrarotti@fbk.eu}

\author[2]{\fnm{Bruno} \sur{Lepri}}\email{lepri@fbk.eu}

\author[1]{\fnm{Riccardo} \sur{Gallotti}}\email{rgallotti@fbk.eu}

\affil*[1]{\orgdiv{CHuB Lab}, \orgname{Fondazione Bruno Kessler}, \orgaddress{\street{Via Sommarive, 18}, \city{Trento}, \postcode{38123}, \state{Italy} \country{}}}

\affil[2]{\orgdiv{MobS Lab}, \orgname{Fondazione Bruno Kessler}, \orgaddress{\street{Via Sommarive, 18}, \city{Trento}, \postcode{38123}, \state{Italy} \country{}}}

\affil[3]{\orgdiv{City Science Lab}, \orgname{Alma Mater Studiorum Università di Bologna}, \orgaddress{\street{Viale Carlo Berti Pichat, 6/2}, \city{Bologna}, \postcode{40127}, \state{Italy} \country{}}}

\maketitle

\section*{Abstract}
In this work, we propose a novel framework for the functional controllability of the ant swarm model \cite{chialvoswarms,patternformation,SwarmPhaseTrans}, a well-known and relevant model of collective behaviour. Our approach introduces a population of controlling stigmergic agents, trained via Reinforcement Learning (RL) \cite{suttonbarto}, that act on the environment to influence the system dynamics and promote the emergence of ordered behaviour. Stigmergic agents are optimized 
in a centralized-training decentralized-execution setting, interacting with ants only through the shared pheromone field. The reward design promotes trail pheromone structures and alignment of ant positions with high-pheromone paths, without requiring control of specific microscopic configurations. Our results demonstrate that the learned policies effectively shift the phase transition line that characterizes the global behaviour of the system, enabling the emergence of trails scenarios in regimes that are typically dominated by randomness. This study provides insights into the potential of RL based control strategies for complex systems, contributing 
to the general understanding of functional controllability in this field. 

\maketitle
\section{Introduction}
\label{sec:Introduction}

From weather and traffic, to swarm behaviour and opinion polarization, many real world phenomena exhibit emergent global behaviours and patterns (beneficial or critical, depending on the cases) that are difficult to understand and predict, even when local properties are well understood and modeled. The need for a unifying theoretical framework to describe such phenomena has led to the formal definition of \textit{complex system} as an ensemble of multiple elements, governed by individual rules and interacting with each other, such that macroscopic patterns and organization spontaneously arise within the system, without an external planner \cite{CSdef}. 
Due to the universality of this definition and the cross-disciplinary nature of such phenomena, the study of complex systems has driven significant advancements across a variety of domains, including human mobility \cite{humanmob}, animal behaviour \cite{BARBOSA20181}, and social networks \cite{JUSUP20221}. Nevertheless, understanding complex systems remains challenging, providing the scope for a rich corpus of literature \cite{challengesCS, foundationsCS, BakSOC}. 
To study complex systems, modeling and simulations via Agent-Based Models (ABMs) \cite{ABM} are often exploited. ABMs represent individual entities with defined rules of interaction, and allow global behaviour to emerge from local dynamics. Such models, while computationally costly and challenging (for instance, with respect to defining model granularity \cite{ABM} and acquiring reliable behavioural data \cite{social_selforg}), can capture heterogeneity, discreteness, and non-linear interactions that are difficult to model analytically. 

Going beyond pure system analysis, researchers have recently focused on controlling complex systems, designing methods to steer their behaviour either towards desired macroscopic states or away from risky ones. In control systems theory, \textit{controllability} refers to the ability to drive a system to any desired state within finite time \cite{reviewCScontrol, controllability}. 
In our work, controllability, intended as driving the system's configuration toward a specific point of the phase space, is not the goal. 
Instead, we aim to drive the system into a region of the phase space, potentially through a phase transition, where a desired global behaviour can emerge. 
In the following, we will refer to this relaxed version of controllability as \textit{functional controllability}. This goal is not only interesting for better understanding the potential and the limits of complex systems, but also relevant for practical reasons: achieving functional controllability in such systems can indeed mitigate undesired collective behaviours, such as cascading failures, congestion and instabilities, hence improving the performances \cite{helbingglobal}, without identifying a specific state, deemed as desirable, when many others can serve the scope. 

In order to investigate how functional control mechanisms can influence emergent outcomes, we study a well known model of collective behaviour: the \textit{ants swarm model} \cite{patternformation}. This model, thanks to both a strong theoretical analysis of its characteristic behaviours \cite{SwarmPhaseTrans}, and an effective simulation framework \cite{chialvoswarms}, proves to be a suitable choice to study the effects of functional control strategies on complex systems.
Our case-study exploits an ABM simulating the self-organization of ants in an unbounded space.  
Individuals are characterized by specific sensory and behavioural parameters, determining their dynamic interactions. According to the model, ants navigate the space by following pheromone traces, while indirectly communicating with each other by reinforcing the traces through a release of pheromone. When local structures (small clusters) or structured queues (pathways) emerge in the population behaviour, the system presents an \textit{ordered phase}. If sensory or behavioural parameters become unfavorable, instead, the individuals lose their ability to organize, undergoing a phase transition and landing into a \textit{disordered phase}. 

In this work we showcase how Reinforcement Learning (RL) \cite{suttonbarto} can be effectively employed to promote the emergence of ordered behaviour in the ants swarm model, and in particular the capability of enhancing the formation of ants trails, i. e., thin elongated connected pathways of pheromone, heavily populated by ants, as defined in Materials and Methods. 
Our proposed approach consists in introducing a population of controlling agents (\textit{smart-agents}) that learn how to act in order to induce the desired collective behaviour in the initial population under control, composed by individuals described by the biologically-inspired ABM dynamics (\textit{ABM-ants}). Pinning control approaches \cite{controlCS}, that apply external inputs to a small subset of agents in order to produce a different behaviour, aim as well at influencing the whole dynamic of the system through the action of a sub-group of agents. However, rather than steering the behaviour of some of the system agents through external stimuli, we introduce an additional set of agents with different intrinsic capabilities, modeled as parametric policies and learned via RL, from data collected in interaction. 
Our smart-agents, trained with the Proximal Policy Optimization (PPO) algorithm \cite{ppo}, influence ABM-ants by acting on the pheromone field, implementing a form of stigmergic interaction \cite{stigmergy} as widely studied in swarm intelligence and ant colony systems \cite{ACP}. We demonstrate that, without any prior knowledge of the dynamics of the system, the smart-agents learned policies are able to shift the phase transition line that characterizes the global behaviour, generating organized structures in regimes that are typically dominated by randomness. 

Much of the existing works on RL-based complex systems control focuses on applying RL agents within the field of active matter \cite{reviewRL4AM}, where the ants swarm model is a natural fit. In particular, relevant examples are represented by \cite{learntotornado} and \cite{learntoflock}.
In \cite{learntotornado}, the idea of learning a single policy to govern the behaviour of all agents in the collective is adopted, with the primary objective of achieving a desired global configuration (such as milling and rotating tornadoes). In \cite{learntoflock}, instead, RL is used to train agents interacting with a population governed by the Vicsek model \cite{vicsek}. Each agent optimizes a local reward in order to promote the emergence of global coordination in motion.
In line with the procedure presented in \cite{learntotornado}, our approach involves training one shared policy that rules how all the smart-agents in the system behave. However, while their formulation considers agents operating in isolation within an otherwise empty environment, our work embeds all the smart-agents in an environment governed by specific model dynamics, as explored by \cite{learntoflock}. As a result, the smart-agents not only should learn to coordinate with one another but must also account for and adapt to the external environmental processes that shape the evolution of the collective state produced by the actions of both smart- and ABM-agents.

In Section \ref{sec:Results} the main experimental results are presented, demonstrating how functional controllability can be achieved in different scenarios and how the phase transition properties of the system are affected. Section \ref{sec:Discussion} contains the implications of our work for the field of complex systems control and summarizes our contributions and future research directions. Details on the swarm model (theory and implementation), the RL optimization framework, and the chosen training scenario can be found in Section \ref{sec:Materials_and_methods}, and are expanded in Supplementary Materials (SM).

\section{Results}
\label{sec:Results}
In this section, we present the effects of introducing a population of controlling agents, trained via RL, on the dynamics of the ant swarm model, evaluating how such control strategy promotes the formation of \emph{trails scenarios}, inducing a shift from disordered to ordered behaviour.

The ant swarm model describes how ants navigate and interact indirectly through a shared pheromone field, that they are also attracted by, as detailed in Materials and Methods, Section 
\ref{sec:Ants_swarmmodel}. 
Individuals deposit pheromone at a constant rate $\eta$ while moving; the pheromone naturally decays over time at rate $\kappa$. The ants response in term of movement is modeled as a non-linear function that is (i) increasing with respect to their sensory capacity $\delta^{-1}$ and (ii) decreasing with respect to the level of 
noise in their signal reception $\beta^{-1}$ (having the meaning of a temperature, in the statistical physics analogy presented in \cite{SwarmPhaseTrans, patternformation}). 
To simulate the model, we consider an ABM obtained by discretizing the ant swarm model, with $N$ ABM-ants moving at every time-step on a two dimensional eight-neighbor square lattice (see Section \ref{sec:Materials_and_methods}.\ref{sec:Ants_swarmmodel} - \ref{sec:discretized_model}).
At each time-step, each of the ABM-ants probabilistically chooses its next site $j$ among the neighboring cells of its current site $i$  according to a probabilistic field $p_{j, i\,}$, described by Eq. \ref{eq:motion-discrete}, that depends on $\left[ \delta ,\,\beta\right]$, the local pheromone concentrations, and the change in direction required to reach the new location. When the model parameters cross specific thresholds (computed analytically in Eq. \ref{Eq:stabcrit} manipulating the model dynamics), the ABM-ants collectively organize into pathways that form and strengthen via a positive feedback loop. For parameter configurations sensibly farther from the mentioned threshold, then, they remain trapped within small clusters. In scenarios characterized by parameters that are under the threshold, instead, the population shows random behaviour, as shown in Figure \ref{fig:phasegrid}, and Section \ref{sec:Materials_and_methods}.\ref{sec:Ants_swarmmodel} - \ref{sec:sim_phasetrans}.

\begin{figure}[t]
    \centering
    \includegraphics[width=0.8\linewidth]{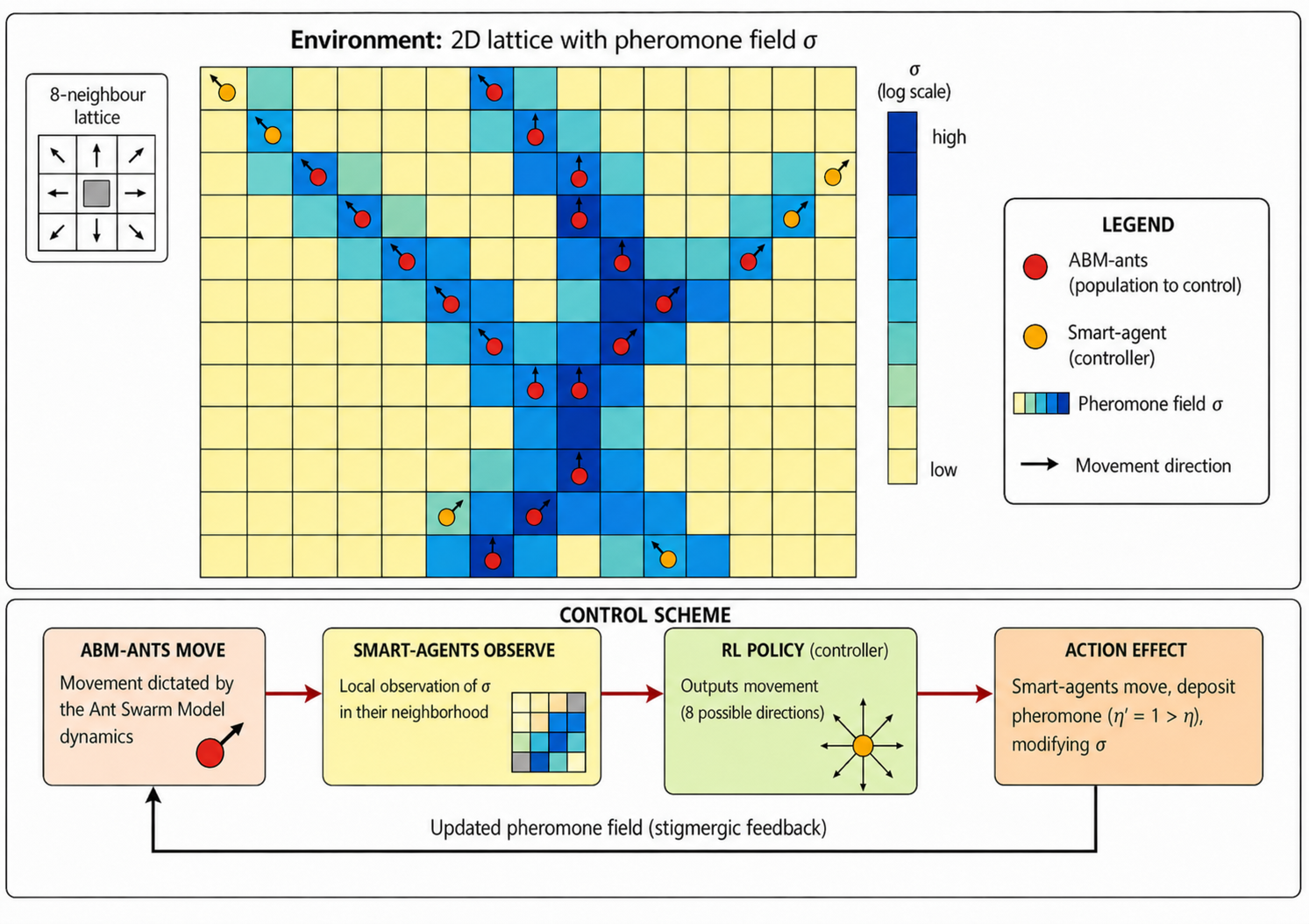}
    
    \caption{\textbf{Ant Swarm Model: stigmergic control scheme with hybrid population}.}
    \label{fig:scheme}
\end{figure}

Our control scheme involves the introduction of a set of $N'$ smart-agents in the environment hosting the $N$ ABM-ants to be controlled, as represented in Figure \ref{fig:scheme}. 
The smart-agents are apparently indistinguishable from ABM-ants, yet fundamentally different in behaviour. Unlike the ant population, their movements are determined by the output of a learned RL policy, and they release a fixed and significant quantity of pheromone $\eta'=1 > \eta$. As detailed in Section \ref{sec:Materials_and_methods}.\ref{sec:RLDesign}, we design a RL scheme, by defining appropriate perception states for the smart-agents, structuring a policy that chooses the agent's next position among the neighboring sites, and defining rewards that guide the system toward \emph{trails scenarios}, that we characterize in Definition \ref{def:ants-trail-scenario} (see Materials and Methods) as scenarios with (i) a bimodal pheromone distribution, with pheromone concentrated along localized high-density traces and nearly absent elsewhere (\textit{bimodal pheromone condition}); (ii) an ant population predominantly distributed on these traces (\textit{population condition}); (iii) pheromone traces forming thin elongated non-branching pathways, rather than extended two-dimensional regions (\textit{trail geometry condition}).

Policies are trained via PPO \cite{ppo} (see SM), in a \textit{centralized-training decentralized-execution} framework. To ensure the statistical significance of our results, we perform $n_{\pi} = 10$ independent policy training procedures in every learning scenario.  
Training proceeds episodically, and the resulting policy is tested at the end of learning by evaluating its performance across the entire phase diagram. In all our training and evaluation scenarios, the smart-agents introduction follows a transient time of $T_{tr}=200$ time-steps in which only the ABM-ants are acting in the environment, and their action continues till the end of the episode, at time-step $T=1200$. Below, we present both qualitative results and quantitative results  on the capability of the trained smart-agents population to control the ABM-ants towards trail-formation. Details on the number $n_{e}$ of evaluation tests for the different scenarios are included in the remaining parts of this section. To evaluate if the action of our policies effectively induces trails scenarios, we observe the following indicators, working as evaluation metrics:
\begin{enumerate}
    \item \textbf{Final individual density.} The order parameter $M(T)\in [-1, \, 1 ]$ (see Eq. \ref{eq:op-M}) measures how the population of ants is distributed at the final step $T$ of a simulation with respect to the pheromone field. It compares how many individuals are located in regions where the pheromone concentration is above average versus below average. Values close to $1$ indicate that most ants are positioned in high-pheromone regions, while values near or below $0$ indicate a more random or uncorrelated spatial distribution. This metric quantifies at which level the \textit{population condition} is verified. The maximum value of $M(T)$ attainable in simulations depends on the $\left[ \delta, \, \beta \right]$-configuration that characterizes the underlying ABM-ants population: when ants have low sensory capacity (governed by $\delta$) and experience high noise in pheromone signal reception (based on $\beta$), values of $M(T)$ close to the maximum cannot be reached. 

    \item \textbf{Final pheromone density.} The order parameter $m(T) \in [-1, \, 1 ]$ (see Eq. \ref{eq:op-m}) measures how the pheromone is distributed spatially in the environment at the terminal step $T$ of the simulation, comparing the total amount of pheromone present in regions above the average concentration with that in regions below it, normalized by the overall pheromone level. High positive values ($\sim 1$) indicate a strongly uneven, structured pheromone field, whereas small positive values or negative ones correspond to a more homogeneous and disordered distribution. Achieving high values of $m(T)$ verify the \textit{bimodal pheromone condition}: $m(T)$ measures how uniform is the distribution of pheromone in the environment. If both $M(T)$ and $m(T)$ achieve values greater than $0.8$ it is noticeable from empirical observation that the system presents an ordered phase, characterized either by pathways or by small clusters.

    \item \textbf{Trails scenario metrics.} To verify the emergence of trails scenarios (verifying the \textit{pheromone}, \textit{population}, and \textit{trail geometry conditions}), we introduce the global metric $\langle \mathcal{TS} \rangle_{50}$, obtained by averaging $\mathcal{TS}(t)$ (see Eq. \ref{eq:global-metric}) on the last $50$ steps of a simulation. $\mathcal{TS}(t)$ evaluates the overall quality of the pheromone and ABM-ants fields at time-step $t$ in terms of trails scenario formation, by combining information provided by $M(t)$ and $m(t)$, with a penalty defined to evaluate negatively the formation of highly branched and periodic structures. Being superiorly bounded by $0$, values of $\mathcal{TS}(t)$ that approach $0$ from left indicate configurations where pheromone is organized into thin, weakly branched, and well-connected trails (rather than dense or highly clustered patterns), favoring pheromone distribution  structured in distinct high- and low-concentration regions, and ABM-ants distribution concentrated along those high-pheromone paths. As noted in  Section \ref{sec:Materials_and_methods}.\ref{sec:Ants_swarmmodel} - \ref{sec:trails-metric}, in simulations where trails scenarios emerge, $\langle \mathcal{TS} \rangle_{50}$ is observed to be greater than an empirical threshold of $-1.6$.
\end{enumerate}

\subsection{Stigmergic functional control steers the system to order}
\label{sec:Learning_rand}
\begin{figure}[t]
    \centering
    \begin{subfigure}[b]{0.45\textwidth}
        \centering        \includegraphics[width=\textwidth]{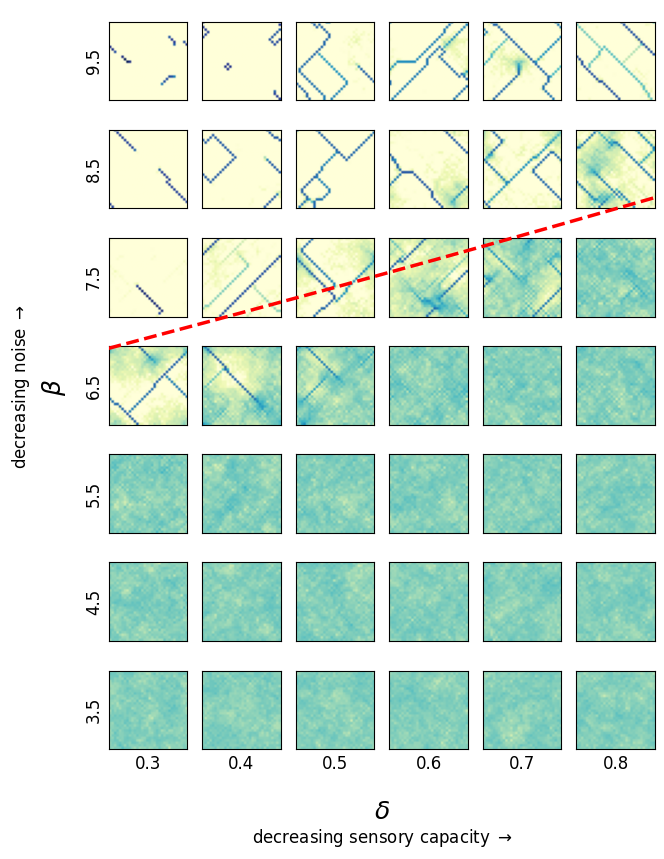}
        \caption{Final pheromone field (log-scale)}        \label{fig:pher_phaseplot}
    \end{subfigure}
    \hfill
    \begin{subfigure}[b]{0.45\textwidth}
        \centering        \includegraphics[width=\textwidth]{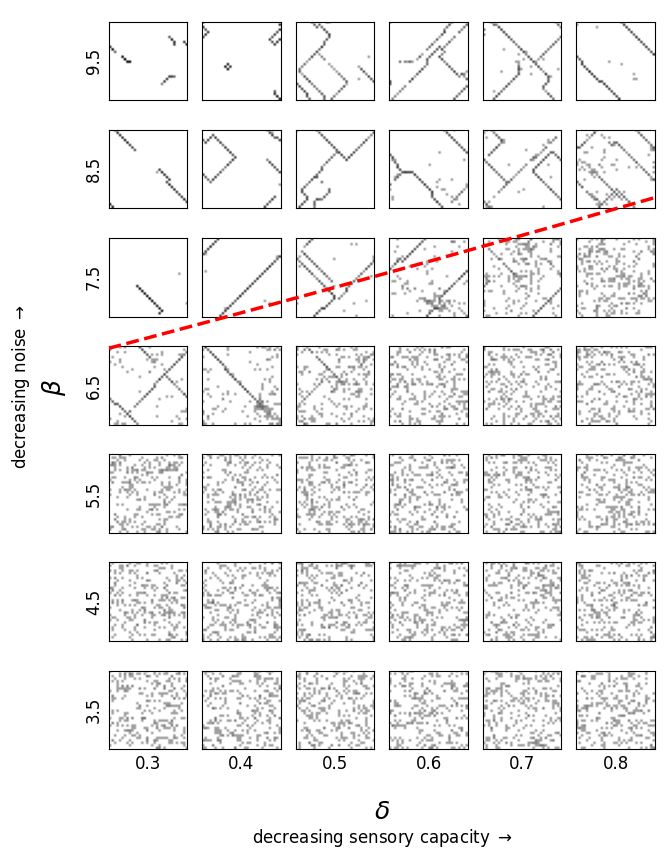}
        \caption{Final density field}        \label{fig:ants_phaseplot}
    \end{subfigure}
    \caption{\textbf{Baseline setup: final pheromone and individual configurations}. In Figure \ref{fig:pher_phaseplot} the final pheromone distributions show a transition from homogeneous to bimodal patterns across the phase transition line (in red). In Figure \ref{fig:ants_phaseplot} the corresponding individual final positions reflect the pheromone traces. Three distinct behaviours are observed: chaotic dynamics (low sensitivity, high noise), clustered aggregations (high sensitivity, low noise), and intermediate pathways formation.}
    \label{fig:phasegrid}
\end{figure}

We present results obtained by deploying a population of smart-agents trained to control a population of $N = 300$ ABM-ants. We consider smart-agents that were exposed during training to ABM-ants characterized by sensory parameters $\left[ \delta, \, \beta \right]$ sampled at each training episode from the predefined sets $\mathsf{s}_\delta$ and $\mathsf{s}_\beta$, defined in Section \ref{sec:Materials_and_methods}.\ref{sec:Ants_swarmmodel} - \ref{sec:sim_phasetrans}, spanning the phase diagram shown in Figure \ref{fig:phasegrid}. 
\begin{figure}[t]
    \centering
    \begin{subfigure}[b]{0.45\textwidth}
        \centering        \includegraphics[width=\textwidth]{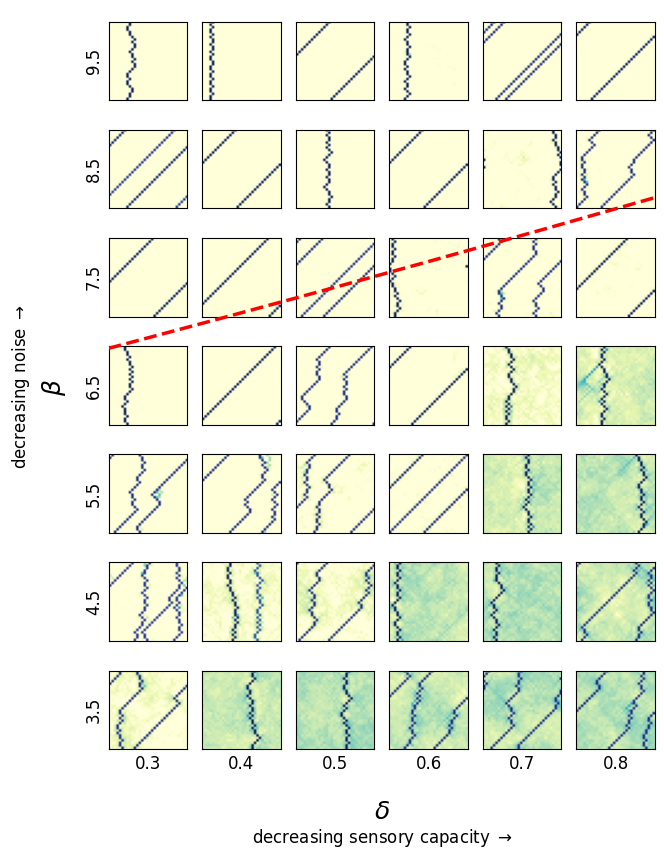}
        \caption{Final pheromone field (log-scale)}        \label{fig:pher_gridplot_rand}
    \end{subfigure}
    \hfill
    \begin{subfigure}[b]{0.45\textwidth}
        \centering        \includegraphics[width=\textwidth]{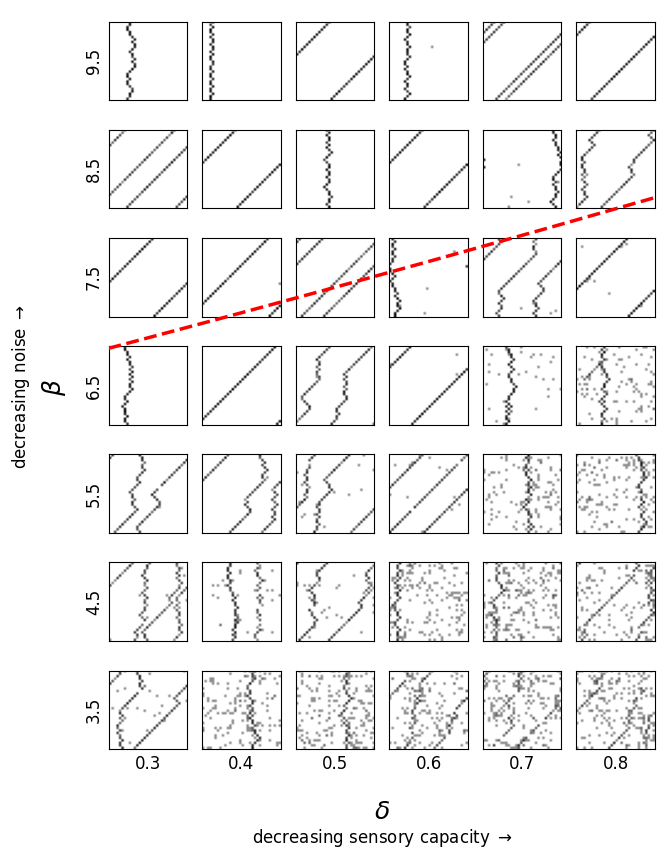}
        \caption{Final density field}        \label{fig:ants_gridplot_rand}
    \end{subfigure}
    \caption{\textbf{Smart setup: final pheromone and individual configurations}.
    In Figure \ref{fig:pher_gridplot_rand} smart-agent intervention shifts the transition boundary, enabling pheromone trails in every configuration. In Figure \ref{fig:ants_gridplot_rand} the ABM-ants follow consistently the pheromone traces even in lower sensitivity and higher noise scenarios (under the original phase transition line of the system, in red, obtained from simulations with only ABM-ants in the environment (see Figure \ref{fig:phasegrid}).}
    \label{fig:rand_gridplot}
\end{figure}
This allows the RL agents to learn from a diverse range of population behaviours (disordered motion, formed pathways, and clustered structures). A successful policy should generalize across different ants behaviours, efficiently guiding ABM-ants along robust trails, as best as possible, considering the underlying population features. However, for certain $\left[ \delta, \, \beta \right]$ pairs the control objective may become entirely infeasible. In the limiting case of very high noise (small $\beta$) and negligible sensory capacity (large $\delta$), ants are effectively unresponsive to the pheromone field and thus uncontrollable through stigmergic pheromone-based guidance, regardless of the policy applied. In less extreme but still not fully controllable scenarios, though, it is still possible to increase the amount of ordered behaviour observed.

 Section \ref{sec:learning_policy_var} contains the average evolution obtained in learning, while training $n_{\pi}$ policies to control the system. Here, instead, we evaluate the effect of our control scheme on the underlying ABM system. To qualitatively characterize such effect, we include visualizations of the final configuration achieved by the system under control across all the possible combinations of $\left[ \delta, \, \beta \right]$ obtainable by selecting values from the sets $\mathsf{s}_\delta$ and $\mathsf{s}_\beta$ defined in Section \ref{sec:Materials_and_methods}.\ref{sec:Ants_swarmmodel} - \ref{sec:sim_phasetrans}. In order to study how the controlled system is affected by the smart-agents embedded in the environment in a quantitative way, instead, we employ the introduced metrics to evaluate the average performance of the $n_{\pi}$ policies learned during training, each of them tested $n_{e}=42$ times (one per each $\left[ \delta ,\,\beta\right]$-configuration considered).

Per each $\left[ \delta, \, \beta \right]$-configuration, we deploy $N' = 30$ smart-agents and observe the behaviour of the new hybrid population. We indicate this experimental setup as \textbf{smart} setup. This enables a direct comparison with the original setup in which only ABM-ants populate the environment (\textbf{baseline} setup). To disentangle the effect of the learned movement policy from eventual consequences of a higher pheromone release from the smart-agents, we include a third simulation experiment, conducted in what we call the \textbf{enhanced} setup, consisting of $N=300$ ABM-ants and $N_{\eta=1} = 30$ ABM-ants that deposit pheromone at a release rate $\eta = 1$, matching the one of the smart-agents in the smart setup, but move according to the underlying ABM dynamics, governed by $\left[ \delta, \, \beta \right]$.

\paragraph{Trail formation.}
One example of the final snapshot of the smart setup test simulation is included in Figure \ref{fig:rand_gridplot}, showing a behaviour characterized by clearly formed pheromone trails in every $\left[ \delta, \, \beta \right]$-configuration and ants trails in a consistent portion of the phase diagram. In some of the scenarios in the bottom-right corner of the phase diagram the difference in pheromone quantity between locations on and off the formed trails is reduced, and a part of the ABM-ants still struggles to follow the path indicated by the trails. This is explainable thinking of the harsher condition of these scenarios, that are less controllable, being characterized by ABM-ants having reduced sensitivity to pheromone and high motion randomness. Notwithstanding this, the obtained behaviour in term of trail formation is sensibly different with respect to the one of the baseline setup shown in Figure \ref{fig:phasegrid}.
Section \ref{sec:additional_res_var} contains the results obtained by all the other policies: in $7$ out of the $10$ cases the learned policy is capable of analogous performance to the ones presented above, two cases are characterized by performance that, although good, occasionally present failures, and no trail formation in just one of the cases. Additionally, Section \ref{sec:learning_policy_var} - \ref{sec:additional_res_var} includes more insights on the learned behaviour.

\begin{figure}[t]
    \centering
    \begin{subfigure}[b]{0.45\textwidth}
        \centering        \includegraphics[width=\textwidth]{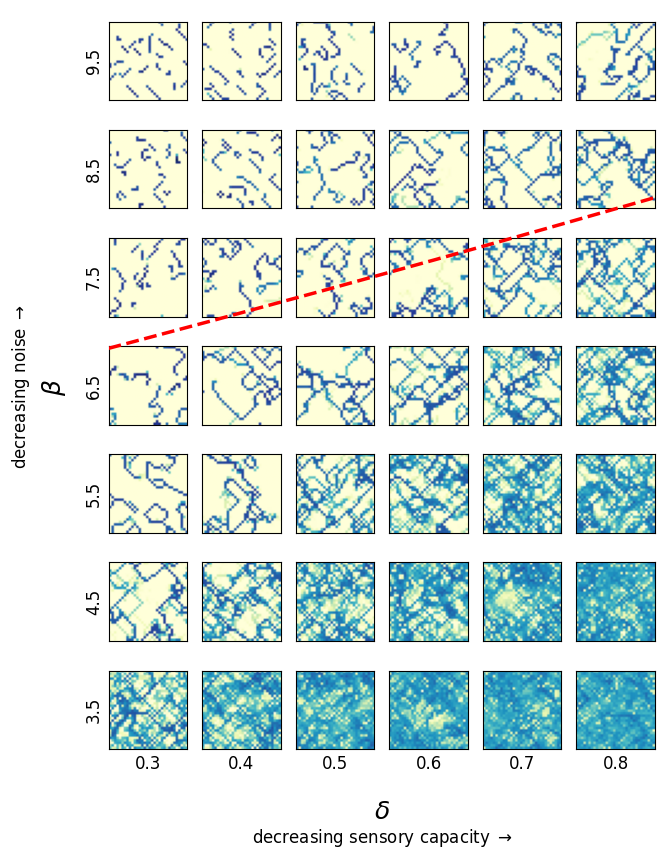}
        \caption{Final pheromone field (log-scale)}        \label{fig:pher_300_30_gridplot_rand}
    \end{subfigure}
    \hfill
    \begin{subfigure}[b]{0.45\textwidth}
        \centering        \includegraphics[width=\textwidth]{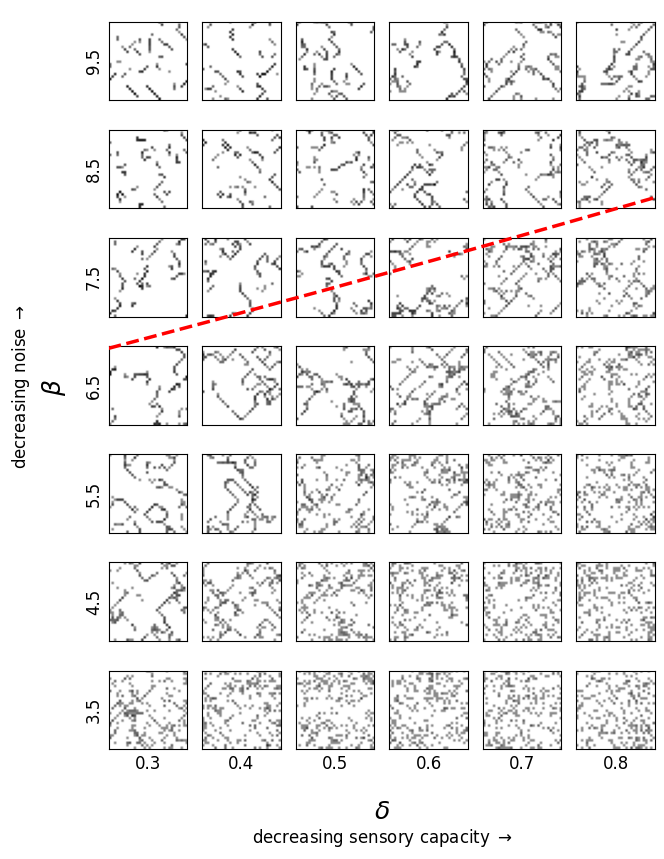}
        \caption{Final density field}        \label{fig:ants_300_30_gridplot_rand}
    \end{subfigure}
    \caption{\textbf{Enhanced setup: final pheromone and individual configurations}. The final pheromone and ant density fields show poorly connected and highly clustered structures.}

    \label{fig:300_30_gridplot}
\end{figure}

The direct effect of the learned movement policy in producing the behaviour shown in Figure \ref{fig:rand_gridplot} can be understood by comparison with the phase diagram associated with the enhanced setup, included in Figure \ref{fig:300_30_gridplot}: the final simulation snapshots in this case present structures of pheromone and ants that are strongly connected but highly clustered, demonstrating that the emergence of trails is not merely a consequence of a higher pheromone release from the smart-agents, but directly related to the chosen moves of the smart-ants in the environment.\\

\begin{figure}[b]
    \centering
    \includegraphics[width=0.7\linewidth]{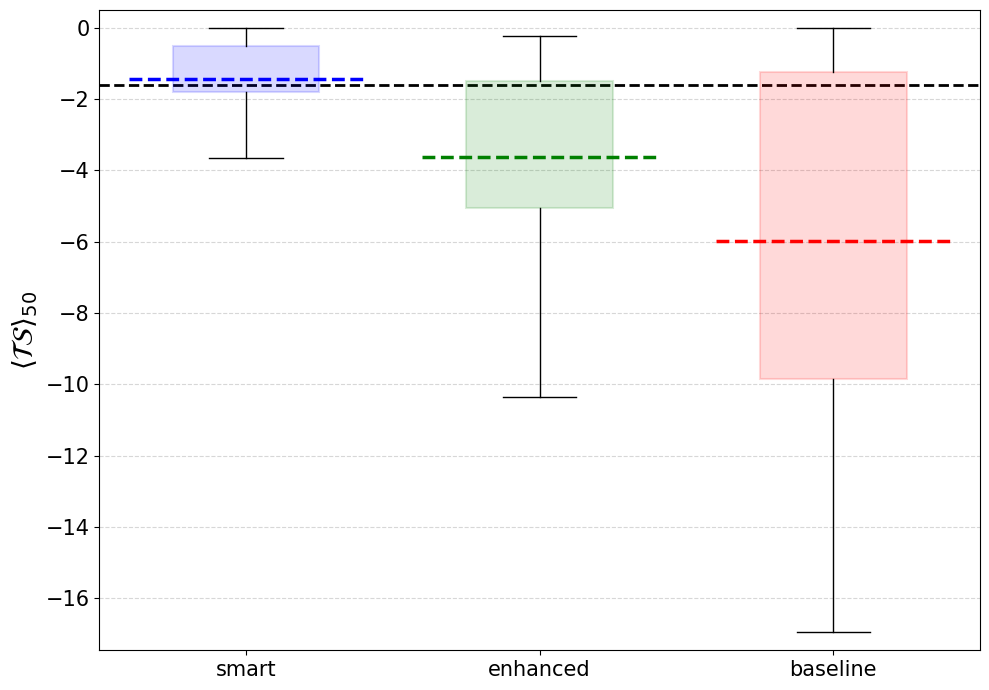}
    \caption{\textbf{Trail scenario metric}. Values of $\langle \mathcal{TS} \rangle_{50}$ over all the simulated evaluations, for the baseline, smart and enhanced setups. The horizontal dashed line represents the empirical threshold of $-1.6$, indicating trail formation.}
    \label{fig:boxplot-aggregated-ST50}
\end{figure}

These observations are quantitatively supported by 
 the values of the trails scenario metric 
 $\langle \mathcal{TS} \rangle_{50}$, averaged over $n_{\pi}$ training campaigns and $n_{e}$ evaluation episodes, with $\left[ \delta, \, \beta \right]$ spanning the phase diagram. 
 The metrics show a substantial increase when moving from the baseline configuration to the smart approach, achieving $\text{avg}\,\,\langle \mathcal{TS} \rangle_{50} = -5.99$ in the baseline setup ($\text{std}\,\,\langle \mathcal{TS} \rangle_{50} = 4.44$), $\text{avg}\,\, \langle \mathcal{TS} \rangle_{50} = -3.63$ in the enhanced one ($\text{std}\,\,\langle \mathcal{TS} \rangle_{50} = 2.71$), and $\text{avg}\,\, \langle \mathcal{TS} \rangle_{50} = -1.43$ in the smart one ($\text{std}\,\,\langle \mathcal{TS} \rangle_{50} = 1.42$). The advantage reflected by the average value of the metric is robustly present across the whole simulation battery performed for this study, as shown in Figure \ref{fig:boxplot-aggregated-ST50}. 
 
The improvement achieved by the smart setup with respect to the baseline is significantly larger than the one observed for the enhanced setup. The average value of $\langle \mathcal{TS} \rangle_{50}$ achieved in the smart setup is the only one that trespasses the empirical threshold of $-1.6$, providing statistically robust evidence for our claim: simply introducing ABM-ants that release additional pheromone, without ensuring optimal movement, does not lead to the formation of trails scenarios in the sense of our definition, as already suggested by the qualitative analysis above, justifying our policy learning effort. 

Section \ref{sec:additional_res_var} includes in Figures \ref{fig:global50_metric} and \ref{fig:boxplot-delta-beta-ST50} the average values of $\langle \mathcal{TS} \rangle_{50}$ achieved separately per each $\left[ \delta, \, \beta \right]$-configuration in the three setups, the associated visual representation in the form of phase diagrams, and the disaggregated results presenting the statistical distribution of $\langle \mathcal{TS} \rangle_{50}$ for each of the $\left[ \delta, \, \beta \right]$-configurations .
\clearpage
\clearpage
\paragraph{Lower phase transition.}
As it is noticeable comparing Figure \ref{fig:phasegrid} and Figure \ref{fig:rand_gridplot}, many parameter configurations that lie below the original transition line (in red in both figures) are reaching a stable ordered phase in the smart setup thanks to the smart-agent intervention. This action establishes a new transition threshold, sensibly lower than the original one. In particular, for high sensory capacity $\delta$, order appears stably for all the considered values of $\beta$.

However, in the ungovernable high noise and low sensitivity region at the bottom right corner of the phase diagram, although smart-agents promote a bimodal distribution of pheromone, the ABM-ants are not provided with full capabilities in following the pheromone trace consistently over time, coherently with the limitations encoded in the $\delta$ and $\beta$ parameters characterizing them. Hence, the configurations located there do not show a fully ordered phase. 

This order-promoting effect of the proposed control strategy can be quantified through the order parameters $m(T)$ and $M(T)$, averaged over $n_{\pi}$ training campaigns and $n_{e}$ evaluation episodes, per each $\left[ \delta, \, \beta \right]$-configuration. The corresponding values are reported in Table \ref{tab:order_par_tab1} (the distribution of the two metrics across the evaluation testbed is included in Section \ref{sec:additional_res_var}, in Figures \ref{fig:boxplot-delta-beta-mT}-\ref{fig:boxplot-delta-beta-MT}, showing the statistical significance of our results). In the table, we highlight in bold the average values of $m(T)$ and $M(T)$ that exceed the empirical threshold of $0.8$. As discussed in Section \ref{sec:Materials_and_methods}.\ref{sec:Ants_swarmmodel} - \ref{sec:sim_phasetrans}, when both $m(T)$ and $M(T)$ are above this threshold, we observe that the system presents an ordered phase. From the table it is visible that the smart setup presents a much higher number of $\left[ \delta, \, \beta \right]$-configurations with both average order parameters above $0.8$ with respect to the baseline ($28/42$ in the smart setup, and $13/42$ configurations in the baseline setup). In particular, the control action exerted by the smart-agents induces an increase of both $m(T)$ and $M(T)$ above the threshold in several configurations that lie below the phase transition line associated with the baseline setup. The smart configuration achieves consistently high values of $m(T)$ across all $\left[ \delta, \, \beta \right]$-configurations, often approaching the maximum value $m(T)=1$. This indicates the emergence of a highly uneven and structured pheromone field, in agreement with the observed formation of well-defined pheromone trails (the \textit{pheromone condition} is satisfied on average in all the configurations). The values of $M(T)$ also raise above $0.8$ in many configurations below the original transition line, suggesting that a larger number of scenarios characterized by the formation of structured pheromone traces are accompanied by a significant presence of ABM-ants following them, compared to the baseline case (the \textit{population condition} is satisfied on average in more configurations than in the baseline case). 
\begin{table}[t]
\centering
\resizebox{0.8\textwidth}{!}{
\begin{tabular}{|l|rrrrrr|rrrrrr|}
\toprule 
$\quad$ & $\quad$ & $\quad$ & $m(T)$ & $\quad$ & $\quad$ & $\quad$ & $\quad$ & $\quad$ & $M(T)$ & $\quad$ & $\quad$ & $\quad$\\
\midrule
\diagbox[innerwidth=0.4cm]{$\beta$}{$\delta$} & 0.3 & 0.4 & 0.5 & 0.6 & 0.7 & 0.8 & 0.3 & 0.4 & 0.5 & 0.6 & 0.7 & 0.8 \\
\midrule
$\quad$ & \multicolumn{2}{l}{baseline} & $\quad$ & $\quad$ & $\quad$ & $\quad$ & $\quad$ & $\quad$ & $\quad$ & $\quad$ & $\quad$ & $\quad$  \\
\midrule
9.5 & \textbf{1.00} & \textbf{1.00} & \textbf{0.98} & \textbf{0.91} & \textbf{0.99} & \textbf{0.91} & \textbf{1.00} & \textbf{1.00} & \textbf{0.95} & \textbf{0.87} & \textbf{0.99} & \textbf{0.81} \\
8.5 & \textbf{0.98} & \textbf{1.00} & \textbf{0.99} & \textbf{0.98} & \textbf{0.85} & \textbf{0.95} & \textbf{0.97} & \textbf{1.00} & \textbf{0.99} & \textbf{0.99} & 0.72          & \textbf{0.96} \\
7.5 & \textbf{1.00} & \textbf{0.86} & \textbf{0.97} & 0.77          & 0.40          & 0.33          & \textbf{0.99} & 0.76          & \textbf{0.99} & 0.69          & 0.18          & 0.25          \\  
6.5 & 0.73          & 0.72          & 0.28          & 0.25          & 0.22          & 0.22          & 0.55          & 0.49          & 0.19          & 0.23          & 0.26          & 0.26          \\
5.5 & 0.22          & 0.21          & 0.21          & 0.19          & 0.18          & 0.18          & 0.16          & 0.14          & 0.17          & 0.24          & 0.27          & 0.21          \\
4.5 & 0.19          & 0.19          & 0.17          & 0.18          & 0.18          & 0.19          & 0.29          & 0.17          & 0.27          & 0.19          & 0.26          & 0.15          \\
3.5 & 0.17          & 0.16          & 0.17          & 0.16          & 0.16          & 0.17          & 0.27          & 0.27          & 0.09          & 0.25          & 0.18          & 0.16          \\
\midrule
$\quad$ & \multicolumn{3}{l}{smart} & $\quad$ & $\quad$ & $\quad$ & $\quad$ & $\quad$ & $\quad$ & $\quad$ & $\quad$ & $\quad$   \\
\midrule
9.5 & \textbf{1.00} & \textbf{0.99} & \textbf{0.99} & \textbf{0.99} & \textbf{1.00} & \textbf{1.00} & \textbf{0.99} & \textbf{0.92} & \textbf{0.94} & \textbf{0.95} & \textbf{0.95} & \textbf{0.94} \\
8.5 & \textbf{0.99} & \textbf{1.00} & \textbf{0.99} & \textbf{0.99} & \textbf{0.99} & \textbf{0.99} & \textbf{0.87} & \textbf{0.99} & \textbf{0.99} & \textbf{0.92} & \textbf{0.95} & \textbf{0.94} \\
7.5 & \textbf{0.99} & \textbf{0.99} & \textbf{1.00} & \textbf{0.99} & \textbf{0.99} & \textbf{0.99} & \textbf{0.94} & \textbf{0.98} & \textbf{0.97} & \textbf{0.92} & \textbf{0.92} & \textbf{0.82} \\
6.5 & \textbf{0.99} & \textbf{0.99} & \textbf{0.99} & \textbf{0.99} & \textbf{0.97} & \textbf{0.97} & \textbf{0.99} & \textbf{0.98} & \textbf{0.94} & \textbf{0.87} & \textbf{0.84} & 0.70          \\
5.5 & \textbf{1.00} & \textbf{0.99} & \textbf{0.99} & \textbf{0.97} & \textbf{0.98} & \textbf{0.96} & \textbf{0.98} & \textbf{0.95} & \textbf{0.90} & 0.72          & 0.63          & 0.44          \\
4.5 & \textbf{0.99} & \textbf{0.99} & \textbf{0.97} & \textbf{0.95} & \textbf{0.94} & \textbf{0.93} & \textbf{0.92} & \textbf{0.84} & 0.64          & 0.51          & 0.27          & 0.15          \\
3.5 & \textbf{0.98} & \textbf{0.97} & \textbf{0.93} & \textbf{0.92} & \textbf{0.93} & \textbf{0.92} & 0.69          & 0.45          & 0.30          & -0.04         & -0.08         & -0.23         \\
 \bottomrule
\end{tabular}
}
\caption{\textbf{Average order parameters}: baseline vs smart setup.}
 \label{tab:order_par_tab1}
\end{table}

The smart setup shows low $M(T)$ values in three extreme configurations, located in the bottom-right region of the phase diagram. This is likely due to the limited controllability of systems with low pheromone sensitivity and high movement randomness. Overall, these results highlight that, in the smart setup, the control action implemented through stigmergic mechanisms leads to the formation of highly defined pheromone structures, as previously observed. However, such control cannot fully compensate for the intrinsic properties of certain configurations. The lower values of $M(T)$ achieved by the smart setup with respect to the  baseline in such configurations can be interpreted considering that, in the baseline setup, lower values of $m(T)$ correspond to a more uniform distribution of pheromone across the environment. When combined with the high randomness in agent movement typical of these configurations, this leads to an increased probability that ABM-ants traverse pheromone-containing locations, thereby artificially inflating $M(T)$.

\subsection{Learning general control from specific configurations}
\label{sec:learning_fix}
The results included in this section are obtained by deploying a population of $N'=30$ smart-agents trained to control a population of $N = 300$ ABM-ants characterized by only one of the configurations of sensory parameters from the phase diagram: $\left[ \delta^*=0.3,\, \beta^*=4.5 \right]$. This scenario is selected to showcase that training smart-agents in a single configuration, if chosen with reasonable characteristics, can lead to learning control capabilities that generalize across configurations spanning the whole phase diagram. Specifically, we target configurations that avoid regimes dominated by noise, even at high pheromone concentrations (bottom-right corner), so that smart-agents can learn from experience in which actual trails scenarios can be formed. Simultaneously, we exclude scenarios where the system's inherent self-organization makes additional control redundant, avoiding to over-expose them to scenarios in which their action results superfluous.

\begin{figure}[t]
    \centering
    \includegraphics[width=0.7\linewidth]{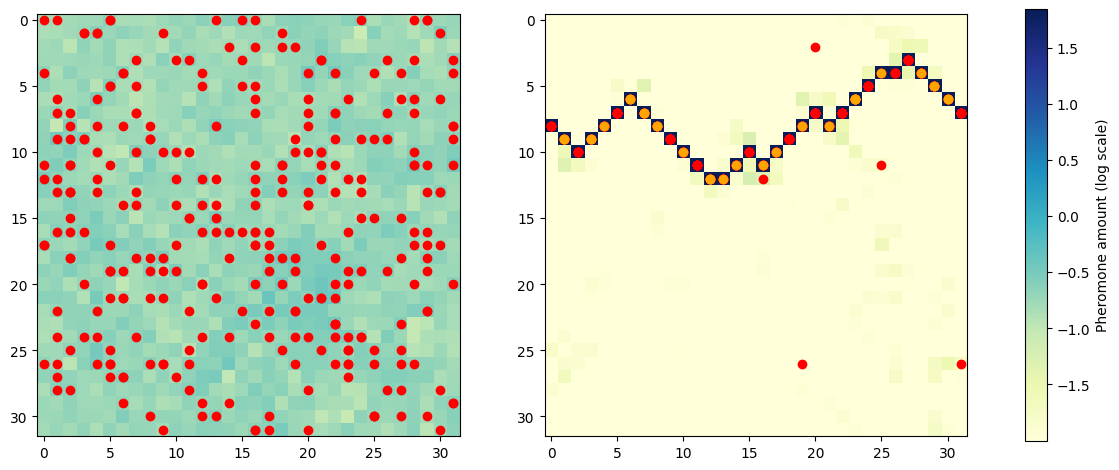}
    \caption{\textbf{System with $\left[ \delta^*,\, \beta^* \right]$ without and with control: example of behaviour}. On the left, ABM-ants (in red) after $t=T$ time-steps are not able to self-organize. With the addition of the smart-agents (in orange), at $t=T$ a trail emerges in the same scenario, containing all the smart-agents and the majority of the ABM-ants.}
    \label{fig:single_control}
\end{figure}

Without smart-agents controlling the system, ABM-ants characterized by $\left[ \delta^*=0.3,\, \beta^*=4.5 \right]$ fail to form trails or clusters and instead exhibit disordered behaviour, as shown on the left side of Figure \ref{fig:single_control}. On the other hand, in this configuration, the ABM-ants present a good sensory capacity (the inverse of $\delta^*$ is small), but highly noisy movements (the inverse of $\beta^*$ is small). Hence, the trail construction failure is probably due to a too weak pheromone field ($\eta= 0.01$) to contrast the level of noise. On the right side of Figure \ref{fig:single_control}, we demonstrate that, by adding to the system the $N'$ smart-agents trained to control it, the population achieves an ordered behaviour, with individuals organized in trails. 

\begin{figure}[t]
    \centering
    \begin{subfigure}[b]{0.45\textwidth}
        \centering
        \includegraphics[width=\textwidth]{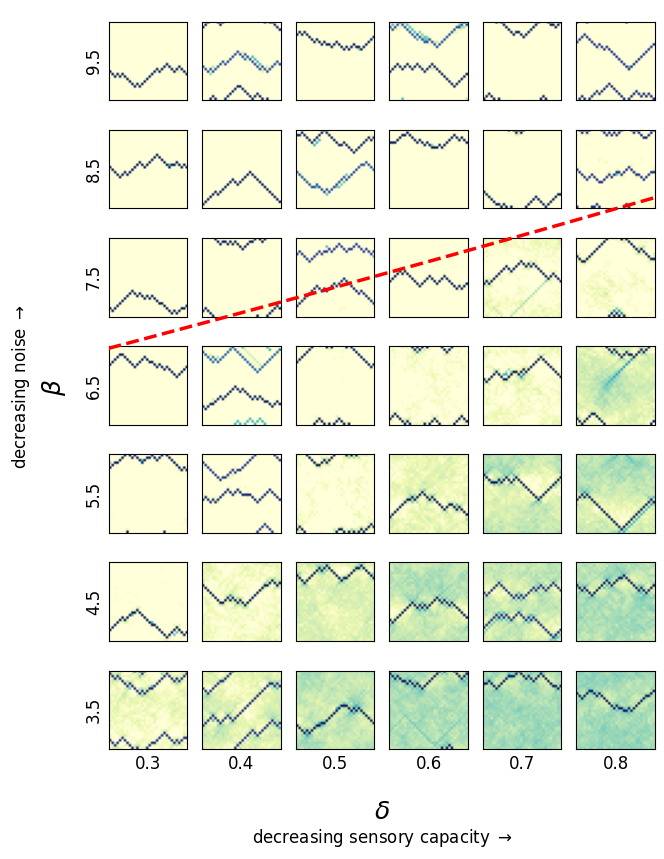}
        \caption{Final pheromone field (log-scale)}
        \label{fig:pher_gridplot}
    \end{subfigure}
    \hfill
    \begin{subfigure}[b]{0.45\textwidth}
        \centering
        \includegraphics[width=\textwidth]{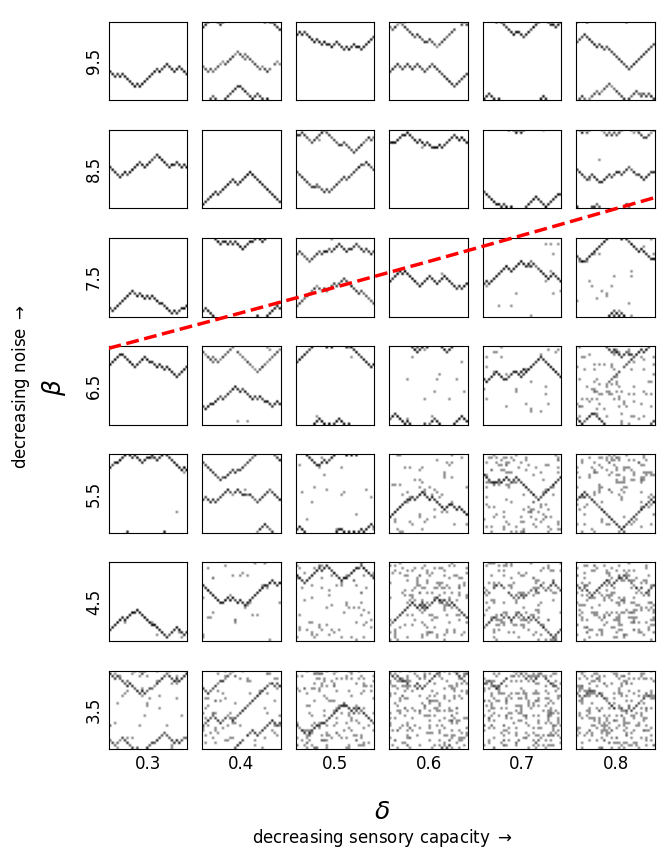}
        \caption{Final density field}
        \label{fig:ants_gridplot}
    \end{subfigure}

    \caption{
    \textbf{Smart setup learned in the $\left[ \delta^*=0.3,\, \beta^*=4.5 \right]$ scenario: final pheromone and individual configurations}. In Figure \ref{fig:pher_gridplot} trails emerge across all parameter settings. By consequence, in Figure \ref{fig:ants_gridplot}, the ABM-ants reliably track existing trails up to a shifted threshold, as observed in Figure \ref{fig:rand_gridplot}.}
    \label{fig:nonrand_gridplot}
\end{figure}

\paragraph{Generalized control capabilities.}
First, we qualitatively analyze the effect of deploying $N' = 30$ smart-agents that move according to a policy learned under the environment characterized by parameters $[\delta^*,\,\beta^*]$. Specifically, we select one of the $n_{\pi}$ trained policy realizations and evaluate its performance across all the different $\left[ \delta, \, \beta \right]$-configurations. The corresponding final snapshot of the test simulations are shown in Figure \ref{fig:nonrand_gridplot}. Section \ref{sec:learning_policy_spec} contains the learning evolution, while Section \ref{sec:additional_res_spec} the results obtained by all the other policies: in $8$ of the $10$ cases the learned policy is capable of analogous performance to the ones presented above, one of the cases is characterized by performance that, although good, occasionally present failures, and no trail formation in only one case. 

Notably, the phase transition between disordered and ordered collective motion is preserved across the parameter space, and the configurations that benefit from the smart-agents intervention are nearly identical to those observed in the previous learning scenario. This suggests that the policy learned at $[\delta^*,\,\beta^*]$ retains sufficient generality to influence the collective dynamics in a consistent manner, even when deployed under environmental conditions that differ from those encountered during training. Furthermore, inspection of the figure reveals that the shape of the trails remains remarkably uniform across the phase diagram, a feature that likely reflects the fact that the underlying policy was learned in a single fixed configuration, since the agents never adapted their behaviour to varying sensory conditions during training. 
As previously done, we can obtain a quantitative evaluation of our control policy by computing the average value of the metrics of interest over $n_{\pi}$ policy realizations and $n_{e} = 42$ evaluations. In this scenario we obtain average trails scenario metrics values $\text{avg}\,\,\langle \mathcal{TS} \rangle_{50} = -1.44$, $\text{std}\,\,\langle \mathcal{TS} \rangle_{50} = 1.90$. These results  are comparable to those observed for ABM-ants and smart-agents trained in full phase space (i.e., $\text{avg}\,\, \langle \mathcal{TS} \rangle_{50} = -1.43$, $\text{std}\,\,\langle \mathcal{TS} \rangle_{50} = 1.42$), as already reported in Section \ref{sec:Results}.\ref{sec:Learning_rand}, confirming that a policy learned in a single fixed configuration can be sufficient to achieve competitive performance across a wide range of environmental conditions, without the need for explicit adaptation during training.
\clearpage
\begin{figure}[b]
    \centering
    \includegraphics[width=0.7\linewidth]{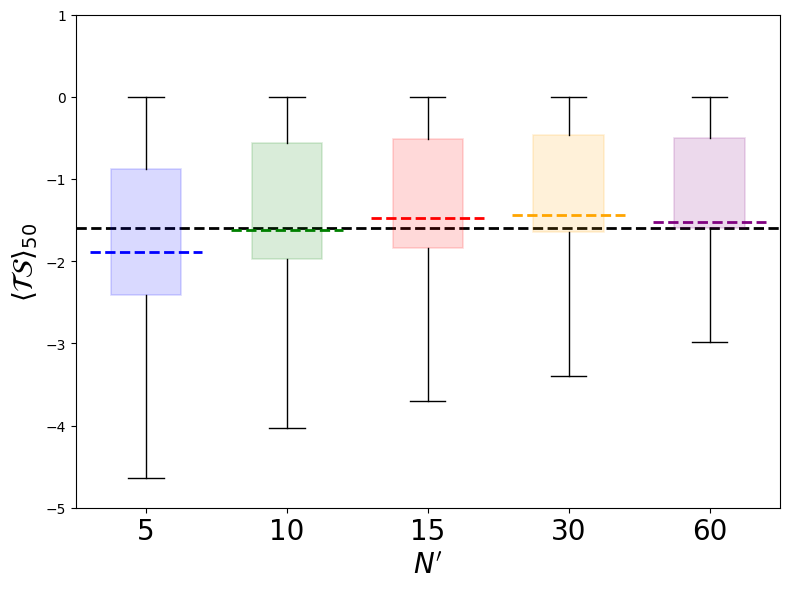}
    \caption{\textbf{Trail scenario metric}. Values of $\langle \mathcal{TS} \rangle_{50}$ in the smart setup with varying number $N'$ of smart-agents. The horizontal dashed line represents the empirical threshold of $-1.6$, indicating trail formation.}
    \label{fig:Nprime_var}
\end{figure}
\clearpage

\paragraph{Varying the number of smart-agents deployed.} In order to study how much the controlled system is affected by the number agents with smart behaviour embedded in the environment, we conduct an ablation study, observing the effect of varying the number $N'$ of smart-agents in the environment. We evaluate each of the $n_{\pi}$ policies learned in the fixed $[\delta^*,\,\beta^*]$ case, on a new batch of $n_{e}$ tests, across all the  $[\delta,\,\beta]$-configurations. The results are presented in Figure \ref{fig:Nprime_var}, indicating that the presence of 
half ($N'=15$) of the smart-agents that were used to learn the smart behaviour in training is sufficient, in order to produce a successful trail. One third ($N'=10$) of smart-agents results on average to be close at achieving trails scenarios, with average value $\text{avg}\,\, \langle \mathcal{TS} \rangle_{50}$ approaching the value of the empirical threshold $-1.6$,  and hence signaling the presence of both successful and not successful scenarios. With $N'=5$ smart-agents, the pheromone gets distributed in trails, but in a quantity that is not sufficient to contrast the pheromone decay, the effect of random pheromone emission by the ABM-ants population, and the stochasticity in the ABM-ants movements. Doubling the number of agents at test time does not lead to a significant improvement in the evaluation metrics, as the performance achieved with $N'=30$ agents is already close to the best observed. This result is reasonable, since the policies were trained on experience collected through the interactions of $N'=30$ agents, and therefore their performance saturates around this number.
\section{Discussion}
\label{sec:Discussion}
Our results, by showing how policies learned through RL can robustly induce collective organization in stochastic swarm environments, aim at contributing to the growing body of work on RL-based control of complex systems, a still relatively recent and sparsely explored research direction. The presented findings suggest that global collective behaviour can be steered by a relatively small number of agents, equipped with a learned control policy relying only on local observations, without requiring explicit knowledge of the underlying system dynamics, or global system information. Such a paradigm is particularly attractive for swarm robotics, where scalable and robust coordination must emerge from local interactions among autonomous agents operating in uncertain environments. Recent advances in RL-based swarm control and collective robotics highlight the growing importance of learning strategies for tasks such as formation control, exploration, and adaptive coordination \cite{BLAIS2023226, drones7110673, sadeghi2024reinforcement, na2023federated}. More broadly, the idea of influencing macroscopic outcomes through localized interventions may also be relevant in socio-technical systems where collective dynamics emerge from interactions among large populations of agents, such as human mobility, social coordination, and distributed adaptive systems. Recent studies have shown that RL-based decision makers can successfully promote cooperation in human groups through targeted modifications of interaction networks \cite{McKee2023}, while related work on information ecosystems demonstrates how interventions at the level of local interactions can substantially alter large-scale collective outcomes \cite{Bak-Coleman2022, Toyokawa2019}.

Recent developments at the intersection of RL and collective dynamics further contextualize our work. A recent but rapidly growing research area is devoted to the control of active matter and swarm systems through learning-based approaches \cite{reviewRL4AM, tovey2024swarmrlbuildingfuturesmart}. Early studies showed that RL can be used to optimize microscopic interaction rules and collective responses in swarming systems, demonstrating the potential of machine-learning methods for controlling emergent behaviour in active matter \cite{palmer2017optimizingcollectivefieldtaxisswarming, ABPEIKAR2022101085, learntoflock, learntotornado}. More recently, RL has been employed to train active particles and colloidal agents to perform collective tasks such as efficient foraging, where cooperative behaviours emerge despite rewards being assigned at the individual level \cite{Loffler2023}. These findings suggest that RL can uncover powerful local strategies that exploit collective effects without requiring explicit programming of interaction rules. Within this broader landscape, our work addresses a distinct problem: rather than generating collective behaviour from scratch, RL agents are embedded within a stigmergic system that can self-organize under specific environment conditions, and learn how to influence its phase behaviour through purely local interactions with both the population and the environmental field. The observed shift of the order–disorder transition and the emergence of trail networks therefore contribute to an emerging line of research investigating how learned controllers based on local observations can exploit and steer the macroscopic dynamics of complex systems.

Beyond RL, methods such as genetic and evolutionary algorithms have been used for coordination and self-organization in complex systems \cite{evolvingrobot, evoactivematter, evostrategies}. These approaches evolve behavioural policies through population-based search, using replication and mutation. However, they often demand significant computational resources and may converge slowly or get stuck in local optima. In contrast, RL is emerging in the field as a framework for learning control policies through direct interaction with the environment, offering new opportunities for managing complexity in high-dimensional and nonlinear domains \cite{controlchaos}. 
Yet, multi-agent RL can face difficulties like sample inefficiency, convergence, and credit assignment, particularly when rewards are shared across agents \cite{MARL}.

Our results show that embedding RL agents in a highly stochastic swarm environment such as the ants swarm model can effectively promote the emergence of collective order. To achieve this, we introduced a centralized-training decentralized-execution multi-agent RL framework together with an individual reward design, based on local information, tailored to trails scenario formation. These two aspects of our method aim at solving convergence and sample inefficiency issues typical of the decentralized training of multi-agent RL systems, and credit assignment problems, respectively. In particular, we show that introducing a limited number of RL-driven smart agents trained within our proposed framework enables the ant swarm system to reach the ordered phase even in regions of parameter space where, in their absence, order does not emerge. Indeed, systematic evaluation across the full phase diagram shows that RL agents shift the phase transition, enlarging the parameter region in which ordered, trail-forming states arise. This demonstrates that the agents actively shape the swarm dynamics, rather than simply amplifying pre-existing ordered states. Importantly, the agents operate without explicit knowledge of either the sensory field or the ABM movement rules, learning effective navigation strategies solely through local interactions and exploration. 

We further show that trail formation cannot be explained solely by increased pheromone deposition. Although allowing a subset of agents to release more pheromones can induce global ordering, it does not generate stable and well-structured trails, as quantified by the proposed global metric. This highlights the importance of the learned navigation policy in producing coherent trail dynamics. Moreover, by testing the effect of a policy trained in a single fixed configuration, we demonstrate that it generalizes successfully across the entire phase diagram without further adaptation. This suggests that it captures transferable principles of navigation and coordination that remain effective under substantially different environmental conditions. Finally, we find that the influence of RL agents is not monotonic with their number: a minimum fraction is required to induce ordering, while performance saturates beyond a threshold. This indicates that large-scale coordination can be achieved through control of only a small subset of the population.
Overall, our framework relies exclusively on local information about the field and neighboring agents, suggesting that it may generalize naturally to other systems (partially or purely stigmergic) based on indirect interactions through a shared environment. Despite these promising results, acknowledging our study's limitations can help guide future research. A primary constraint is that the smart agents currently operate within a simulated environment characterized by a fixed lattice topology, predefined pheromone dynamics, and a constant population size. Investigating variations in these foundational assumptions, such as introducing agent heterogeneity, dynamic environmental conditions, or alternative interaction mechanisms, represents an essential step to assess the boundaries of the learned policies. Furthermore, while the learned strategies demonstrate remarkable generalization across a wide portion of the phase diagram, our findings highlight specific parameter regimes where functional controllability remains fundamentally limited. These regions, characterized by very low pheromone sensitivity and high motion randomness, present severe challenges for macroscopic steering; establishing the precise theoretical limits of controllability in such high-noise regimes remains an open question requiring deeper investigation.
Additionally, although the ant swarm model provides a robust benchmark with analytically characterized phase transitions, the extent to which these learned control strategies transfer to other classes of complex systems remains to be fully validated. A natural extension of this framework would involve its application to alternative systems governed by indirect interactions, including active matter models. Future work could also explore alternative reward formulations and richer optimization schemes targeting collective behaviours beyond trail formation, such as system resilience, exploration efficiency, or robustness against external perturbations. From a theoretical perspective, a more comprehensive characterization of the interplay between learned control policies and underlying phase transitions could offer profound new insights into the functional controllability of complex distributed systems.

\section{Materials and Methods}
\label{sec:Materials_and_methods}

\subsection{Ants Swarm Model}
\label{sec:Ants_swarmmodel}
Swarm models represent systems in which individual agents co-exist in an environment, and interact indirectly through a shared field \cite{patternformation}. These models are specific instances of stigmergic systems \cite{stigmergy}, in which coordination among agents emerges from indirect interactions, mediated by modifications of a shared environment rather than from direct communication or centralized control. Examples of stigmergic dynamics span a diverse range of domains, from natural environments 
\cite{herdability, slimemold} to digital ones 
\cite{wikistigmergy}. 

In swarm models, the agents' motion is influenced by a scalar field $\sigma(\pmb{x}, t)$, defined over space $\pmb{x} \in \mathbb{R}^d$ and time $t \in \mathbb{R}_{0}^{+}$. The field evolves dynamically, due both to agents performing field reinforcement, and environmental processes, such as temporal decay, as represented by the following decay differential equation with agent contributions
\begin{equation}
\label{Eq:sigma_xt}
    \frac{\partial \sigma(\pmb{x},t)}{\partial t} = \eta \,\rho(\pmb{x},t) - \kappa \, \sigma(\pmb{x},t),
\end{equation}
where $\rho(\pmb{x},t)$ is the density of individuals at time $t$ in location $\pmb{x}$, $\eta$ represents the rate at which each agent reinforces the field $\sigma(\pmb{x}, t)$, and $\kappa$ is the decay rate, encoding the rate at which $\sigma(\pmb{x}, t)$ fades in time. The agents' dynamics respect the Langevin equation below
\begin{equation}
\label{Eq:langevin}
    \ddot{\pmb{x}} = - \Gamma \dot{\pmb{x}} - \nabla U[\sigma(\pmb{x},t), \, \delta] + \pmb{\xi}(\beta,t),
\end{equation}
where $\Gamma$ is a friction coefficient, and $U[\sigma(\pmb{x},t), \, \delta]$ is a potential encoding the dependence of the agents' motion on $\sigma(\pmb{x}, t)$ (and possibly on agents' sensory parameter $\delta$). The term $\pmb{\xi}(\beta,t)$ is Gaussian noise with variance determined by the noise strength $1/\beta$.\\

The ants swarm model \cite{chialvoswarms} is a swarm model representing the dynamics of population of real ants observed in laboratories, while they follow and release pheromone traces. In this case, the field $\sigma(\pmb{x}, t)$ represents the pheromone field that is used by ants populations to coordinate. The potential function characterizing the ants motion has been modeled as
\begin{equation}
    U[\sigma(\pmb{x},t), \, \delta] = -\ln\left(1 + \frac{\sigma(\pmb{x},t)}{\, 1 + \, \delta \, \sigma(\pmb{x},t)\, }\right),
\end{equation}
where $\delta$ represents the inverse of the sensory capacity of the ants. For lower values of $\delta$, the ants are encouraged to follow pheromone traces, while larger values dampen this effect.

As shown in \cite{patternformation, SwarmPhaseTrans}, by studying the system in the thermodynamic limit (with infinite volume and number of ants, but fixed global density of ants $\langle\rho\rangle$), the following stability criterion for the disordered phase has been derived analytically:
\begin{equation}
\label{Eq:stabcrit}
    \beta -(1 + \frac{1}{\sigma_0} + 2\delta + \delta \sigma_0 + \delta^2 \sigma_0) > 0
\footnote{This criterion is derived perturbing the equilibrium solution $\sigma_0$ of the pheromone field dynamics \eqref{Eq:sigma_xt}, under the assumption that the evolution of this field is slow in comparison to the relaxation time of the particle density, which instead can be considered in its equilibrium as $\rho_{eq}(\pmb{x},t) \propto \exp(-\beta \, U[\sigma(\pmb{x},t), \, \delta]).$
}.
\end{equation}
In Eq. \ref{Eq:stabcrit}, $\sigma_0 = \langle\rho\rangle\frac{\eta}{\kappa}$ is the average pheromone quantity in the system at the equilibrium. This criterion translates in a phase transition line in the hyper-parameters space that divides the random behaviour from the emergence of collective behaviour, with individuals spontaneously organizing into preferred paths. By varying the model parameters, it is possible to observe the associated model transitioning from a disordered phase, characterized by a nearly uniform pheromone distribution, to an ordered phase, where the field exhibits a bimodal distribution. 

\subsubsection{Discretized model}
\label{sec:discretized_model}
We simulate the ants dynamics in $\mathbb{R}^2$ by employing the discrete ABM introduced in \cite{chialvoswarms}, characterized by discretized space and time.
The space is partitioned as a two dimensional eight-neighbor square lattice $V$ with side length $L_V$ and periodic boundary conditions. Per each site $i = (x_i, y_i)$ on the lattice, we indicate as $\mathcal{N}(i)$ the set of its $8$ neighbors, and with $\sigma(i, t)$ the value of discretized pheromone field on site $i$ at iteration $t$.  The lattice is initialized with a set number of ants $N$, each of them provided of a random direction of motion $\vec{h}$, and randomly distributed across the sites. Each site holds a null pheromone quantity. After initialization, at each discrete time-step $t \in \lbrace 1, \dots , T \rbrace$, each ant is positioned on a site $i$, and selects a neighboring site $j \in \mathcal{N}(i)$, by sampling according to probability:
\begin{equation}
\label{eq:motion-discrete}
    p_{ji} = \frac{\exp(-\beta \,  U[\sigma(j, t), \, \delta])\cdot w(\Delta_{ji})}{\sum\limits_{k \in \mathcal{N}(i)} \exp(-\beta \, U[\sigma(k, t), \,\delta])\cdot w(\Delta_{ki})} = \frac{\left(1 + \frac{\sigma(j, t)}{1 \, + \,\delta \, \sigma(j, t)}\right)^\beta \cdot w(\Delta_{ji})}{\sum\limits_{k \in \mathcal{N}(i)}\left(1 + \frac{\sigma(k, t)}{1 \, + \,\delta \, \sigma(k, t)}\right)^\beta \cdot w(\Delta_{ki})},
\end{equation}
where $\Delta_{ji}(t)$ is the turning degree (the angle between the current direction of the ant, and the one needed to reach cell $j$). The values for $w(\Delta_{ji})$, included in Table \ref{tab:direction_weights}, are tuned to have a realistic motion, where sharp turns are strongly penalized to reflect the effects of inertia and ants' biomechanics, consistently with Eq. \ref{Eq:langevin}.

After sampling their next site, all the ants simultaneously update their current heading of motion $\vec{h}$ with the vector pointing from their current site to the chosen next one, and move according to their new direction. Then, each ant deposits an amount $\eta$ of pheromone in the new location.
Finally, the pheromone field is exponentially decayed in every site $i$ according to $\sigma(i, t+1) = \sigma(i, t) -\kappa\sigma(i, t)$.
\begin{table}[htbp]
    \centering
    \begin{tabular}{|c|c|c|c|c|c|}
        \hline
        \(\Delta_{ji}\) & $0$ & $\pi/4$ & $\pi/2$ & $3\pi/4$ & $\pi$ \\
        \hline
        \(w(\Delta_{ji})\) &  $1$ & $1/2$ & $1/8$ & $1/12$ & $1/50$ \\ 
        \hline
    \end{tabular}
    \caption{Values of turning degree \(\Delta_{ji}\) and associated weight \(w(\Delta_{ji})\).}
    \label{tab:direction_weights}
\end{table}
\subsubsection{Simulations, phase transitions, and order parameters}
\label{sec:sim_phasetrans}
To investigate different regimes of swarm behaviour, simulations are performed with fixed environmental values of $[L_V=32, \, N=300, \, \eta=0.01, \, \kappa=0.015]$ and varying the sensory parameters $\beta$ and $\delta$.\footnote{The transition can also be induced by varying other parameters while keeping $\beta$ and $\delta$ fixed, provided the stability criterion \eqref{Eq:stabcrit} is satisfied. However, following the approach in \cite{chialvoswarms}, we vary only the sensory parameters to focus on ants behaviour.} In particular, we consider discretized pairs of $\beta$ and $\delta$ taken respectively from the sets
$\mathsf{s}_\delta = \{0.3,\, 0.4,\, 0.5,\, 0.6,\, 0.7,\, 0.8\}$ and $\mathsf{s}_\beta = \{3.5,\, 4.5,\, 5.5,\, 6.5,\, 7.5,\, 8.5,\, 9.5\}$. 

The final configurations reached in simulations for different $[\delta, \beta ]$ couples are shown in Figure \ref{fig:phasegrid}: in the left side of the Figure, it can be observed that the final pheromone distribution transitions from an homogeneous to a bimodal pattern as the parameters cross the phase transition line (in red in the figure) defined by Eq. \ref{Eq:stabcrit}. Notably, the final positions of the individuals, shown in the right grid plot, reflect the pheromone traces above this line.
From the plots in Figure \ref{fig:phasegrid}, three distinct behaviours emerge. The chaotic regime is identified by low sensitivity and high noise (bottom-right corner). Instead, at high sensitivity and low noise, ants aggregate into small clustered regions (top-left corner). Finally, the intermediate region, is characterized by the formation of a connected network of thin pathways. 

Suitable order parameters, defined in \cite{SwarmPhaseTrans}, can be measured to evaluate if the system presents ordered and disordered phases: for instance, the order parameter related to the individuals density at time $t$, defined as
\begin{equation}
    M(t) = \frac{\mu_+(t) - \mu_-(t)}{N},
    \label{eq:op-M}
\end{equation}
where we consider $\mu_{\pm}(t) = \sum_{i;\sigma(i, t) \gtrless \langle\sigma\rangle_t} \mu(i, t)$, indicating with $\langle\sigma\rangle_t$ 
the average pheromone quantity in the environment, and $\mu(i, t)$ the number of individuals at site $i$ of the lattice. $M(t)$ takes values in $[-1,\,1]$: it approaches $1$ when the simulated ants are predominantly located within regions of high pheromone concentration (ordered phase), and it is lower than $0$ when they are randomly distributed with respect to the pheromone traces. The pheromone counterpart of Eq. \ref{eq:op-M} is defined as
\begin{equation}
    m(t) = \frac{\sigma_+(t) - \sigma_-(t)}{L_V^2 \langle\sigma\rangle_t},
    \label{eq:op-m}
\end{equation}
where $\sigma_{\pm}(t) = \sum_{i\,;\,\sigma(i, t) \gtrless \langle\sigma\rangle_t} \sigma(i, t)$.\footnote{We remark that the order parameters introduced here exhibit minor differences from those in \cite{SwarmPhaseTrans}. Specifically, we employ a time-dependent formulation restricted to the interval $\left[-1,1\right]$.} Similar considerations to those made for the values of $M(t)$ apply here: $m(t)$ approaches $1$ when the pheromone distribution is bimodal, and goes lower than $0$, when instead it is homogeneous. 

\begin{figure}[t!]
    \centering
    \begin{subfigure}{0.45\textwidth}
        \centering        \includegraphics[width=\linewidth]{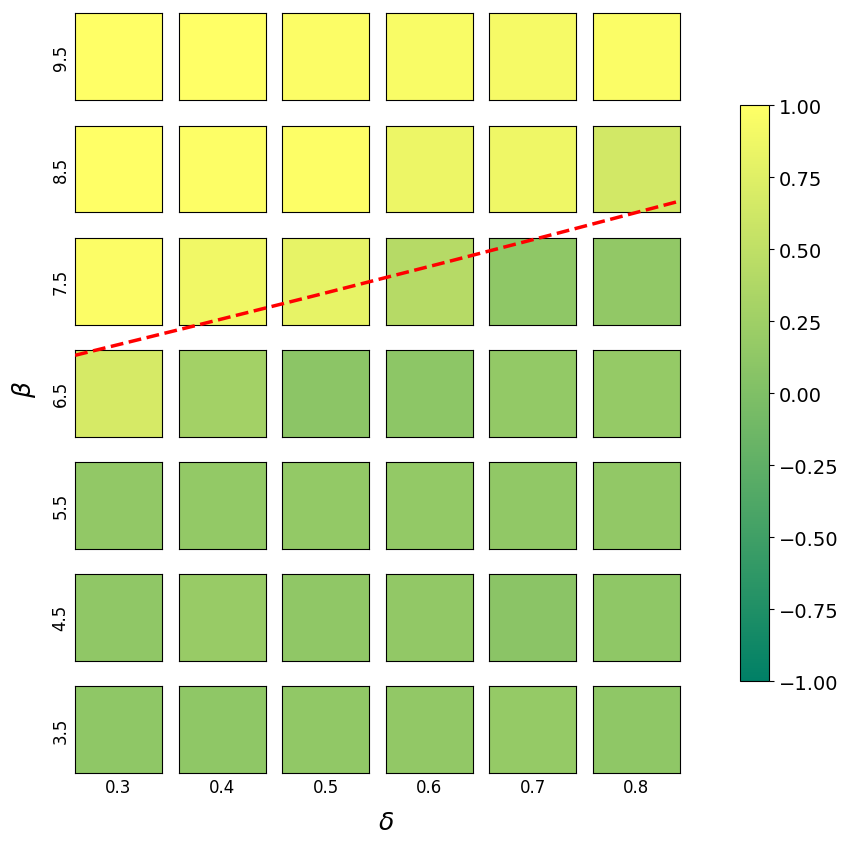}
    \end{subfigure}
    \hfill
    \begin{subfigure}{0.45\textwidth}
        \centering        \includegraphics[width=\linewidth]{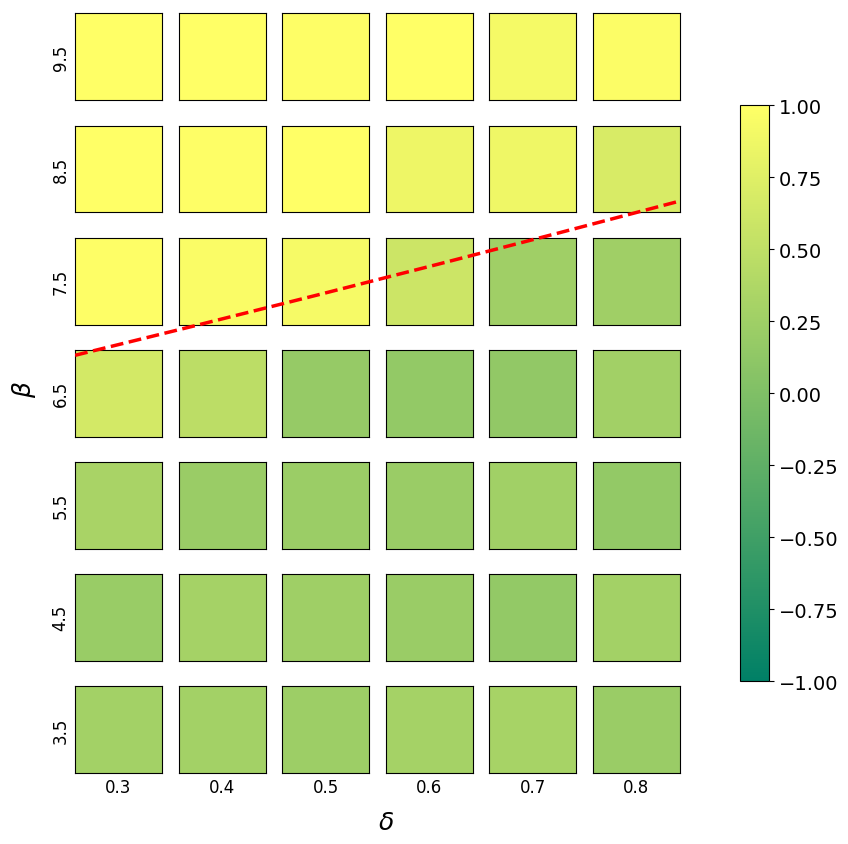}
    \end{subfigure}
    \caption{\textbf{Order parameters across the phase diagram in the baseline setup}. Values of $m(T)$ are presented on the right side and $M(T)$ on the left. The two metrics obtained at the end of the simulations show a clear separation between ordered and chaotic regimes. An empirical threshold of $0.8$ can be used as indicator for the presence of an order phase, but with limited discriminative power between pathway and cluster formations.}
    \label{fig:order_params}
\end{figure}

In Figure \ref{fig:order_params} we showcase the order parameter values associated with the simulation scenarios included in Figure \ref{fig:phasegrid}. We can observe that the order parameters' theoretical limit values of $-1$ and $1$ are not attained in simulation, due to finite-size effects and discretizations. Instead, based on the inspection of Figure \ref{fig:phasegrid} and Figure \ref{fig:order_params}, we identify an empirical threshold of $0.8$ for both order parameters as a reliable indicator that the system is in a stable ordered phase. These measures are useful to distinguish order from chaos, reflecting well the phase line boundaries. However, they are not sufficient to reliably discern between scenarios with pathways or clusters formation. 

\subsubsection{Trails Scenarios}
\label{sec:trails-metric}
Among the three types of behaviour resulting from the simulation of the ants swarm model described in the previous section, we focus  on the intermediate region, characterized by the formation of a connected network of thin pathways.
As highlighted by \cite{patternformation}, this type of behaviour presents a balance between order and fluctuations, making the network adaptable and resilient. Thanks to these features, two properties can emerge: \textit{memory}, the ability of the system to recover previously established patterns even after large perturbations, and \textit{bootstrapping}, the capability to amplify weak initial signals or trails into fully formed structures. 

Taking inspiration from this, we set our functional control goal as steering the system toward a behaviour enhancing as much as possible these characteristics. We individuate as optimal candidate what we call the \emph{trails scenario}, defined as follows:
\begin{definition}
\label{def:ants-trail-scenario}
Considering a pheromone distribution $\sigma$ and a distribution $\rho$ of individuals over the environment, we say that $(\sigma, \, \rho)$ is a \emph{trails scenario} if the following conditions are satisfied:
\begin{enumerate}
    \item (\textit{bimodal pheromone condition}): the pheromone distribution $\sigma$ divides the environment in regions with high pheromone concentration, and regions where the pheromone concentration is negligible;

    \item (\textit{population condition}): the population distribution $\rho$ closely approximates $\sigma$, with most individuals on the high-concentration regions;

    \item (\textit{trail geometry condition}): the pheromone traces, created by the combination of $\sigma$ and $\rho$ evolving in time, form thin elongated connected pathways, characterized by the absence of extended two-dimensional regions and branching structures.
\end{enumerate}
\end{definition}
These conditions identify precisely the dynamical regime in which the swarm exhibits a balance between stability and fluctuations described in \cite{patternformation}. In particular, a bimodal pheromone distribution ensures the existence of persistent collective structures that can act as distributed spatial memories, while the concentration of ants along the trails reinforces these structures through positive feedback. At the same time, the requirement that trails remain thin, connected, and non-branching prevents the system from collapsing into rigid saturated clusters or diffuse disordered configurations, due to low ABM-ants pheromone sensitivity and noisy movements. In this regime, localized pheromone traces can survive partial perturbations and guide the reconstitution of previously established pathways, thus supporting memory effects. Similarly, weak initial pheromone inhomogeneities or sparse ant flows can be selectively amplified into coherent trails, enabling bootstrapping phenomena. 

While the level at which a ant swarm model scenario at time $t$ satisfies the population condition (or the bimodal pheromone condition) could be quantified by observing the associated order parameter $M(t)$ introduced in Eq. \ref{eq:op-M} (or the order parameter $m(t)$ introduced in Eq. \ref{eq:op-m}), the trail geometry condition requires a metric that captures the spatial distribution of pheromone across the environment. Below, we exploit basic elements from graph theory together with the order parameters' definitions to define a metric that captures how much a simulation scenario is close to be a trails scenario, considering the entire definition with its three conditions.

\paragraph{Trail scenario metric.} We start by constructing a spatial graph representing the pheromone traces on $V$ at time $t$.
Considering the scenario at instant $t$, we define a binary mask $M_t \in \lbrace 0, 1\rbrace^{L_V  \times L_V}$ of the discretized pheromone field $\sigma(\, :\,\, , \, t \, )$, containing value $1$ in the entries $(x_i, y_i)$ such that site $i$ presents pheromone value above the mean, i.e,  
\begin{equation}
M_{t}(x_i, \, y_i) \, = \,
\begin{cases}
\,\,\, 1 & \text{if } \,\, \sigma(i, t) \, > \,  \langle\sigma\rangle_t, \\[6pt]
\,\,\, 0 & \text{otherwise}.
\end{cases}
\label{eq:mask-V}
\end{equation}
This mask individuates the locations with the majority of pheromone in the whole environment $V$ at time $t$. 

From this mask, we construct a spatial undirected graph $\mathrm{G}_{t} = (\mathrm{N}_{t}, \mathrm{E}_{t})$, where $\mathrm{N}_{t}$ contains  a node $n_i$ per each site $i = (x_i, y_i)$ such that $M_{t}(x_i, \, y_i) = 1$. Given two nodes $n_i , \, n_j  \in N_{t}$, the edge $(n_i, \, n_j)$ connecting them exists in $ E_{t}$ if a smart-ant positioned in the site $i$ can reach with a single action the site $j$ in $V$, i.e., if $(x_i, y_i)$ and $(x_j, y_j)$ are adjacent in the lattice $V$. Thanks to this procedure, we build a graph encoding the spatial structure of the most pronounced pheromone traces on $V$. 

Let $\mathrm{C}(\mathrm{G}_{t})$ be the set of all connected components of $\mathrm{G}_{t}$, i.e., the subgraphs $\mathrm{C} = (\mathrm{N}_{\mathrm{C}}, \mathrm{E}_{\mathrm{C}})$ such that $\mathrm{N}_{\mathrm{C}} \subseteq \mathrm{N}_t$ and $\mathrm{E}_{\mathrm{C}} \subseteq \mathrm{E}_t$. We consider the set of connected components of $\mathrm{G}_t$ with at least two nodes as 
\begin{equation*}
    \mathrm{C}_{2}(\mathrm{G}_{t}) = \{ \mathrm{C}_t^k = (\mathrm{N}_{t}^k, \mathrm{E}_{t}^k) \in \mathrm{C}(\mathrm{G}_{t}) : |\mathrm{N}_{t}^k| \geq 2 \}\footnote{Connected components with $|\mathrm{N}_{t}^k| = 1$ (and hence $|\mathrm{E}_{t}^k| = 0$ are excluded here, given that both ABM-agents and smart-ants, occupying a new site at every time-step $t$, and not having the possibility to station onto the the same site for two steps consecutively, are de facto incapable of reinforcing or maintaining this type of pheromone trace.)},
\end{equation*}
and define the following metric
\begin{equation}
   \mathcal{TS}(t) = \left[\, \sum_{k=1}^{|\mathrm{C}_{
   2}(\mathrm{G}_{t})|} \mathcal{P}(\mathrm{C}_t^k) \, \right] + \frac{|\mathrm{C}_{
   2}(\mathrm{G}_{t})|}{2}\left(m(t)\cdot \frac{\min(\, \text{avg}_k|\mathrm{N}_{t}^k|\, ,\, L_V)}{L_V} + M(t)\right),
   \label{eq:global-metric}
\end{equation}
where $M(t)$ and $m(t)$ are defined in Eq. \ref{eq:op-M} and Eq. \ref{eq:op-m}, respectively. The term $\mathcal{P}(\mathrm{C}_t^k)$ is a penalty, 
defined as 
\begin{equation}
  \mathcal{P}(\mathrm{C}_t^k) = -\frac{|\mathrm{E}_{t}^k|}{|\mathrm{N}_{t}^k| - 1}
  \label{eq:global-metric-penalty}
\end{equation}
proportional to the number $|\mathrm{E}_{t}^k|$ of edges of the connected component $\mathrm{C}_t^k$. The denominator normalizes the number of edges by the number of nodes $|\mathrm{N}_{t}^k|$, penalizing components with high average degree. This penalization evaluates negatively a connected component $\mathrm{C}_t^k$ with highly branched or periodic structures (like chessboard configurations), while assigning high value to thinner trails. Under this formulation, the optimal connected component (i.e., the one with the highest value of $\mathcal{P}(\mathrm{C}_t^k)$ achievable) is a connected chain, for which $\mathcal{P}(\mathrm{C}_t^k) = -1$.

The function $\mathcal{TS}(t)$ defined in Eq. \ref{eq:global-metric}, hence, assigns higher values to those scenarios in which (i) connected components are composed by thin pheromone trails with a low branching degree, thanks to the first term, proportionally penalizing each connected component with such characteristics; (ii) a high proportion of target individuals is positioned on the pheromone traces, due to $M(t)$; (iii) the pheromone is distributed in a bimodal fashion in the environment, as stated by $m(t)$. This value is weighted by the average size of the connected components (saturated to the environment lattice size) in order to reward bimodal pheromone distribution only in the presence of relatively big connected components. This avoids to evaluate positively the formation of small clusters (as we discussed, $m(t)$ alone was not sufficient as a metric to distinguish between the scenarios with small clusters formation, and the ones with long connected pathways emergence). Both the saturation of the average connected component size, and the multiplicative factor of $|\mathrm{C}_{2}(\mathrm{G}_{t})| / 2$ serve to maintain comparable orders of magnitude between the different terms inside the metric.\\

\begin{figure}[t]
    \centering
    \begin{subfigure}[b]{0.45\linewidth}
        \centering
        \includegraphics[width=\linewidth]{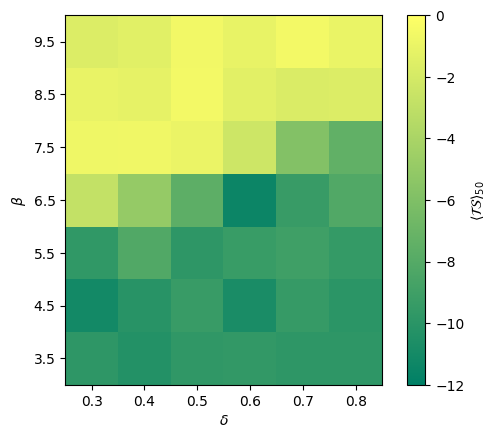}
        \caption{}
        \label{fig:avgTS50_phase_diag_cont}
    \end{subfigure}
    \begin{subfigure}[b]{0.45\linewidth}
        \centering
        \includegraphics[width=\linewidth]{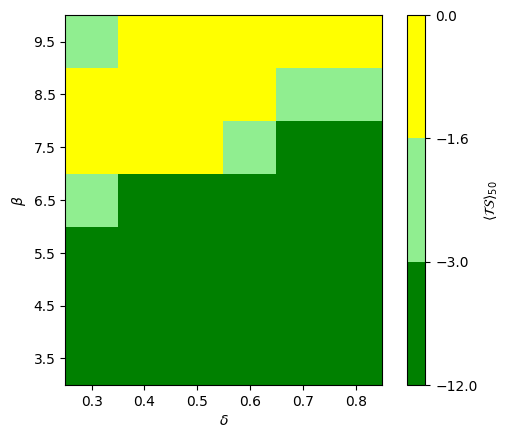}
        \caption{}
        \label{fig:avgTS50_phase_diag_disc}
    \end{subfigure}
     \caption{\textbf{Trail scenario metric $\langle \mathcal{TS} \rangle_{50}$ across the phase diagram}: (a) continuous values; (b) discretized values.}
    \label{fig:avgTS50_phase_diag}
\end{figure}

To evaluate if a trails scenario emerges in a simulated episode of length $T$, we analyze $\langle \mathcal{TS} \rangle_{50}$, obtained by averaging $\mathcal{TS}(t)$ exclusively on the last $50$ steps of the episode. This averaging is considered in order to contrast the highly stochastic nature of $\mathcal{TS}(t)$, that is sensitive to fluctuations, and hence less informative if evaluated, for instance, on the single final step step $T$. Figure \ref{fig:avgTS50_phase_diag} analyzes the empirical behaviour of $\langle \mathcal{TS} \rangle_{50}$ for the different $[\delta, \beta]$-configurations. On the left, it shows the average values of $\langle \mathcal{TS} \rangle_{50}$ obtained over $n_{e} = 10$ realizations, one per each seed employed to simulate the Ants Swarm Model in each $[\delta, \beta]$-configuration. The same values are represented on the right, discretized \emph{ad-hoc} to highlight how values of $\langle \mathcal{TS} \rangle_{50} \geq -1.6$ are associated with scenarios where connected networks of thin pathways form. Such discetization clearly clusters the area that in Figure \ref{fig:phasegrid} shows the emergence of such behaviour.

\subsection{Reinforcement Learning}
\label{sec:RL}
Reinforcement Learning (RL) is a Machine Learning paradigm based on agents learning a behaviour policy through interaction with their own environment, with the goal of maximizing a reward accumulated along a long horizon. In RL, an agent explores the environment by implementing actions in it, and receiving an associated feedback (reward) from it. The core idea is that the agent improves its decision-making policy through trial-and-error, by observing the direct consequences of its actions, and using them to estimate the long-term effects of its own strategies \cite{suttonbarto}.

The classic RL problem is typically modeled using a Markov Decision Process (MDP) \cite{mdp}, that is, a tuple $(\mathcal{S}, \mathcal{A}, p, r, \gamma)$ such that:
\begin{itemize}[leftmargin=1cm]
\item $\mathcal{S}$ is the set of all the possible states $s$ of the system;
\item $\mathcal{A}$ is the set of all the actions $a$ implementable by an agent;
\item $p: \mathcal{S} \times \mathcal{A} \times \mathcal{S}  \longrightarrow [0, \, 1 ]$ is a transition probability function, such that $p(s,\, a, \, s') = \mathbb{P}( s' \, \vert \,  s, \, a )$, is the probability of the environment moving to state $s'$ from state $s$, due to the implementation of action $a$;
\item $r: \mathcal{S} \times \mathcal{A}  \times \mathcal{S} \longrightarrow \mathbb{R}^{+}_{0}$ is the reward function. In particular, $r(s, a, s')$ is a real number, representing the reward an agent receives by taking action $a$ in state $s$ and reaching state $s'$;
\item $\gamma \in [0, 1]$ is a discount factor, weighting the importance of future rewards.
\end{itemize}
At every time-step $t$, a state $s_t=s \in \mathcal{S}$ is measured from the environment, and the agent takes an action $a_t=a \in \mathcal{A}$. The system moves to a new state $s_{t+1} = s' \in \mathcal{S}$ according to $p$, and the agent receives a reward $r(s_t, a_t, s_{t+1})$. The discounted cumulative reward associated with an infinite-horizon trajectory $\tau_t = (s_t, \, a_{t}, \, s_{t+1}, \, a_{t+1},  \dots )$ of states and actions of the system, such that $s_{t+1} \sim p(s_t, \, a_t, \, \cdot )$ for each $t$, is defined as 
\begin{equation}
\label{eq:disc-cum-rwd}
    \mathcal{G}(\tau_t) = \sum_{k=t}^{\infty} \gamma^{t-k} \, r(s_k, \, a_k, \, s_{k+1})\, .
\end{equation}
Considering an agent acting in the environment, its decision-making process is modeled as a stochastic policy function $\pi: \mathcal{S} \times \mathcal{A}\longrightarrow [0, \, 1 ]$, such that $\pi(s, \, a)$ represents the probability of the agent implementing action $a$ while in state $s$. For any policy $\pi$ it is possible to define its value function $V^{\pi}: \mathcal{S} \longrightarrow \mathbb{R}^{+}_{0}$ as the expectation (with respect to $p$ and $\pi$) of the cumulative discounted rewards $G$ defined in Eq. \ref{eq:disc-cum-rwd} obtained by $\pi$ over time, starting from any state $s$, i.e.,
\begin{equation}
\label{eq:2.2}
     V_{\pi}(s) = \mathbb{E}\Big[ \,  \mathcal{G}(\tau_0) \,\, \vert \,\, s_0 = s ,\; s_{t+1} \sim p(s_t, \, a_t, \, \cdot ) ,\;  a_t \sim \pi(s_t) \, \Big].
\end{equation}
Solving the RL problem means learning a policy $\pi$ that maximizes the expected discounted cumulative reward for each state $s \in S$ on average, i.e., such that 
\begin{equation}
\pi^{*} = \text{arg}\max_{\pi \in \Pi} \mathbb{E}_{s}[V_{\pi}(s)],
\label{eq:RL-pb}
\end{equation}
with $\Pi$ being the set of all the possible policies.
In this work, we adopt a multi agent RL algorithm where agents progressively collect data and transmit it to a central learner. This learner updates the value function and distributes the updated policy back to the agents, who collaborate toward the common goal.
Full algorithm and training details are reported in Sections \ref{sec:ppo} - \ref{sec:training_res}.

\subsection{Ant Swarm Model Control via RL stigmergic agents}
\label{sec:RLDesign}
In this section, we describe our framework for achieving functional control of the discrete ant swarm model described in Section \ref{sec:discretized_model}, guiding the emergence of a desired behaviour without requiring direct intervention or centralized coordination. The core idea is to influence the collective behaviour of the system individuals (the mentioned ABM-ants) by embedding, within the same environment, additional agents (indicated in this work as smart-agents) dedicated to achieving the desired control task through modifications of the environment. Indeed, rather than acting directly on the elements to be controlled, the controlling population acts in a stigmergic manner, specifically by modifying shared environmental variables that mediate indirect interactions and influence the dynamics of the controlled population. 

In order to obtain agents that are autonomously capable of accomplishing functional control, RL techniques can be employed: the agents are trained through direct experience, attempting at controlling the system, and their control policy is progressively refined according to the RL paradigm. In the following sections, we describe the specific design choices that we made to enable functional control of the discretized ant swarm model, in order to favor trail formation as an emerging behaviour. 

A key aspect of our framework is the interplay between the stigmergic nature of the system under control and the stigmergic capabilities of the controlling agents. Since interactions among ABM-ants are mediated through the pheromone field $\sigma$, the environment itself acts as an information substrate through which agents indirectly influence one another. Exploiting this property, the smart-agents implement control not by directly modifying the behaviour of the ants, but by locally altering the same stigmergic field, thus steering the collective dynamics through environmental interventions.
This perspective naturally enables the definition of graph-based reward functions grounded in the local structure of the pheromone landscape. In particular, for each smart-agent, we define a local interaction graph induced by the distribution of $\sigma$ within its neighborhood, where nodes and weighted connections encode the spatial organization emerging from stigmergic interactions. Quantities extracted from the local graphs provide measurable indicators of the desired functional behaviour. These graph-derived metrics are then employed as reward signals for RL, allowing smart-agents to optimize policies that promote the emergence of globally ordered phases with coherent trail formation.

We consider a population of $N'>1$ smart-agents, that depose pheromone in the environment with rate $\eta'=1 > \eta$, and whose motions do not follow the model dynamics, but implement a movement policy that is optimized via RL in order to influence the behaviour of the controlled population, promoting the emergence of trails. Formulating the learning problem associated to the described scenario, requires careful design choices, including the definition of the state representation, the set of available actions, and the reward function. 

\subsubsection{State and action spaces design}
The state representation of a smart-agent contains the signals that the agent perceives from the surrounding environment. Each of the smart agents at any time-step is provided with a local vision, containing information related to the sites surrounding its current position in the environment, within a certain distance.  We define the set of the sites included in the local vision of smart agents positioned in site $i =(x_i, y_i) \in V$ as the $9 \times 9$ neighborhood centered in $\chi$, i.e., $\mathcal{N}(\chi) = \lbrace \, \, \chi + (x, y)  \,\, : \,\, i, j \in \mathbb{Z} \, , \,  -4 \leq x, y \leq 4 \, \, \rbrace \subset V$.
Given an agent on the lattice $V$, we indicate as $p_t$ the site on which the agent is positioned in $V$ at time $t$. We consider as associated local state the collection of matrices $s_t =[s_t^1 \, , \, s_t^2 \, , \, s_t^3 \, , \, s_t^4]$. Each of the matrices $s_t^k \in \mathbb{R}^{ 9 \times 9}$ for $k \, = \, 1, \, 2, \, 3, \, 4$ contains a different signal characterizing the local state, measured at each site located on the $9 \times 9$ neighborhood $\mathcal{N}(p_t)$ of $p_t$. For $u, v \in \lbrace 1, 2, \dots, 9 \rbrace$, the value $\chi_t(u, v) = (p_t^1 + i \, , \, p_t^2 + j) = (p_t^1 + (u - 5) \, , \, p_t^2 + (v - 5))$ connects the cell $(u, v)$ indicating a position on the the $9 \times 9$ neighborhood $\mathcal{N}(p_t)$ with the associated site with respect to the reference system of $V$. In the following, the matrices $s_t^k$  for $k \, = \, 1, \, 2, \, 3, \, 4 $  are described in detail:
\begin{enumerate}
    \item The first slice, indicated as $s_t^1$, encodes information about the local pheromone field. It corresponds to the the pheromone distribution on the sites in $\mathcal{N}(p_t)$, namely $s_t^{1}(u, v) = \sigma(\chi_t(u, v) \, , \, t)$.
    \item The second matrix $s_t^2$ represents the spatial distribution of ABM-ants in the vicinity of the agent. Each entry records the number of ABM-ants occupying the corresponding site within $\mathcal{N}(p_t)$, that is $s_t^{2}(u, v) = \mu(\chi_t(u, v) \, , \,  t)$.
    \item The third matrix $s_t^3$ captures the presence of other smart-agents around the reference agent, i.e., $s_t^{3}(u, v) = \mu_{(a)}(\chi_t(u, v) \, , \, t)$, where $\mu_{(a)}(i, \, t)$ denotes the number of smart-agents at a given site $i$ at time-step $t$.

    \item The fourth matrix $s_t^4$ provides directional information by indicating the previous direction of motion of the smart-agent inside the local neighborhood, that is the direction towards which the agent is facing, after the previous action, represented through a one-hot encoding over the local grid, as
    \[
    s_t^{4}(u,v) \, = \, 
    \begin{cases}
        \,\,\, 1, & \text{if} \,\,\, \chi_t(u,v) = p_t + (p_t - p_{t-1}), \\[6pt]
        \,\,\, 0, & \text{otherwise}.
    \end{cases}
    \]
\end{enumerate}
At each time-step $t$, a smart-agent selects an action from a discrete action space $\mathcal{A}  = \{a_t^1, a_t^2, \dots, a_t^8\}$, where each action represents the choice of the lattice site to occupy at the next step. In particular, $a_t^k$ with $k = 1,\dots,8$, moves the agent from its current position $p_t \in V$ to the corresponding adjacent site $x_t^k$, updating its position to $p_{t+1} = x_t^k$.

\subsubsection{Graph-based reward design}
\label{sec:reward}

We design a reward function that encourages the learning of optimal movements of the smart-agents in the environment, in order to obtain the emergence of elongated, thin, and non branching pathways, densely populated with ABM-ants, through the smart-agents' combined stigmergic action in the environment. 

Training a multi-agent control population (we focus on the case $N'>1$), we aim at avoiding issues due to credit assignment \cite{MARL}. Global rewards, i.e., rewards that are shared by all the smart-agents present in the environment, might fail to distinguish the agents whose actions are truly responsible for good (or bad) state transitions, and evaluating uniformly all the actions performed at the same time-step, based on their combined effect. Hence we define an individual step-reward, that evaluate the effect of a single agent's action over its local state. 

We consider a smart-agent at time-step $t$, positioned in site $p_t$ of $V$, perceiving a local state $s_t$ and performing an action $a_t$. As a consequence of the actions at time $t$, the agent position at the next time-step $t+1$ shift to $p_{t+1}$ and a new state $s_{t+1}$ is perceived. Our reward is based on metrics associated with a graph, built from information contained in $s_{t+1}$ and representing the local spatial distribution of pheromone and ABM-ants. Below, the three steps necessary to define our graph-based reward are described in detail.

\paragraph{Step 1. Construction of the local spatial graph of the pheromone traces.}
In analogy with Section \ref{sec:trails-metric}, we define a local binary mask $M^{\text{loc}}_{t+1} \in \lbrace 0, 1\rbrace^{9\times 9}$ of the first component of $s_{t+1}$, containing the pheromone field $\sigma$ in the local neighborhood $\mathcal{N}(p_{t+1})$ of the agent. The local mask contains value $1$ in the entries that in $s_{t+1}^1$ present values above the average pheromone quantity $\langle s_{t+1}^1 \rangle$ in the sites of the neighborhood, that is, 
\begin{equation}
M^{\text{loc}}_{t+1} (x, \, y) \, = \,
\begin{cases}
\,\,\, 1 & \text{if } \,\, s_{t+1}^1(x, y) \, > \, \langle s_{t+1}^1 \rangle \\[6pt]
\,\,\, 0 & \text{otherwise}.
\end{cases}
\label{eq:mask-loc}
\end{equation}

This mask individuates the locations with the majority of pheromone in the neighborhood of the agent at time $t+1$. 
We alter $M^{\text{loc}}_{t+1}$ by artificially assigning value $1$ to the central location $\bar{i} = (5, 5)$, corresponding to the new position $p_{t+1}$ of the agent in $V$, i.e., $M^{\text{loc}}_{t+1}(5, \, 5) = 1$. By doing this, we are exaggerating the effect of action $a_t$ on the pheromone presence registered by the local state $s_{t+1}^1$. Given the smart-agent transition dynamics, deciding to move from $p_{t}$ to $p_{t+1}$ implies deciding to release pheromone in $p_{t+1}$, enforcing there the pheromone trace. Even if doing so one time might not be sufficient to raise $s_{t+1}^1(5, 5)$ enough to trespass $\langle s_{t+1}^1 \rangle$ and naturally achieve $M_{t+1}(5, \, 5) = 1$, we encode the intention behind this move, interpreting the smart-agent's move $a_t$ as a decision to increase the amount of pheromone in $p_{t+1}$.

We exploit $M^{\text{loc}}_{t+1}$ to build a graph $\mathrm{G}(s_{t+1}) = (\mathrm{N}(s_{t+1}), \mathrm{E}(s_{t+1}))$ encoding the spatial structure of the most pronounced pheromone traces on the local neighborhood $\mathcal{N}(p_{t+1})$, surrounding the new position $p_{t+1}$ assumed by the agent, following the same procedure described in Section \ref{sec:trails-metric} to build the global graph $\mathrm{G}_t$ from $M_t$. 


\paragraph{Step 2. Computation of relevant metrics on the graph.}
We denote by $\mathrm{C}(s_{t+1}) = (\mathrm{N}_{\mathrm{C}}(s_{t+1}), \, \mathrm{E}_{\mathrm{C}}(s_{t+1}))$ the connected component of the graph $\mathrm{G}(s_{t+1})$ that contains the node corresponding to the central site $\bar{i} = (5,5)$ and define
\begin{equation}
    \mu_{+}(s_{t+1}) = \sum_{m \, \in \, \mathrm{N}_{\mathrm{C}}(s_{t+1})} \mu(m \, , \,  t) \, , 
\qquad \mu_{-}(s_{t+1}) = \sum_{m \, \in \,  \overline{\mathrm{N}_{\mathrm{C}}(s_{t+1})}} \mu(m \, , \,  t),
\label{eq:graph_metric1}
\end{equation}
where $\overline{\mathrm{N}_{\mathrm{C}}(s_{t+1})}= \mathrm{N}(s_{t+1})  \setminus \mathrm{N}_{\mathrm{C}}(s_{t+1})$. The total number of ABM-ants on nodes in $\mathrm{N}(s_{t+1})$ is $\mu(s_{t+1})  = \mu_{+}(s_{t+1}) + \mu_{-}(s_{t+1})$. Analogously we set 
\begin{equation}
    \sigma_{+}(s_{t+1}) = \sum_{m \, \in \, \mathrm{N}_{\mathrm{C}}(s_{t+1})} \sigma(m \, , \,  t) \, , 
\qquad \sigma_{-}(s_{t+1}) = \sum_{m \, \in \,  \overline{\mathrm{N}_{\mathrm{C}}(s_{t+1})}} \sigma(m \, , \,  t),
\label{eq:graph_metric2}
\end{equation}
from which we derive the total amount of pheromone on nodes in $\mathrm{N}(s_{t+1})$ as $\sigma(s_{t+1}) = \sigma_{+}(s_{t+1}) + \sigma_{-}(s_{t+1})$.

From Eq.s \ref{eq:graph_metric1} - \ref{eq:graph_metric2} it is possible to define local versions of the global order parameters $M(t)$ and $m(t)$ defined in Eq.s \ref{eq:op-M} - \ref{eq:op-m}. The local versions characterize the ABM-ants and pheromone density on the local connected component $\mathrm{C}(s_{t+1})$, in the following way
\begin{align}
M(s_{t+1}) & = \frac{\mu_{+}(s_{t+1}) - \mu_{-}(s_{t+1})}{\mu(s_{t+1})}, \label{eq:op-loc-M}\\
m(s_{t+1}) & = \frac{\sigma_{+}(s_{t+1}) - \sigma_{-}(s_{t+1})}{\sigma(s_{t+1})}. \label{eq:op-loc-m}
\end{align}
Eq. \ref{eq:op-loc-M} measures the relative concentration of target individuals within the agent's connected component compared to those outside it, making it a local version of the order parameter $M$. If maximized as a reward, it encourages smart-agents to form coherent paths that are easy to follow for the population to control. 
Eq. \ref{eq:op-loc-m}, instead, is as a local version of $m$, capturing the relative variability of the pheromone within the component, with high values indicating a local bimodal distribution.

\paragraph{Step 3. Reward definition based on the graph metrics.}
We propose a step-wise reward function that locally captures the same properties encoded in the trails scenario metric of Eq. \ref{eq:global-metric}. By maximizing the expected long-horizon discounted return, the RL-based optimization process encourages the emergence of these characteristics consistently along the smart ants’ trajectories, effectively promoting their presence throughout the entire environment. In particular, we set
\begin{equation}
   r(s_{t+1}) = \mathcal{P}(s_{t+1}) + \frac{1}{2}\left(m(s_{t+1})\cdot \frac{\min(\vert \mathrm{N}_{\mathrm{C}}(s_{t+1}) \vert ,\, L_V)}{L_V} + M(s_{t+1})\right),
   \label{eq:rwd_tot}
\end{equation}
with $L_V=9.$

The term $\mathcal{P}(s_{t+1})$ in \eqref{eq:rwd_tot}, is a penalty on the connected component $\mathrm{C}(s_{t+1})$, defined analogously to \eqref{eq:global-metric-penalty} employed in Eq. \ref{eq:global-metric}, i.e., 
\begin{equation}
    \mathcal{P}(s_{t+1}) = 
\begin{cases}
\frac{- \vert \mathrm{E}_{\mathrm{C}}(s_{t+1})\vert}{\vert \mathrm{N}_{\mathrm{C}}(s_{t+1}) \vert - 1} & \text{if} \, \, \vert \mathrm{N}_{\mathrm{C}}(s_{t+1})\vert > 1 \\[6pt]
\,\,\,\,\,\, 0 & \text{if} \, \, \vert \mathrm{N}_{\mathrm{C}}(s_{t+1})\vert = 1.
\end{cases} 
\label{eq:rwd_penalty}
\end{equation}
 In Eq. \ref{eq:rwd_penalty} we choose not to penalize single-node components, as they typically occur only at the initial time-step of an episode. To avoid division by zero, their contribution is explicitly set to zero.

\clearpage
\section*{Acknowledgements}
Our research adheres to strict ethical standards. No human participants were involved in our experiments, and no deception or manipulation was applied. After thorough assessment, we do not anticipate any additional ethical concerns or risks related to our work.

This study did not rely on pre-collected datasets; all results were generated through the computational simulations and Reinforcement Learning experiments described in the manuscript. The code used to replicate the dynamics and findings of this study is available from the corresponding author upon reasonable request.

The authors declare that they have no competing interests.

Funded by the European Union. Views and opinions expressed are however those of the author(s) only and do not necessarily reflect those of the European Union or the European Health and Digital Executive Agency (HaDEA). Neither the European Union nor the granting authority can be held responsible for them. AP, AG and RG were funded by the Digital Twin Bologna Project (CUP: F39I23000940007), co-funded by the European Union (ERDF – ESF 2021-2027 - Priority 1 – Digital Agenda and Urban Innovation). Additionally, AG and RG acknowledge the financial support received from the PNRR ICSC National Research Centre for High Performance Computing, Big Data and Quantum Computing (CN00000013), under the NRRPMUR program funded by the NextGenerationEU. LF and BL were funded by Grant Agreement no. 101120763 - TANGO and Grant Agreement no. 101120237 - ELIAS. Moreover, LF and BL acknowledge the financial support supplied by Ministero delle Imprese e del Made in Italy (IPCEI Cloud DM 27 giugno 2022 – IPCEI-CL-0000007) and European Union (Next Generation EU).

RG thanks Marco Pistore for the useful discussion.

\bibliography{sn-bibliography}
\newpage
\clearpage


\begin{center}
    {\LARGE \bf Supplementary Materials}
    
    \vspace{0.5cm}
    
    \vspace{1cm}
\end{center}

\setcounter{page}{1}
\setcounter{section}{0}
\setcounter{figure}{0}
\setcounter{table}{0}
\setcounter{equation}{0}

\renewcommand{\thesection}{SM\arabic{section}} 
\renewcommand{\thepage}{SM\arabic{page}}
\renewcommand{\thefigure}{SM\arabic{figure}}
\renewcommand{\thetable}{SM\arabic{table}}
\renewcommand{\theequation}{SM\arabic{equation}}

\makeatletter
\renewcommand{\theHsection}{SM\arabic{section}}
\renewcommand{\theHfigure}{SM\arabic{figure}}
\renewcommand{\theHtable}{SM\arabic{table}}
\renewcommand{\theHequation}{SM\arabic{equation}}
\makeatother

\hypersetup{pageanchor=true}

\section{RL algorithm: Proximal Policy Optimization}
\label{sec:ppo}
Proximal Policy Optimization (PPO) \cite{ppo} is a policy gradient actor-critic algorithm for solving RL problems in an online, model-free setting. The algorithm employs an \textit{actor policy} with parameters $\theta$, which defines the stochastic policy $\pi_{\theta}$ modeling the decisions of the agent, and a \textit{critic policy} with parameters $\phi$, representing the parameterized value function $V_{\phi}$. PPO optimizes the expected return by directly updating the actor parameters $\theta$ while using the critic $V_{\phi}$ to provide advantage estimates. Importantly, it ensures that each policy update remains within a trust region around the current policy, preventing excessively large updates that could destabilize learning. This is achieved by updating $\theta$ in order to maximize the following clipped objective function:
\begin{equation}
\label{eq:ppo-objective}
    L(\theta) = \mathbb{E}_{t} \left[ \min \left( \, \dfrac{\pi_{\theta}(a_t \mid s_t)}{\pi_{\theta_{\text{old}}}(a_t \mid s_t)} \cdot \hat{A}_{t}, \; \text{clip}\left( \, \dfrac{\pi_{\theta}(a_t \mid s_t)}{\pi_{\theta_{\text{old}}}(a_t \mid s_t)}, \, 1 - \epsilon, \, 1 + \epsilon \, \right) \cdot \hat{A}_{t} \,\right) \right], 
\end{equation}
where $\pi_{\theta}$ is the current policy, $\pi_{\theta_{\text{old}}}$ is the previous policy, $\epsilon$ is a small hyperparameter, and $\hat{A}_t$ is an estimator of the advantage function at time $t$. The advantage function is typically estimated using Generalized Advantage Estimation (GAE), i.e.,
\begin{equation}
\label{eq:gae}
    \hat{A}_t = \sum_{l=0}^{\infty} (\gamma \lambda)^l \, \delta_{t+l}, \quad \text{with } \delta_t = r_t + \gamma \, V_{\phi}(s_{t+1}) - V_{\phi}(s_t),
\end{equation}
where $\lambda \in [0,1]$ is the GAE parameter \cite{ppo} controlling the bias–variance trade-off in the advantage estimate. An additional entropy regularization term can be added to encourage exploration and prevent premature deterministic policies. The entropy of the policy at state $s_t$ is defined as
\begin{equation}
    \mathcal{H}(s_t; \,  \theta) = - \sum_{a} \pi_\theta(a \mid s_t) \, \log \pi_\theta(a \mid s_t),
\end{equation}
and the final actor objective becomes
\begin{equation}
    \mathcal{L}_{\text{a}}(\theta) = \mathbb{E}_{t} \Big[ L(\theta) + c \, \mathcal{H}(s_t; \theta) \Big],
    \label{eq:ppo-objective-final}
\end{equation}
where $c$ is a hyperparameter controlling the strength of the entropy bonus. The optimization of Eq. \ref{eq:ppo-objective-final} is carried out via  multiple epochs of stochastic gradient ascent on mini-batches sampled from trajectories of states and actions collected by employing the current policy $\pi_{\theta}$ to generate the actions, and measuring the next state from the environment.
The critic parameters $\phi$ are instead updated by minimizing the loss $\mathcal{L}_{\text{c}}$ between the values $V_{\phi}(s_t)$ predicted by the parameterized value function and the returns $\mathcal{G}(\tau_t)$ computed according to Eq. \ref{eq:disc-cum-rwd}.


\section{RL training: tuning and evolution of the performance}
\label{sec:RLtraining}

\subsection{Training}
\label{sec:training_res}

The employed training procedure involves a number $N_{epi}$ of repeated episodes. Each episode consists of a simulation of the environment such that, in addition to the ABM-ants, $N' = 30$ smart-agents are introduced into the environment after $t = T_{tr} = 200$ time-steps and remain active until the end of the episode, at $t = T = 1200$. Each smart-agent releases $\eta' = 10 \cdot \eta = 1$ unit of pheromone at every time-step, performs an action, and collects data. Such data are then used to perform a learning step via RL, after the end of the episode and before the next. 

The interaction with the environment, reflected in the collected data, follows the classic RL scheme: a local state $s_t^{(n)}$ is observed by an agent $n$ at time $t$, which then performs an action $a_t^{(n)}$ sampled from the discrete probability distribution produced by actor network given as input $s^{(n)}_t$. It then observes a new state $s^{(n)}_{t+1}$, and receives a reward $r_t^{(n)}$. The experience collected in one episode is then saved in dataset made by tuples $(s_t^{(n)}, \, a_{t}^{(n)}, \, r_{t}^{(n)}, \, s_{t+1}^{(n)})$, one per each time-steps $t \in \left[ T_{tr}, \, T_{tr}+1, \, \cdots, \, T-1, \,T\right]$ and per each agent $n\in \left[1, \,  \cdots, \, N'\right]$.

After the end of the episode, the collected data are all sampled once in batches of size $b_{size}=60$, and used to optimize the actor and the critic parameters, employing the PPO algorithm. Our PPO-based scheme follows the \textit{centralized training, decentralized execution} paradigm, where all agents transmit their collected data to a central node hosting the actor and critic policies, that are shared. Both the actor and critic policies are parameterized using neural networks. The networks are initialized at random, and consist of two convolutional layers with $32$ and $16$ filters respectively, each using a $3 \times 3$ kernel and a stride of 1, followed by two fully connected layers with $256$ and $128$ units, and a final linear output layer. All layers, except the output, employ a ReLU activation function and batch normalization is applied after every layer.
The actor network outputs $8$ logits $z$, which define a categorical distribution via
\[
\pi_\theta(a_t^i \mid s_t) = 
\frac{\exp(z_i)}{\sum_{j=1}^8 \exp(z_j)}, 
\quad i=1,\dots,8.
\]
We augment the critic loss $\mathcal{L}_{\text{c}}$ described in Section \ref{sec:ppo} with a term defined by the Smooth L1 function \cite{smoothl1}, defined as
\[
\text{SmoothL1}(x) =
\begin{cases}
\frac{1}{2}x^{2}, & \text{if } |x| < 1,\\[5 pt]
|x| - \frac{1}{2}, & \text{otherwise},
\end{cases}
\]
where $x = V_\phi(s_t) - \mathcal{G}(\tau_t)$, with $\mathcal{G}(\tau_t)$ the return (Eq. \ref{eq:disc-cum-rwd}) and $V_\phi(s_t)$ the parametrized value function. This is usually done to improve stability, by making the loss quadratic for small errors but linear for large errors, reducing the impact of outliers.
The actor optimization is performed using a clipped surrogate loss $\mathcal{L}_{\text{a}}$ (see Eq. \ref{eq:ppo-objective-final}) with clipping parameter $\epsilon = 0.2$, augmented by an entropy bonus ($c=0.05$) to encourage exploration. Additionally, GAE (see Eq. \ref{eq:gae}) is employed with parameter $\lambda = 0.95$.
\begin{algorithm}[t]
\caption{Centralize-Training Decentralize-Execution PPO}
\label{alg:ppo}
\begin{algorithmic}[1]
\Require actor network $\pi_\theta$, critic network $V_\phi$
\Require clip parameter $\epsilon$, entropy coefficient $c_{\mathrm{ent}}$, discount $\gamma$, GAE parameter $\lambda$
\Require episodes $N_{epi}$, mini-batch size $b_{size}$
\State Initialize parameters $\theta, \phi$
\Repeat
    \State Collect trajectories $\{(s_t^{(n)},a_t^{(n)},r_t^{(n)},s_{t+1}^{(n)})\}_{t=0}^{T-1}$ using $\pi_\theta \,\forall$ agent $n$
    \State Compute returns $R_t^{(n)}$ and advantages $\hat{A}_t^{(n)}$ with GAE
    \State Store old log-probabilities $\ell_{\theta_{\mathrm{old}}}(a_t^{(n)}\mid s_t^{(n)})=\log(\pi_{\theta_{\text{old}}}(a^{(n)} \mid s_t^{(n)})$
    \For{episode = 1 to $N_{epi}$}
        \State Divide trajectories into mini-batches of size $b_{size}$
        \For{each mini-batch}
            \State \textbf{Actor update:}
            \begin{align*}
            \psi_t^{(n)}(\theta) 
                &= \frac{\ell_{\theta}(a_t^{(n)} \mid s_t^{(n)})}
                        {\ell_{\theta_{\mathrm{old}}}(a_t^{(n)} \mid s_t^{(n)})}, \\
            L_t(\theta) 
                &= \min\!\Big( \psi_t^{(n)}(\theta)\hat{A}_t^{(n)}, \; 
                   \mathrm{clip}(\psi_t^{(n)}(\theta), 1-\epsilon, 1+\epsilon)\hat{A}_t^{(n)} \Big), \\
            \mathcal{H}(s_t^{(n)} ; \theta)
                &= - \sum_{a} \pi_\theta(a \mid s_t^{(n)}) 
                   \log \pi_\theta(a \mid s_t^{(n)}), \\
            \mathcal{L}_{\mathrm{a}}(\theta) 
                &= - \frac{1}{b_{size}}\sum_{t \in b_{size}}\Big( L_t(\theta) 
                   + c_{\mathrm{ent}} \,\mathcal{H}(s_t^{(n)} ; \theta) \Big).
            \end{align*}

            \State Update actor parameters $\theta \leftarrow \theta - \eta_\theta \nabla_\theta \mathcal{L}_{\mathrm{a}}(\theta)$
            \State \textbf{Critic update:}
            \[
            \mathcal{L}_{\mathrm{c}}(\phi) = \frac{1}{b_{size}}\sum_{t\in b_{size}} \mathrm{SmoothL1}\big(V_\phi(s_t^{(n)}), R_t^{(n)}\big)
            \]
            \State Update critic parameters $\phi \leftarrow \phi - \eta_\phi \nabla_\phi \mathcal{L}_{\mathrm{c}}(\phi)$
        \EndFor
    \EndFor
    \State Set $\theta_{\mathrm{old}} \leftarrow \theta$ and $\phi_{\mathrm{old}} \leftarrow \phi$ for next iteration
\Until{convergence}
\end{algorithmic}
\end{algorithm}
We use the Adam optimizer \cite{Adam} to update both the actor and critic networks, with separate learning rates $\text{lr}_a = 5 \times 10^{-6}$ and $\text{lr}_c = 5 \times 10^{-5}$, respectively. The Adam momentum parameters are set to their default values, namely $\beta_1 = 0.9$, $\beta_2 = 0.999$. Algorithm \ref{alg:ppo} contains full details on the employed PPO steps. At the end of the actor and critic update phase, the data of the current episode are discarded, hence the learning at the end of the episode happens exclusively by exploiting recent experience, collected by the last version of the actor policy. To ensure statistical robustness of our observations we conducted $n_{\pi} = 10$ independent training procedures (differently seeded for reproducibility).
\subsection{Basic training scenario}
\label{sec:learning_policy_var}
\begin{figure}[t]
    \centering
    \begin{subfigure}[b]{0.48\linewidth}
        \centering
        \includegraphics[width=\linewidth]{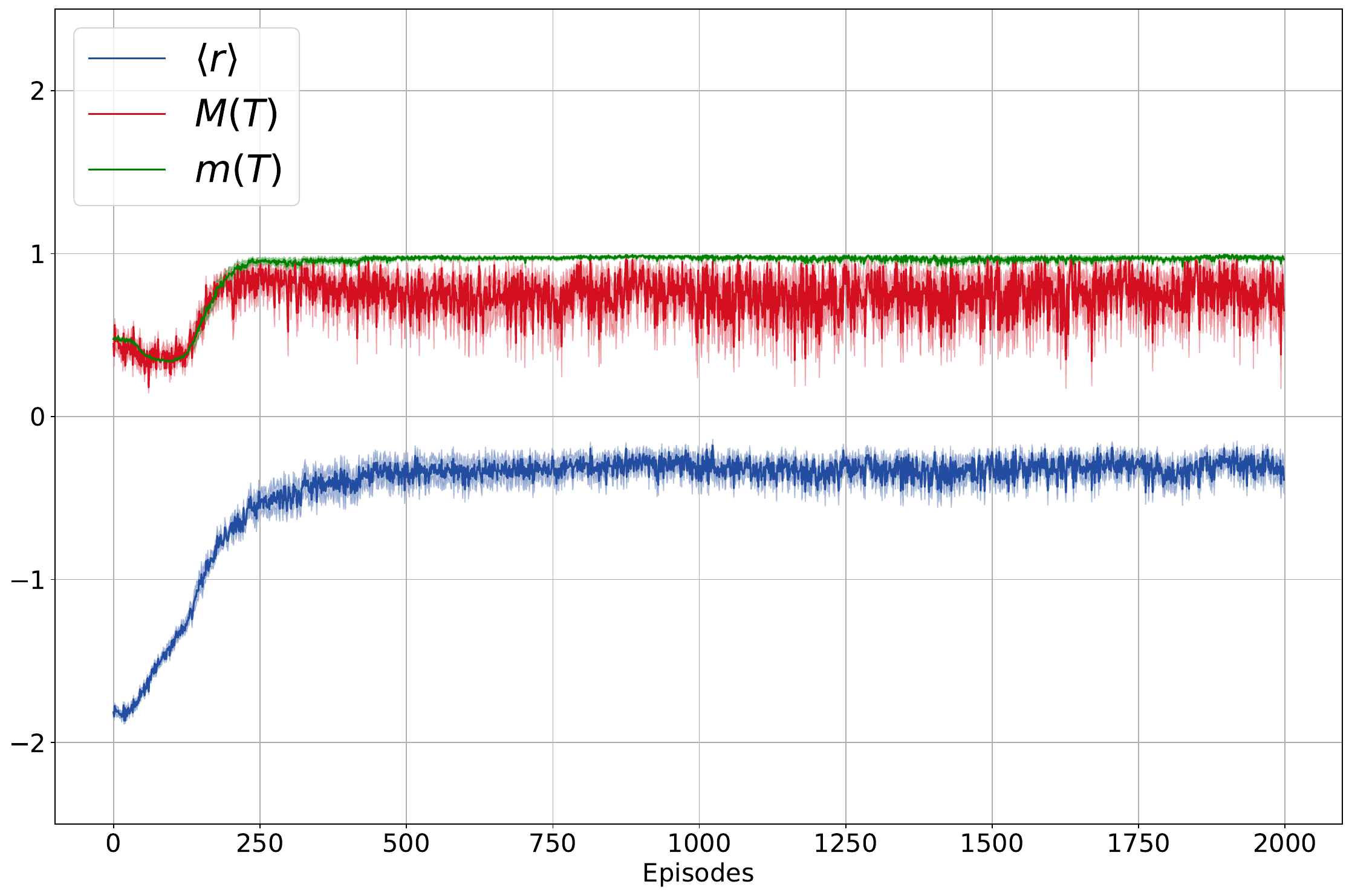}
        \caption{}
        \label{fig:avg_res_rand}
    \end{subfigure}
    \hfill
    \begin{subfigure}[b]{0.48\linewidth}
        \centering
        \includegraphics[width=\linewidth]{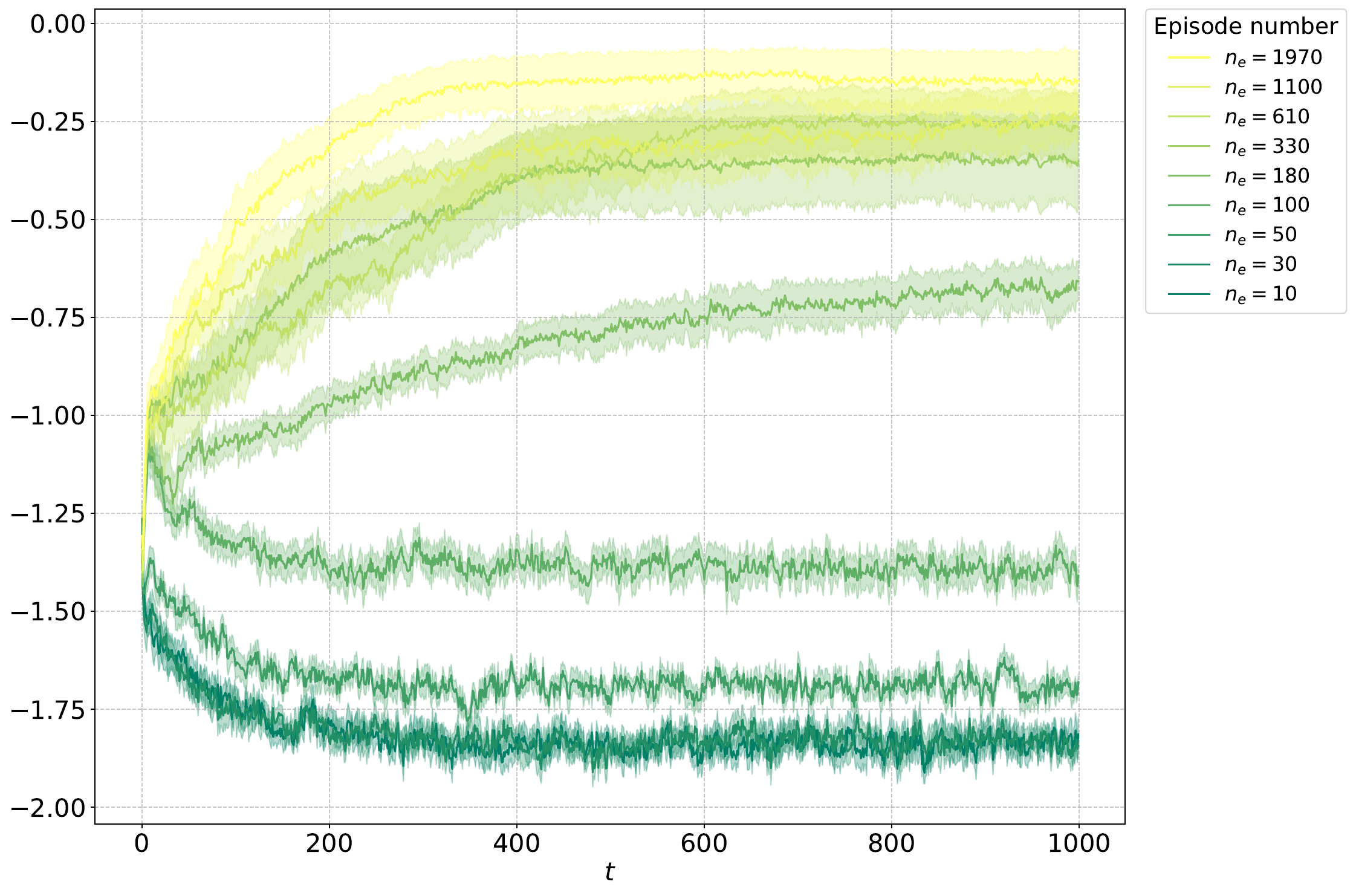}
        \caption{}
        \label{fig:avg_epi_rand}
    \end{subfigure}
    \caption{\textbf{Basic training scenario: metrics in training, averaged over $n_{\pi}=10$ learning realizations.} (a) The evolution of the average reward $\langle r \rangle$ across episodes exhibits noticeable variability due to randomly sampled ABM-ant sensory parameters. The order parameters $M(T)$ and $m(T)$ fluctuate accordingly, reflecting population-dependent differences in trail formation efficiency. (b) The averaged reward trajectories along an episode show gradual improvement over training episodes despite the realization variability. Later episodes tend to achieve higher rewards than earlier ones, although convergence presents small fluctuations, indicating the presence of stochasticity. Early progress is fast, then the curve stabilize, hence rewards have been plotted on episodes corresponding to a geometric scale.}
    \label{fig:combined_rand_figures}
\end{figure}

\paragraph{Learning evolution.}
In the first training scenario, constituted by $N_{epi} = 3000$, ABM-ants are characterized by sensory parameters $\left[\delta, \, \beta \right]$ that are randomly sampled at the beginning of each episode from the predefined sets $\mathsf{s}_\beta$ and $\mathsf{s}_\delta$ defined in Section \ref{sec:Materials_and_methods}.\ref{sec:Ants_swarmmodel} - \ref{sec:sim_phasetrans}. The evaluation of the order parameters (Eqs. \ref{eq:op-M} - \ref{eq:op-m}) during training provides insight into the phase of the system, (and in particular the value achieved at the final step of the episode, when the scenario is at equilibrium), while the average reward serves as an indicator of the emergence of a populated trail. 
In Figure \ref{fig:avg_res_rand} (that contains the evolution of these indicators across successive episodes, averaged among $n_{\pi} = 10$ learning realizations) both the average rewards $\langle r \rangle$ and the density order parameter per episode  initially increase, but they eventually converge to lower values in this case, exhibiting stronger fluctuations. This reflects the presence in training of $\left[\delta , \beta \right]$-configurations that are more difficult (or even impossible) to control, at high noise and low sensitivity (bottom-right corner of Figure \ref{fig:rand_gridplot}). To further assess the evolution of the policy performance throughout learning, per each of the $n_{\pi}$ training procedures we conduct intermediate evaluations of the last updated version of the actor every $10$ episodes. The corresponding averaged reward trajectories are shown in Figure \ref{fig:avg_epi_rand}. The improvement of the smart-agents is evident, as rewards obtained in early episodes (green curves) are consistently lower than those accumulated in later episodes (yellow curves), indicating that the smart-agents on average improve their performance as training progresses, even if the couple $\left[\delta , \beta \right]$, that determined the test configuration, is chosen at random from the sets. 

\paragraph{Policy analysis.}
Gaining a detailed understanding of how the smart-agents, using a shared policy, are able to generate and maintain a stable trail that is also followed by the ABM-ants is particularly challenging due to the high dimensionality of the state space. Rather than attempting a full characterization of the policy, we therefore focus on alternative forms of analysis. Specifically, we examine a set of representative or critical scenarios, as well as the history of actions executed during the testing phase, in order to extract qualitative insights into the emergent behaviours and decision-making strategies learned by the agents.

\begin{figure}[t]
    \centering
    \begin{subfigure}[t]{0.32\textwidth}
        \centering
        \includegraphics[width=\linewidth]{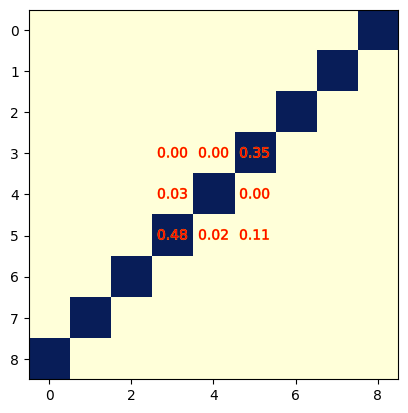}
        \caption{}
        \label{fig:policy1}
    \end{subfigure}
    \hfill
    \begin{subfigure}[t]{0.32\textwidth}
        \centering
        \includegraphics[width=\linewidth]{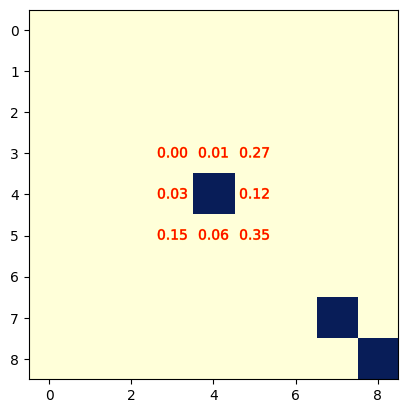}
        \caption{}
        \label{fig:policy2}
    \end{subfigure}
    \hfill
    \begin{subfigure}[t]{0.32\textwidth}
        \centering
        \includegraphics[width=\linewidth]{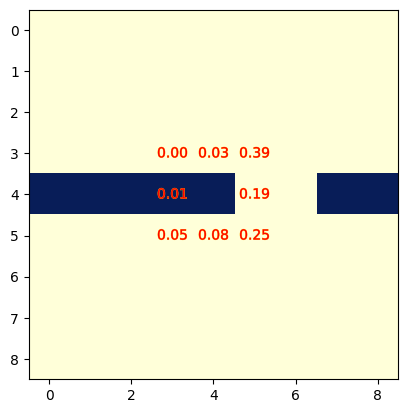}
        \caption{}
        \label{fig:policy3}
    \end{subfigure}
    \caption{\textbf{Examples of policy in synthetic states.} In blu the sites with more pheromones than the average seen by the smart-agent in the center, in red the probabilities to jump to neighbors sites.}
    \label{fig:policies}
\end{figure}

In particular, we analyze the choices made by the policy in synthetic states, showcasing scenarios of interest. The synthetic states shown in Figure \ref{fig:policies} are constructed as follows. In the first channel, a baseline pheromone value of $10$ is assigned to desired sites. In the second channel, one ABM-ant is placed at the same locations, while the third channel contains a single smart-agent positioned at the center. For each synthetic state, we generate eight samples, one for each possible direction of motion (populating the fourth channel). These states are then passed through the actor, obtained in the basic training scenario, to compute the probabilities of the eight possible actions for each of the $n_\pi$ policy trained in the random configuration scenario. Finally, the probabilities are averaged over all directions and the policies, to characterize the learned behaviour in an unbiased manner. In the first example (Figure \ref{fig:policy1}), the smart-agent located at the center continues to reinforce the existing trail with high probability ($\sim 80\%$). Non-diagonal moves are discouraged, as they would increase connectivity, and only $\sim 10\%$ of the time does the agent attempt to create a new diagonal path. In the second example (Figure \ref{fig:policy2}), the smart-agent is attracted to the pheromone trail in the bottom-right area of its field of view (with probability $\sim 35\%$), while still maintaining some tendency to explore. Finally, in the third example (Figure \ref{fig:policy3}), the smart-agent is strongly driven to fill the gap. Diagonal actions remain more likely, even though the existing trail is horizontal. This behaviour has also been observed in the original ant swarm model, where diagonal trails emerge without any bias introduced by a learned policy.\\

\begin{figure}
    \centering
    \includegraphics[width=0.8\linewidth]{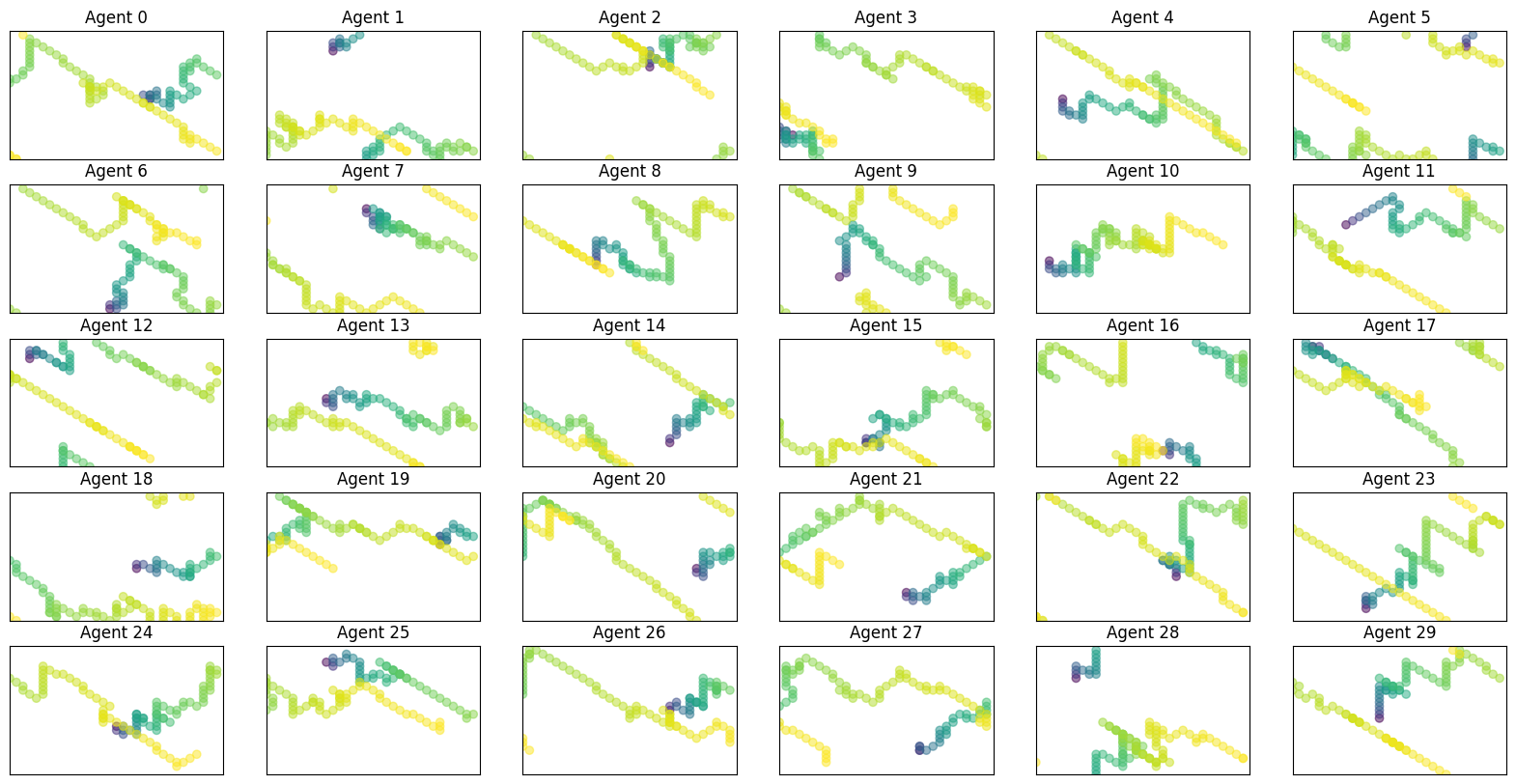}
    \caption{\textbf{Trajectories of smart-agents in the initial phase.} Shown are the paths of $N' = 30$ smart-agents during the first $t = 100$ time-steps in a test using ABM-ants $\left[\delta = 0.5, \, \beta = 6.5\right]$. Trajectories are colored from blue to yellow to indicate the passage of time. This early phase highlights the formation of main trails, as smart-agents initially explore the environment rapidly and subsequently converge on a single path that maximizes their reward.
}
    \label{fig:initial-phase}
\end{figure}

In Figure \ref{fig:initial-phase}, we show the trajectories of the $N' = 30$ smart-agents during the first $t = 100$ time-steps of a test using one of the candidate policies trained in the basic training scenario, alongside a population of ABM-agents with randomly chosen sensory parameters $\left[\delta = 0.5, \, \beta = 6.5\right]$. This initial phase is particularly interesting to analyze, as it allows us to observe how the main trails are formed. From Figure \ref{fig:initial-phase} we can observe that, initially, the smart-agents explore the search space more stochastically, shaping the pheromone field through their exploration. Over time, they progressively concentrate on the most promising trails, reinforcing them and guiding the collective behaviour toward stronger solutions. After this period, a high-reward configuration emerges and also the ABM-ants begin to follow it consistently. 

\subsection{Limited training scenario: ABM with $\left[ \delta^* = 0.3, \,\beta^* = 4.5 \right]$}
\label{sec:learning_policy_spec}

\begin{figure}[t]
    \centering
    \begin{subfigure}[b]{0.48\linewidth}
        \centering
        \includegraphics[width=\linewidth]{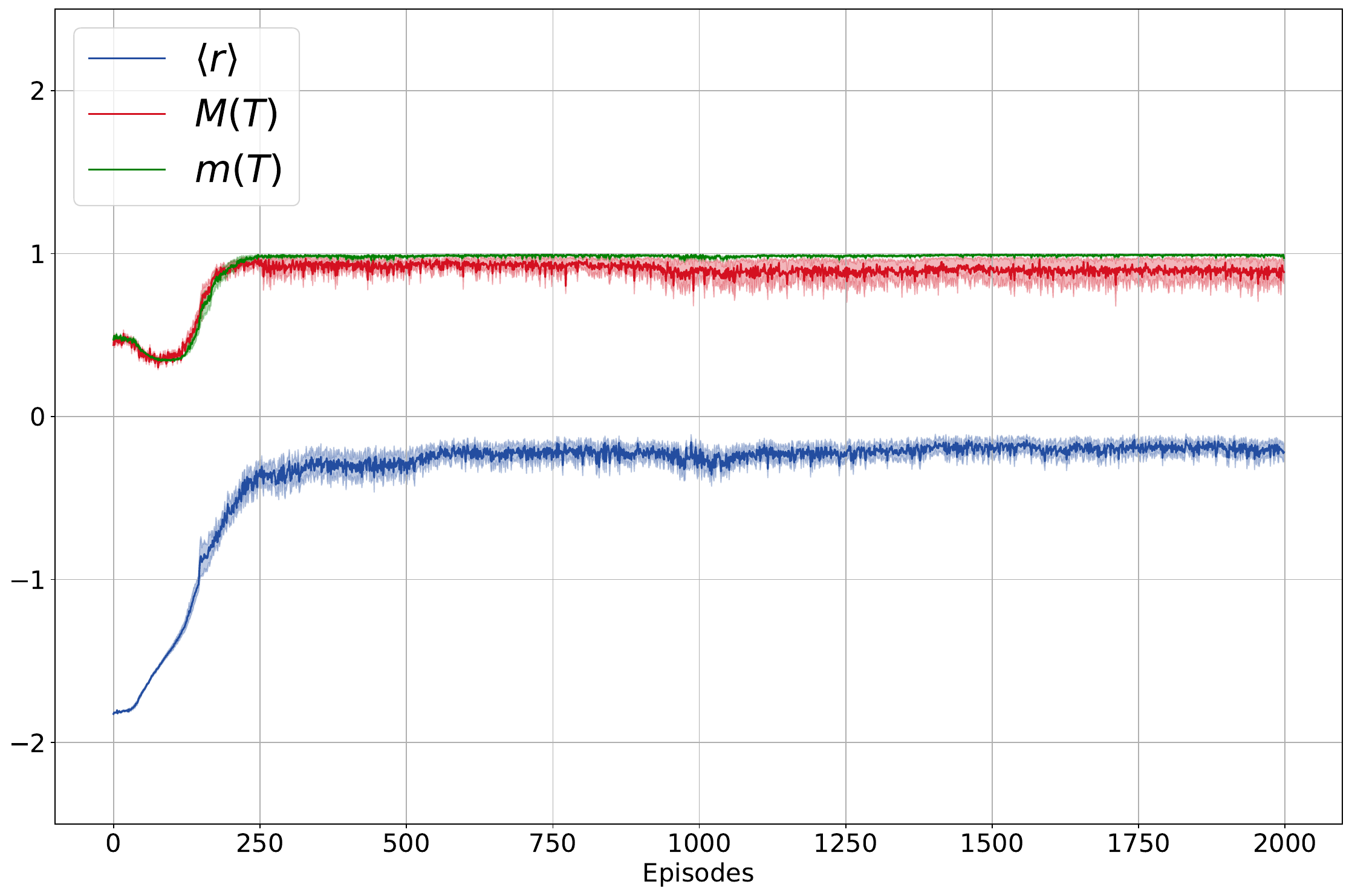}
        \caption{}
        \label{fig:avg_res_nonrand}
    \end{subfigure}
    \hfill
    \begin{subfigure}[b]{0.48\linewidth}
        \centering
        \includegraphics[width=\linewidth]{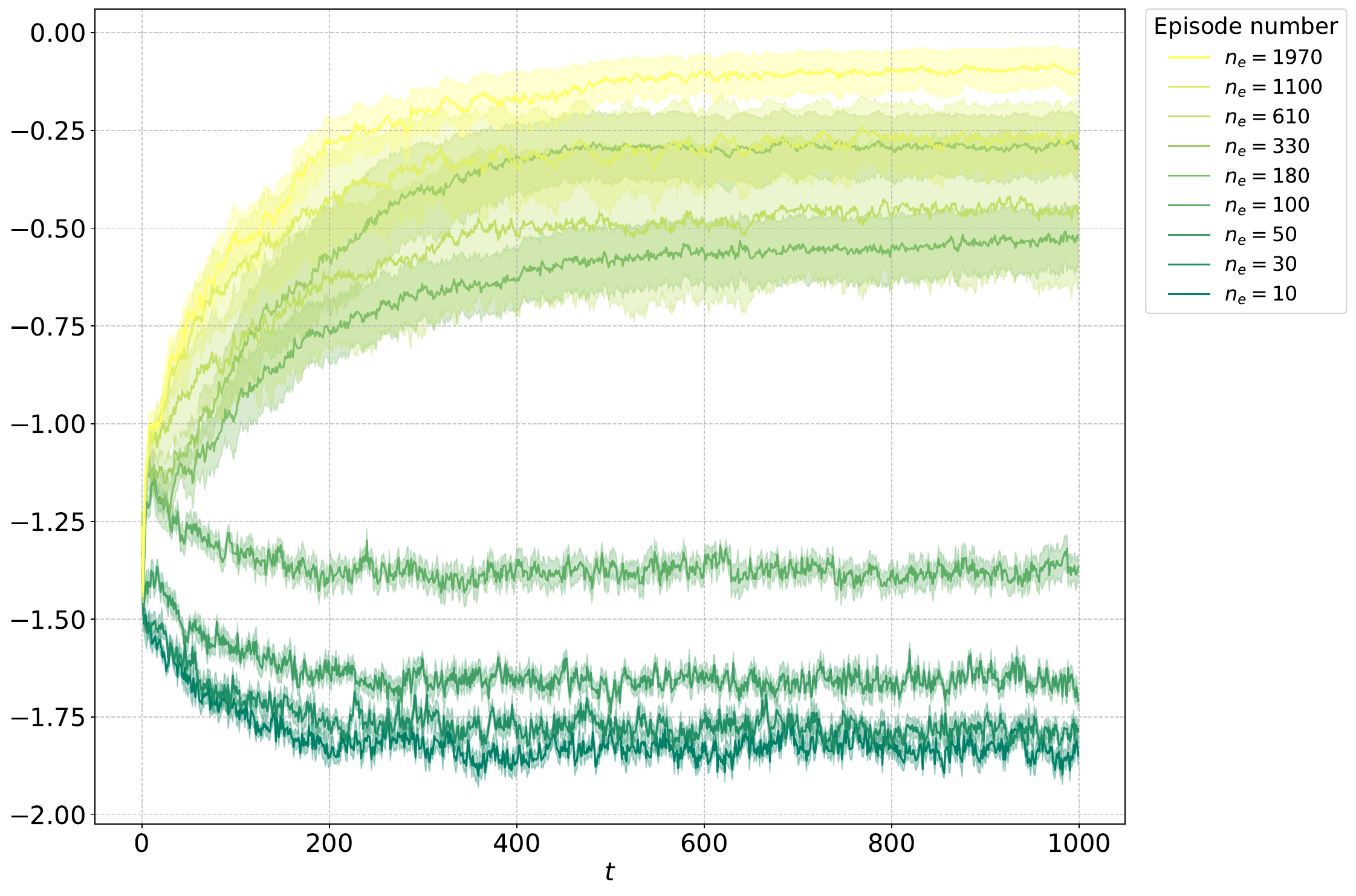}
        \caption{}
        \label{fig:avg_epi_nonrand}
    \end{subfigure}
    \caption{\textbf{Limited training case: metrics in training, averaged over $n_{\pi}=10$ learning realizations.} (a) The evolution of the average reward $\langle r \rangle$ across episodes shows a steady improvement. This, jointly with the convergence of the order parameters, at the final time-step $t=T$, $M(T)$ and $m(T)$ toward $1$, indicate the evolution of the policy towards behaviours promoting the emergence of a pheromone trail, consistently populated by ABM-ants. (b) The averaged reward trajectories along an epsode show consistently higher rewards in later episodes (yellow curves) compared to early ones. Moreover, the individual curves appear to become gradually more monotonically increasing, indicating that the policy learns behaviours that enhance trail formation and does so increasingly faster within each episode. Early progress is fast, then the curve stabilize, hence rewards have been plotted on episodes corresponding to a geometric scale.}
    \label{fig:combined_abm_figures}
\end{figure}

\paragraph{Learning evolution.}
The second training scenario considers $N_{epi} = 2000$ episodes with ABM-ants characterized, in each episode, by sensory parameter pair $\left[ \delta^* = 0.3, \,\beta^* = 4.5 \right]$. As shown in Figure \ref{fig:avg_res_nonrand}, the average reward $\langle r \rangle$ per episode progressively increases over training and approaches values closer to $0$ compared to the initial episodes. A similar trend is observed for the order parameters, which converge toward values close to their maximum of $1$. Together, these behaviours indicate the emergence of a stable trail structure, consistent with the observations discussed in Section \ref{sec:Results}, and analogous to what seen in the basic training scenario. The corresponding averaged reward trajectories are shown in Figure \ref{fig:avg_epi_nonrand}.  The improvement of the smart-agents, as in the previous case, is evident.

\section{Additional results}
\label{sec:additional_res}
\subsection{Basic training scenario: testing}
\label{sec:additional_res_var}
\begin{figure}[t]
    \centering
    \begin{subfigure}[b]{0.30\textwidth}
        \includegraphics[width=\textwidth]{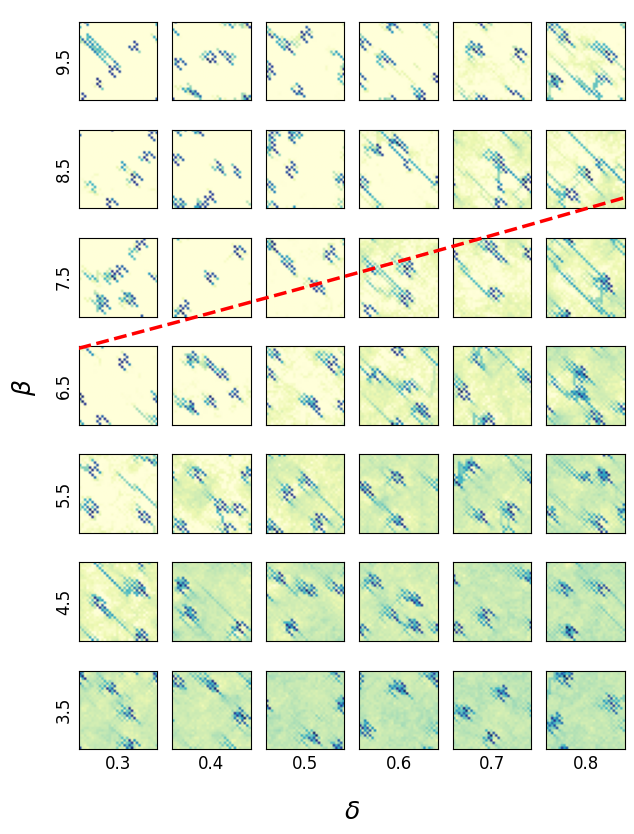}
        \caption{}
        \label{sfig:l1_pher}
    \end{subfigure}
    \hspace{10pt}
    \begin{subfigure}[b]{0.30\textwidth}
        \includegraphics[width=\textwidth]{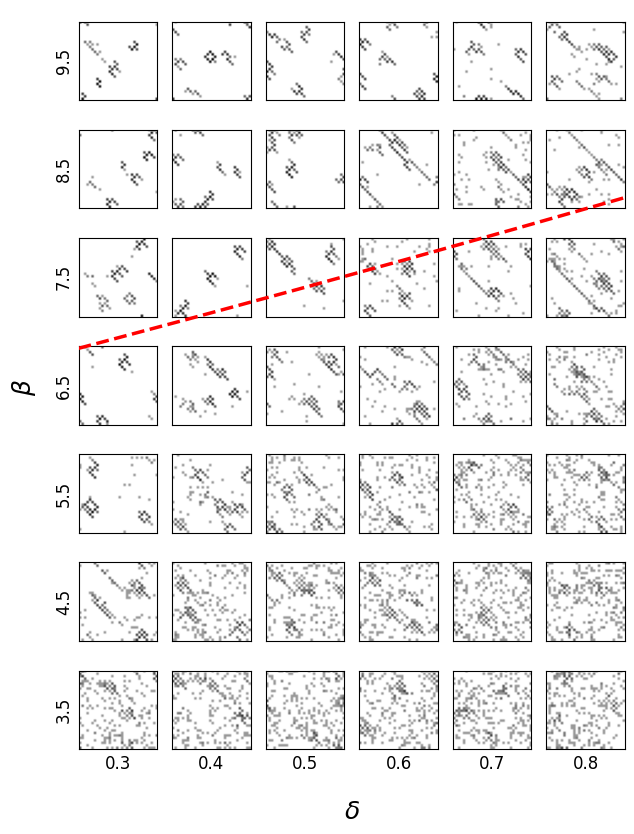}
        \caption{}
        \label{sfig:l1_ants}
    \end{subfigure}
    \vspace{0.2cm} 

    \begin{subfigure}[b]{0.30\textwidth}
        \includegraphics[width=\textwidth]{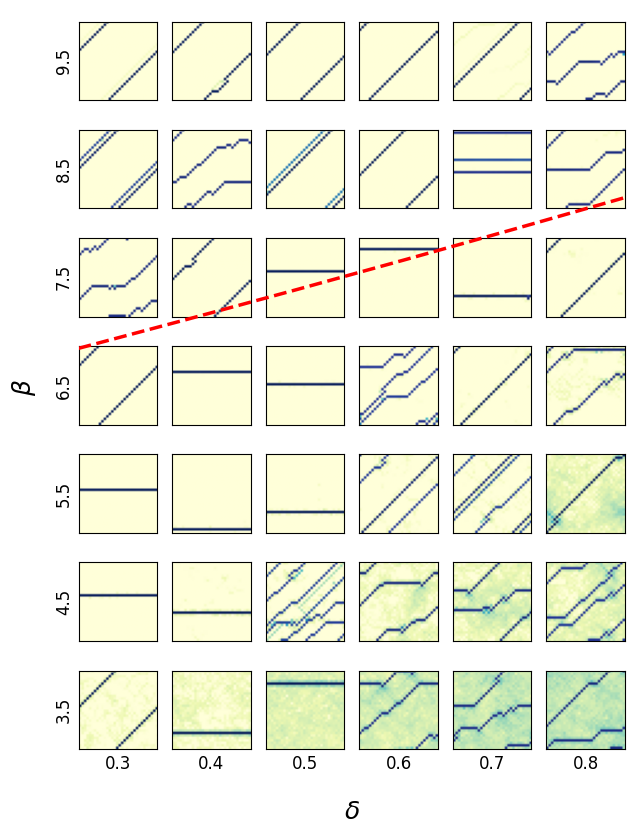}
        \caption{}
        \label{sfig:l2_pher}
    \end{subfigure}
    \hspace{10pt}
    \begin{subfigure}[b]{0.30\textwidth}
        \includegraphics[width=\textwidth]{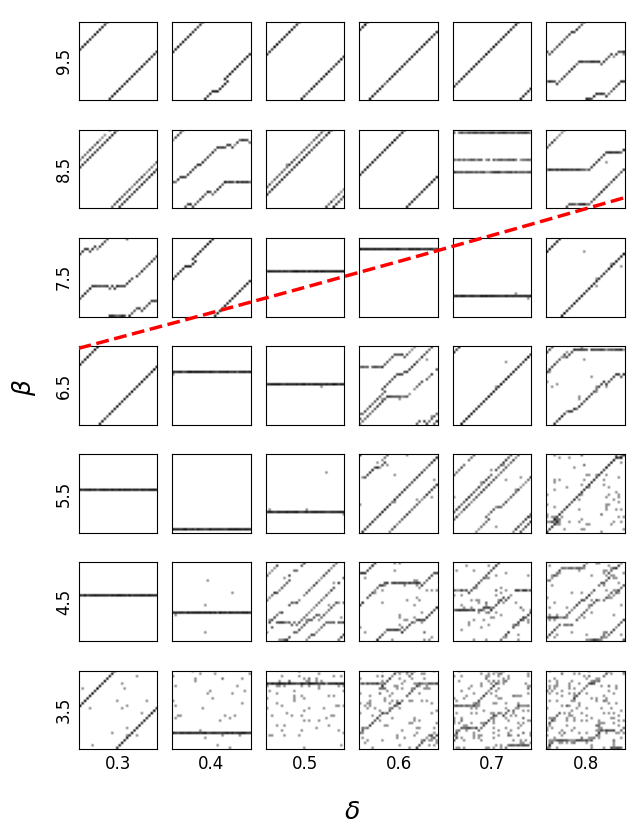}
        \caption{}
        \label{sfig:l2_ants}
    \end{subfigure}
    \vspace{0.2cm}

    \begin{subfigure}[b]{0.30\textwidth}
        \includegraphics[width=\textwidth]{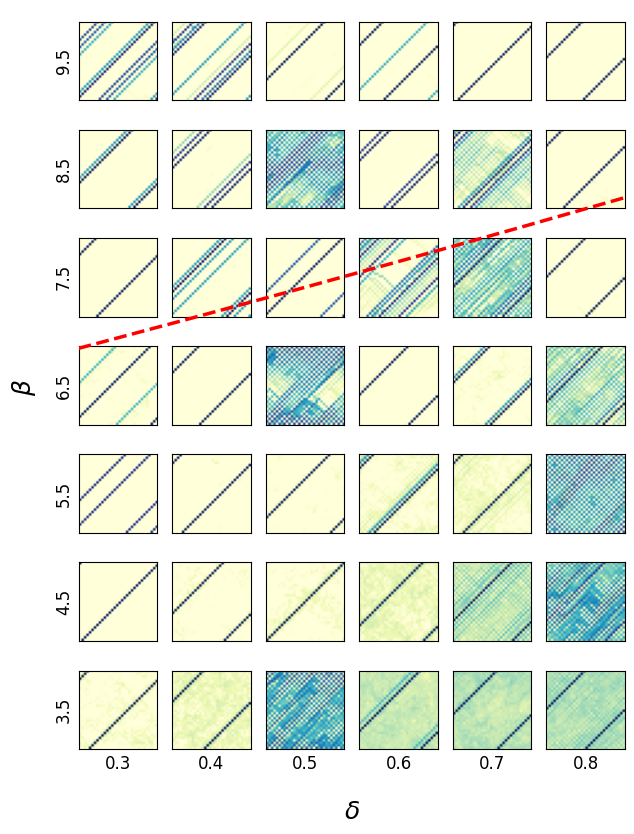}
        \caption{}
        \label{sfig:l3_pher}
    \end{subfigure}
    \hspace{10pt}
    \begin{subfigure}[b]{0.30\textwidth}
        \includegraphics[width=\textwidth]{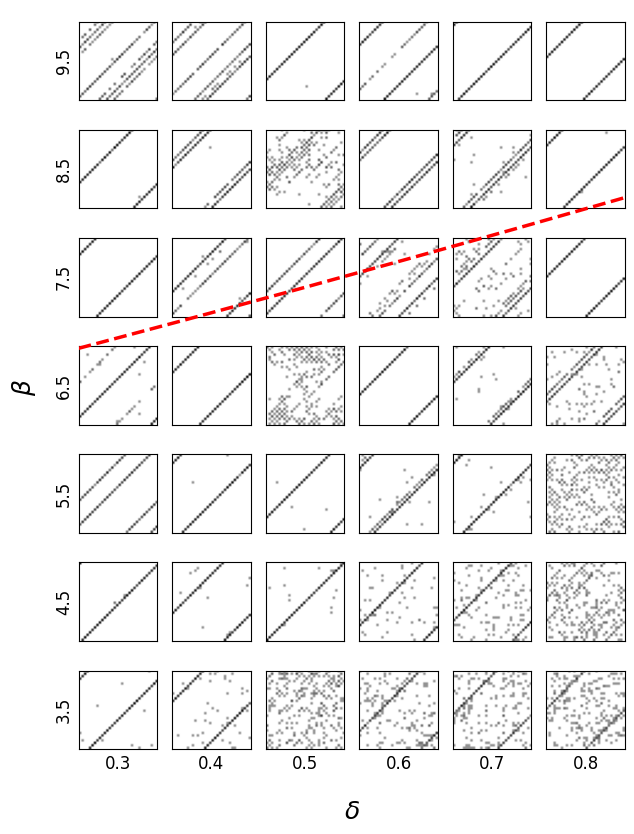}
        \caption{}
        \label{sfig:l3_ants}
    \end{subfigure}

    \caption{\textbf{Basic training scenario: testing}. Phase diagrams of pheromone (left) and ABM-ants density (right) for policies $\pi=1$ to $\pi=3$.} 
    \label{fig:grid_1to3}
\end{figure}

\begin{figure}[H]
    \centering
    \begin{subfigure}[b]{0.30\textwidth}
        \includegraphics[width=\textwidth]{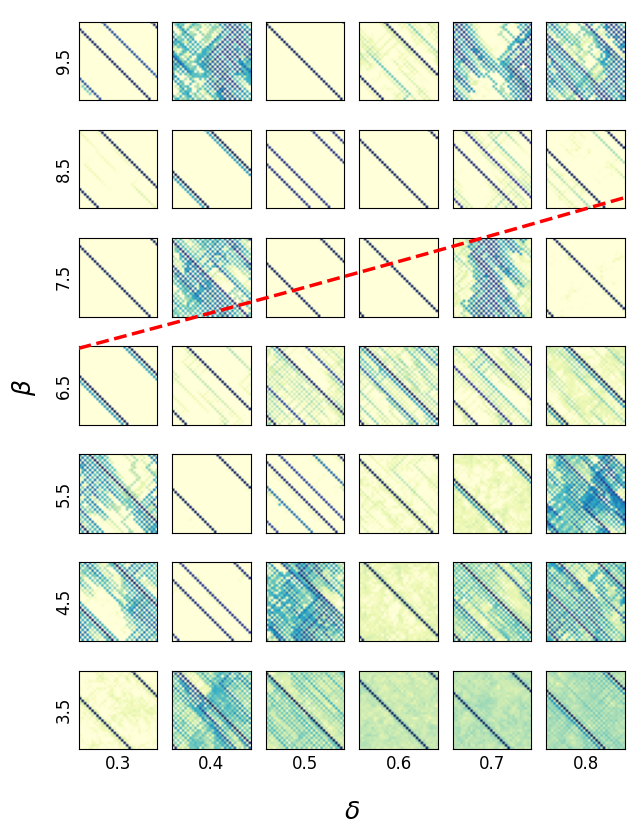}
        \caption{}
        \label{sfig:l4_pher}
    \end{subfigure}
    \hspace{10pt}
    \begin{subfigure}[b]{0.30\textwidth}
        \includegraphics[width=\textwidth]{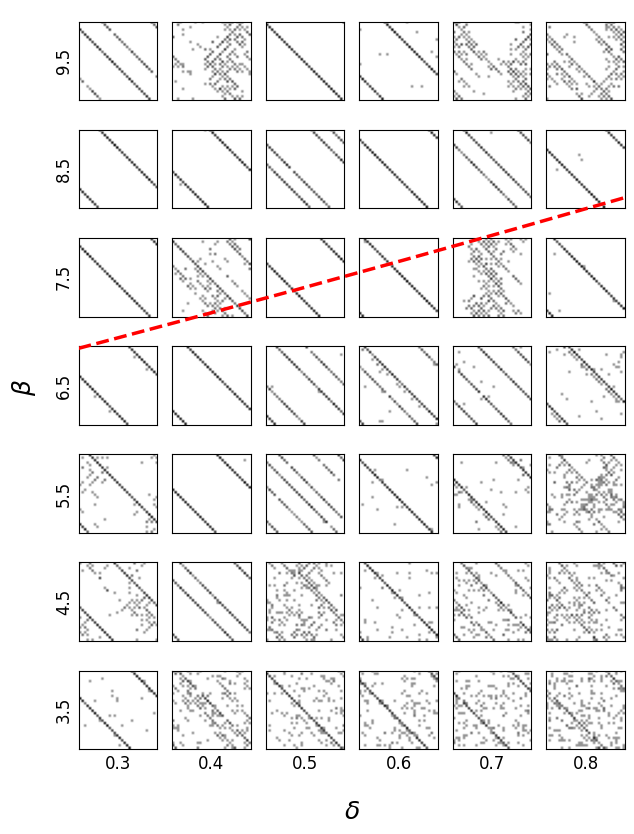}
        \caption{}
        \label{sfig:l4_ants}
    \end{subfigure}
    \vspace{0.2cm}

    \begin{subfigure}[b]{0.30\textwidth}
        \includegraphics[width=\textwidth]{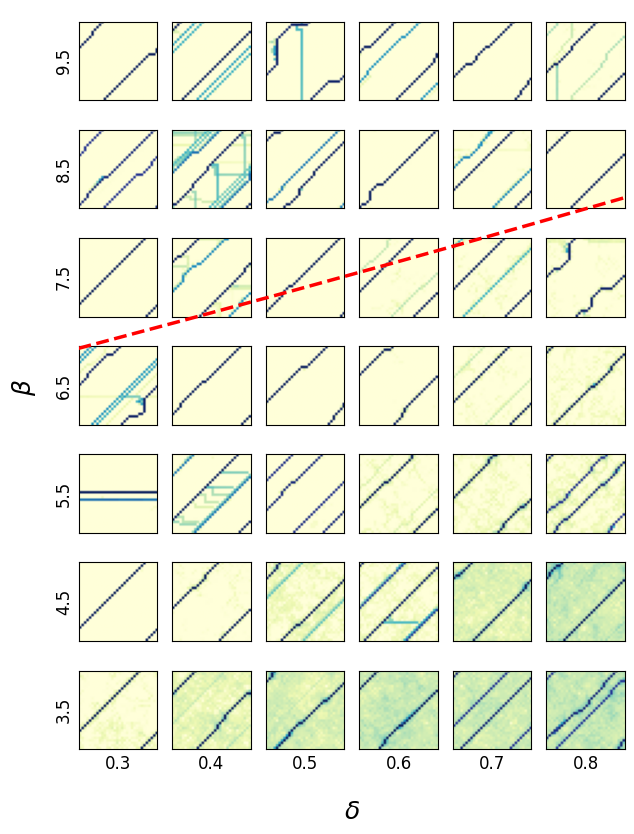}
        \caption{}
        \label{sfig:l5_pher}
    \end{subfigure}
    \hspace{10pt}
    \begin{subfigure}[b]{0.30\textwidth}
        \includegraphics[width=\textwidth]{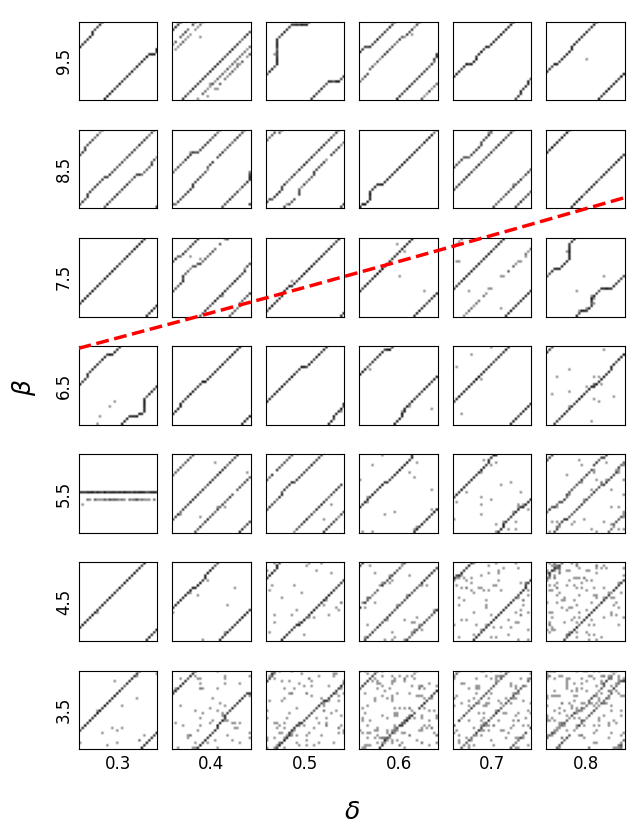}
        \caption{}
        \label{sfig:l5_ants}
    \end{subfigure}
    \vspace{0.2cm}

    \begin{subfigure}[b]{0.30\textwidth}
        \includegraphics[width=\textwidth]{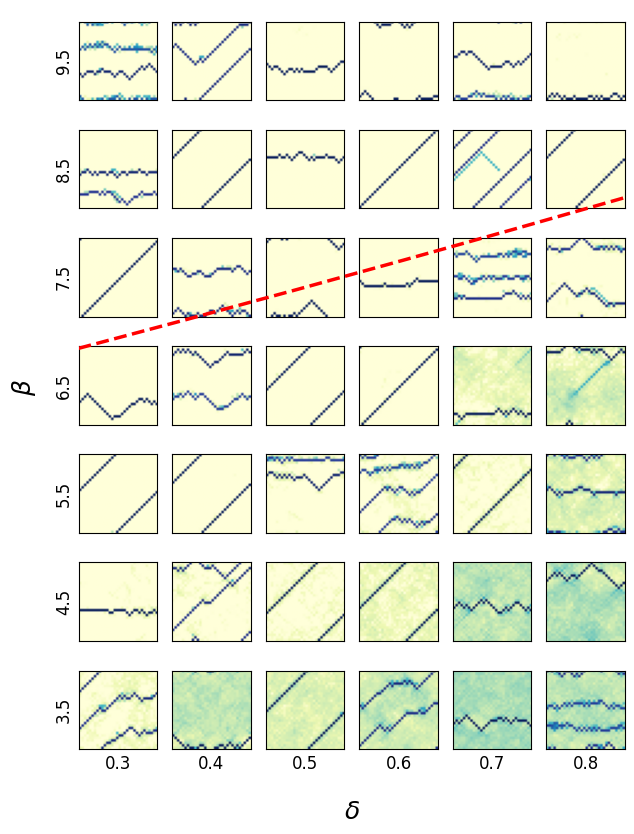}
        \caption{}
        \label{sfig:l6_pher}
    \end{subfigure}
    \hspace{10pt}
    \begin{subfigure}[b]{0.30\textwidth}
        \includegraphics[width=\textwidth]{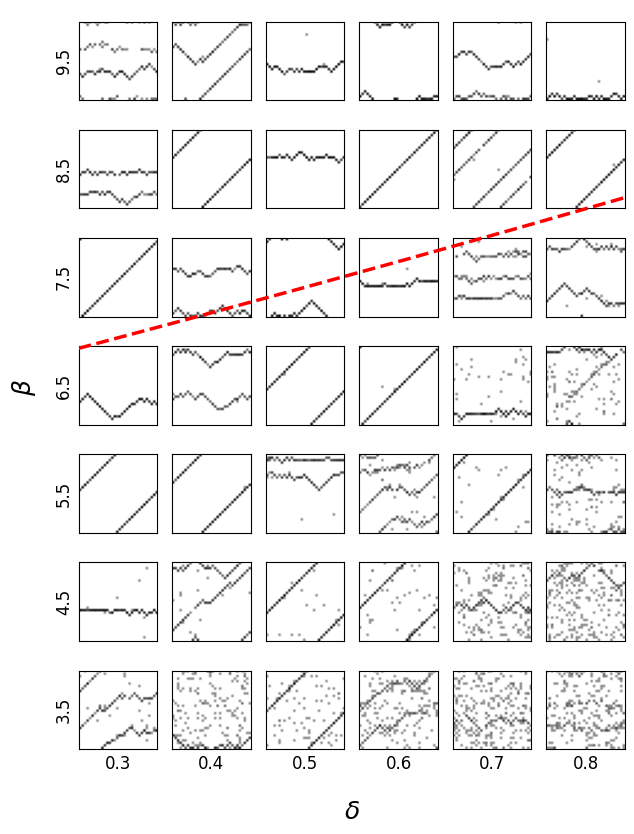}
        \caption{}
        \label{sfig:l6_ants}
    \end{subfigure}

    \caption{\textbf{Basic training scenario: testing}. Phase diagrams of pheromone (left) and ABM-ants density (right) for policies $\pi=4$ to $\pi=6$.} 
    \label{fig:grid_4to6}
\end{figure}

\paragraph{Final pheromone and individual concentrations.}
Figures \ref{fig:grid_1to3}\subref{sfig:l1_ants} and \ref{fig:grid_1to3}\subref{sfig:l1_pher}, associated at the policy $\pi=1$, exhibit qualitatively different
    behaviour from the others, consistent with its poor training outcome in terms of average reward.
    Figures \ref{fig:grid_1to3}\subref{sfig:l3_ants}, \ref{fig:grid_1to3}\subref{sfig:l3_pher}, \ref{fig:grid_1to3}\subref{sfig:l4_ants}
    and \ref{fig:grid_1to3}\subref{sfig:l4_pher} (policies $\pi=3,4$) occasionally fail to form trails, though the success
    rate appears to depend on the stochasticity of the system rather than on the sensory parameters.
    All remaining policies consistently reproduce the desired behaviour observed in the Section \ref{sec:Results}
    with policy $\pi=0$, characterized by a shift of the phase line. \\

\newpage

\begin{figure}[H]
    \centering
    \begin{subfigure}[b]{0.30\textwidth}
        \includegraphics[width=\textwidth]{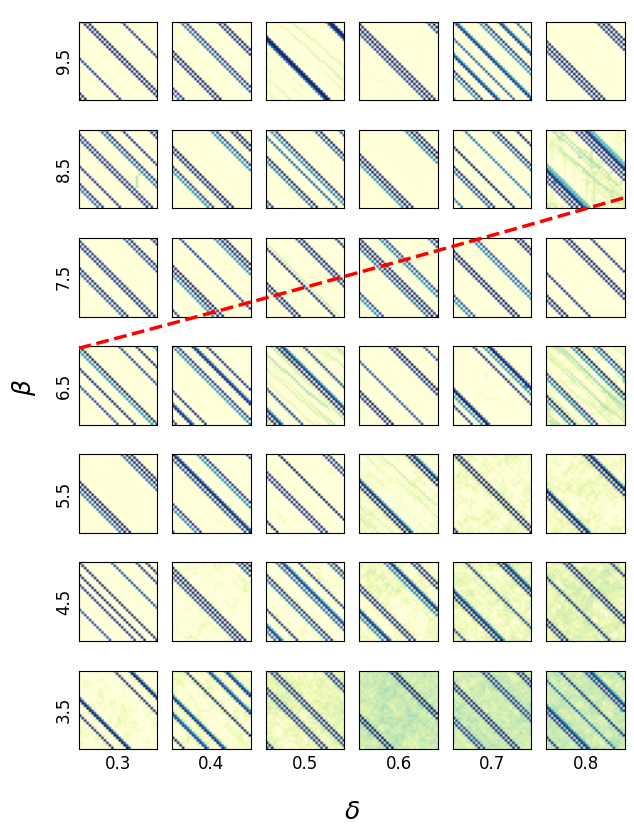}
        \caption{}
        \label{sfig:l7_pher}
    \end{subfigure}
    \hspace{10pt}
    \begin{subfigure}[b]{0.30\textwidth}
        \includegraphics[width=\textwidth]{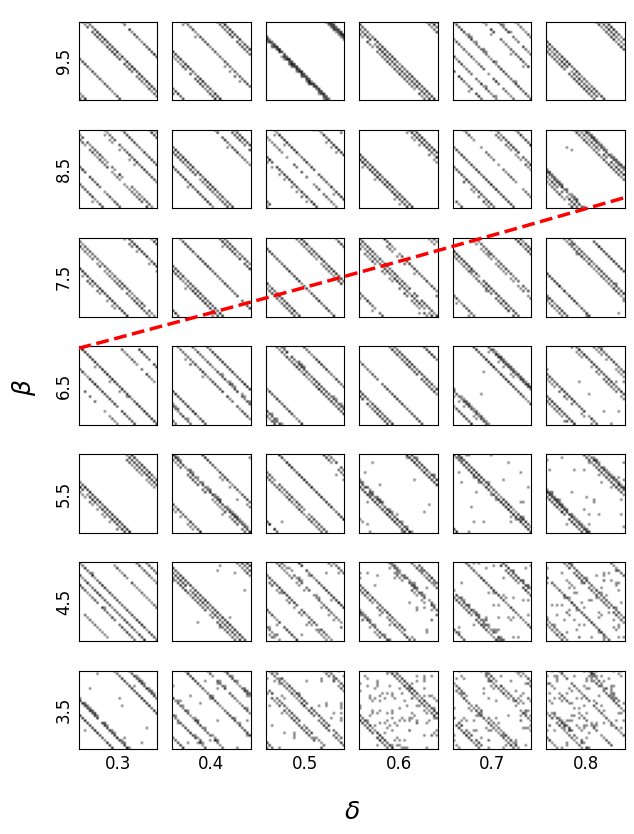}
        \caption{}
        \label{sfig:l7_ants}
    \end{subfigure}
    \vspace{0.2cm}

    \begin{subfigure}[b]{0.30\textwidth}
        \includegraphics[width=\textwidth]{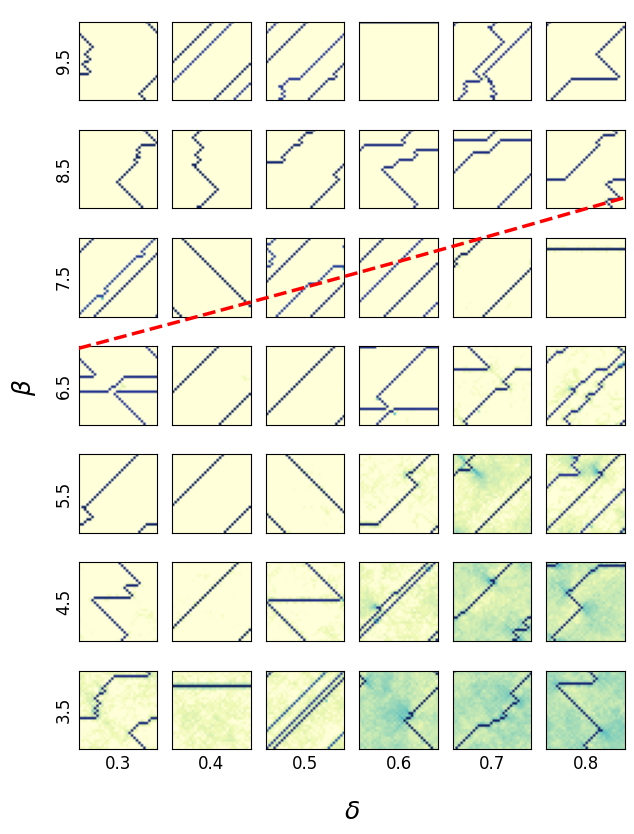}
        \caption{}
        \label{sfig:l8_pher}
    \end{subfigure}
    \hspace{10pt}
    \begin{subfigure}[b]{0.30\textwidth}
        \includegraphics[width=\textwidth]{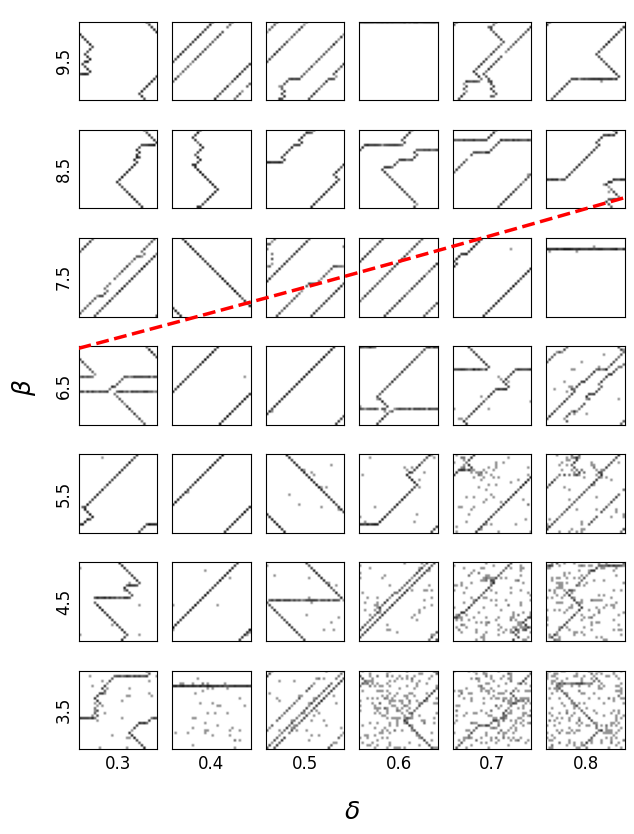}
        \caption{}
        \label{sfig:l8_ants}
    \end{subfigure}
    \vspace{0.2cm}

    \begin{subfigure}[b]{0.30\textwidth}
        \includegraphics[width=\textwidth]{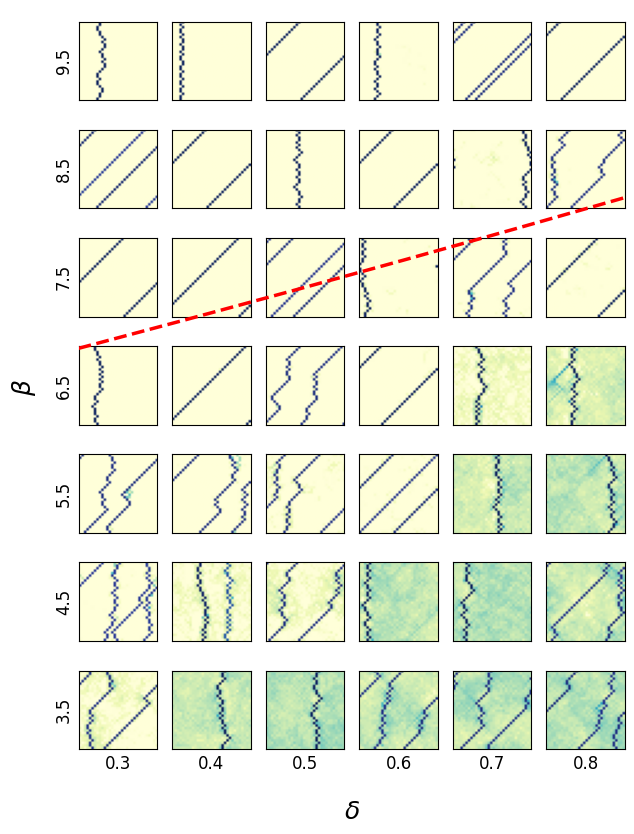}
        \caption{}
        \label{sfig:l9_pher}
    \end{subfigure}
    \hspace{10pt}
    \begin{subfigure}[b]{0.30\textwidth}
        \includegraphics[width=\textwidth]{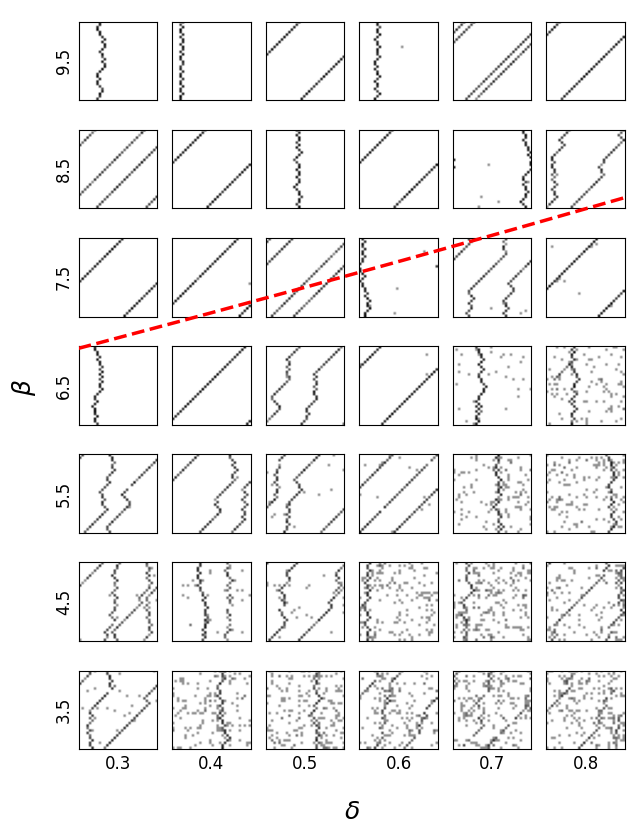}
        \caption{}
        \label{sfig:l9_ants}
    \end{subfigure}

    \caption{\textbf{Basic training scenario: testing}. Phase diagrams of pheromone (left) and ABM-ants density (right) for the remaining 3 policies ($\pi=7$ to $\pi=9$). (Ref: Original 9-policy grid.)} 
    \label{fig:grid_7to9}
\end{figure}

\begin{figure}[H]
    \centering
    \includegraphics[width=0.94\linewidth]{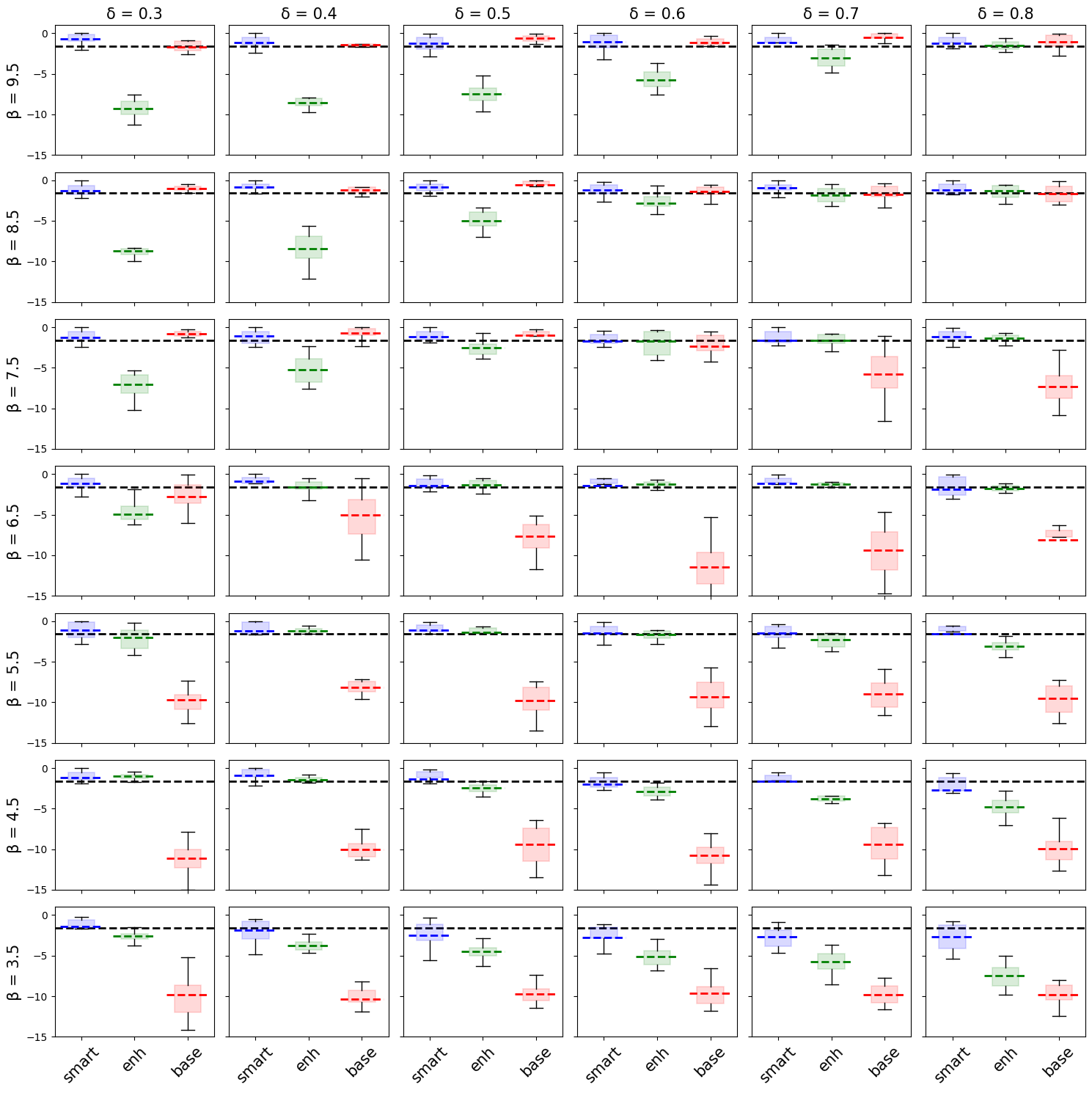}
    \caption{\textbf{Trail scenario metric}. Values of $\langle \mathcal{TS} \rangle_{50}$ over all the simulated evaluations, divided per each $\left[ \delta, \, \beta \right]$-configuration, for the three considered settings: (1) $N$ ABM-ants (base); (2) baseline population plus $N_{\eta=1}$ ABM-ants releasing higher pheromone (enh); (3) baseline population plus $N'$ smart-agents (smart). The horizontal dashed line represents the empirical threshold of $-1.6$, observed in relation with trail formation.}
    \label{fig:boxplot-delta-beta-ST50}
\end{figure}

\clearpage
\paragraph{Trails scenario metric across the phase diagram.} Figure \ref{fig:boxplot-delta-beta-ST50} shows that the average values included in the main part of the paper for the baseline, enhanced and smart setups present a trend that is robustly replicated across all the evaluation campaign, spanning all the $[\delta, \, \beta]$-configurations.

\begin{figure}[t]
    \centering
    
    \begin{minipage}{0.6\textwidth}
        \centering
        \resizebox{0.9\textwidth}{!}{
        \begin{tabular}{|l|rrrrrr|}
        \toprule 
        $\quad$ & $\quad$ & $\quad$ & $\langle\mathcal{TS}\rangle_{50}$ & $\quad$ & $\quad$ & $\quad$\\
        \midrule
        \diagbox[innerwidth=0.4cm]{$\beta$}{$\delta$} & 0.3 & 0.4 & 0.5 & 0.6 & 0.7 & 0.8 \\
        \midrule
        $\quad$ & \multicolumn{2}{l}{baseline} & $\quad$ & $\quad$ & $\quad$ & $\quad$ \\
        \midrule
        9.5 & -1.66           & \textbf{-1.43}  & \textbf{-0.62} & \textbf{-1.11}  & \textbf{-0.55} & \textbf{-1.02} \\
        8.5 & \textbf{-1.03}  & \textbf{-1.21}  & \textbf{-0.60} & \textbf{-1.36}  & -1.76          & -1.66 \\
        7.5 & \textbf{-0.79}  & \textbf{-0.72}  & \textbf{-0.97} & -2.39           & -5.78          & -7.36 \\  
        6.5 & -2.79           & -5.00           & -7.63          & -11.41          & -9.36          & -8.15 \\
        5.5 & -9.67           & -8.18           & -9.81          & -9.31           & -9.02          & -9.49 \\
        4.5 & -11.09          & -10.04          & -9.36          & -10.76          & -9.43          & -9.94 \\
        3.5 & -9.83           & -10.33          & -9.74          & -9.62           & -9.83          & -9.84 \\
        \midrule
        $\quad$ & \multicolumn{3}{l}{smart} & $\quad$ & $\quad$ & $\quad$   \\
        \midrule
        9.5 & \textbf{-0.73} & \textbf{-1.18} & \textbf{-1.24} & \textbf{-1.02} & \textbf{-1.16} & \textbf{-1.22} \\
        8.5 & \textbf{-1.28} & \textbf{-0.82} & \textbf{-0.85} & \textbf{-1.22} & \textbf{-0.95} & \textbf{-1.17} \\
        7.5 & \textbf{-1.25} & \textbf{-1.11} & \textbf{-1.17} & -1.68          & -1.62          & \textbf{-1.17} \\
        6.5 & \textbf{-1.16} & \textbf{-0.86} & \textbf{-1.44} & \textbf{-1.38} & \textbf{-1.11} & -1.87          \\
        5.5 & \textbf{-1.08} & \textbf{-1.19} & \textbf{-1.14} & \textbf{-1.50} & \textbf{-1.50} & \textbf{-1.51} \\
        4.5 & \textbf{-1.18} & \textbf{-0.90} & \textbf{-1.33} & -1.97          & -\textbf{1.59} & -2.66          \\
        3.5 & \textbf{-1.41} & -1.86          & -2.45          & -2.75          & -2.71          & -2.70          \\
         \bottomrule
        \end{tabular}
        }
    \vspace{0.8cm}
    \end{minipage}
    \hfill
    \begin{minipage}{0.3\textwidth}
        \centering
        
        \begin{subfigure}[b]{1\linewidth}
            \centering
            \includegraphics[width=\linewidth]{images/cavgbase50.png}
            \caption{Baseline}
            \label{fig:bas50_global}
        \end{subfigure}
        \begin{subfigure}[b]{1\linewidth}
            \centering
            \includegraphics[width=\linewidth]{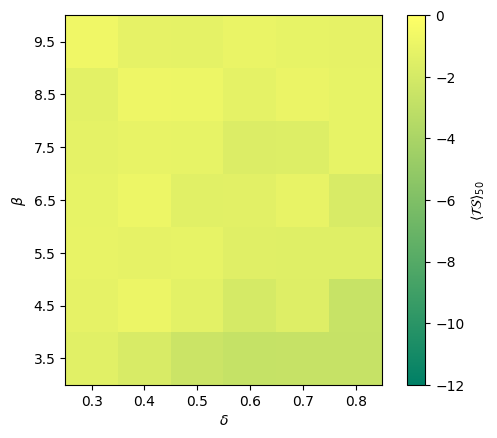}
            \caption{Smart}
            \label{fig:rand50_global}
        \end{subfigure}
     
    \end{minipage}
    \vspace{0.2cm}
    \caption{\textbf{Average trails scenario metric per each $[\delta, \beta]$-configuration}: (a) $N$ ABM-ants (base); (b) baseline population plus $N'$ smart-agents trained across the full phase space (smart).}
    \label{fig:global50_metric}
\end{figure}

Figure \ref{fig:global50_metric}, instead, shows the average values of the global metric $\langle \mathcal{TS} \rangle_{50}$ achieved in the two setups. These results are obtained by averaging over $n_e = 10$ evaluation runs for each configuration. Specifically, in the smart-agents setup, these runs correspond to $n_{\pi} = 10$ distinct policies (one evaluation per policy), whereas in the baseline ABM-ants scenarios they represent $10$ independent stochastic realizations (one per random seed). The behaviour of the metric across the phase diagram is consistent with what is shown in Figures \ref{fig:phasegrid} and \ref{fig:rand_gridplot} in terms of trail formation and presence of clustered and random configurations. In particular, in the baseline setup, the metric exhibits high values in the region above the phase transition line, consistently with its definition, as this is the only region where trails can occur. The smart setup, instead, presents values of the  metric that show improvements across all regions of the phase plot, consistently with Figure \ref{fig:rand_gridplot}, and slightly worse values in the chaotic region (lower-right).

\clearpage
\end{comment}

\begin{figure}[t]
    \centering
    \includegraphics[width=1\linewidth]{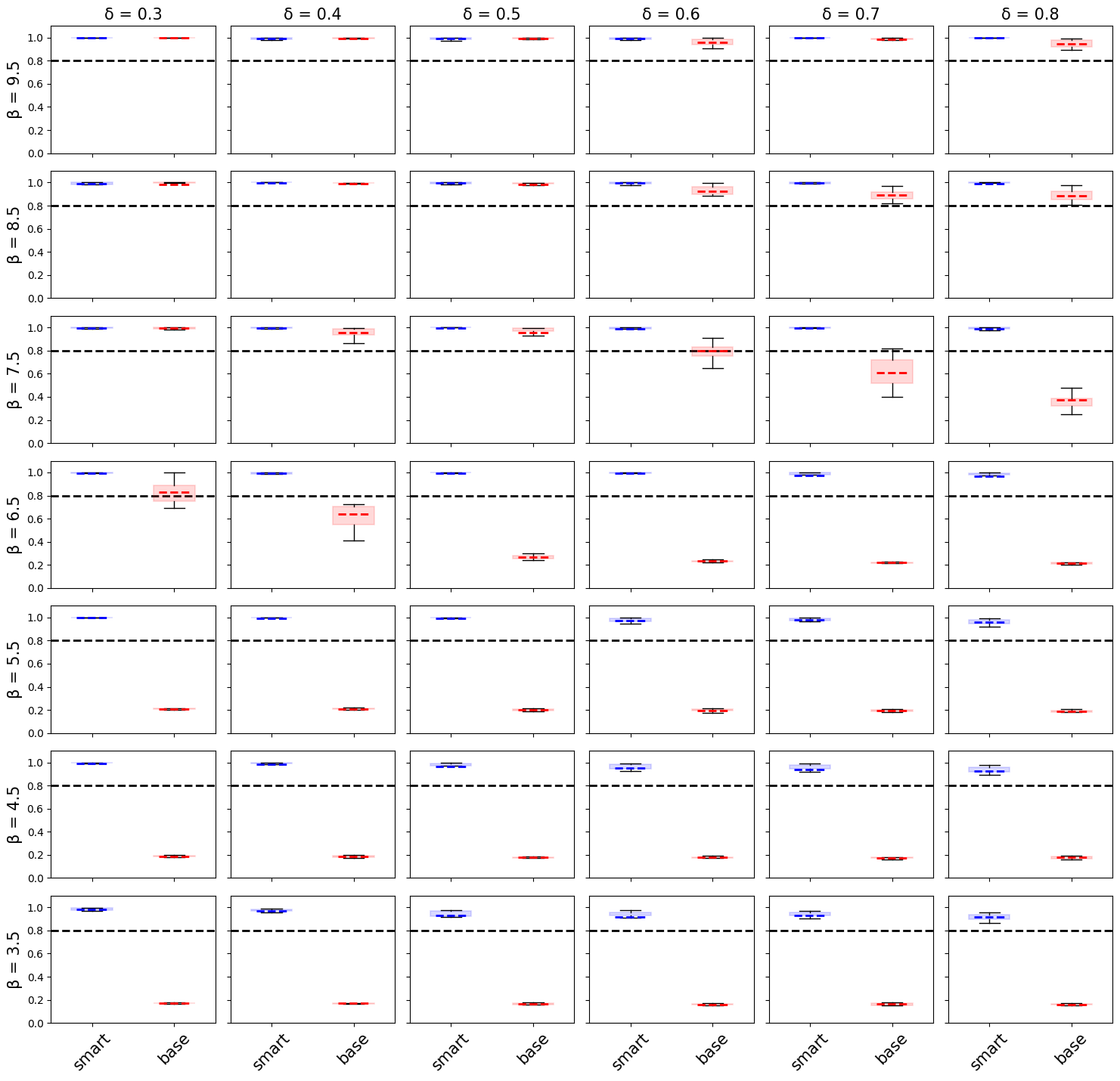}
    \caption{\textbf{Distribution of $m(T)$ per each $\left[ \delta, \, \beta \right]$-configuration}. Simulation settings included: (1) $N$ ABM-ants (base), (2) baseline population plus $N'$ smart-agents trained across the full phase space (smart).}
    \label{fig:boxplot-delta-beta-mT}
\end{figure}

\begin{figure}
    \centering
    \includegraphics[width=1\linewidth]{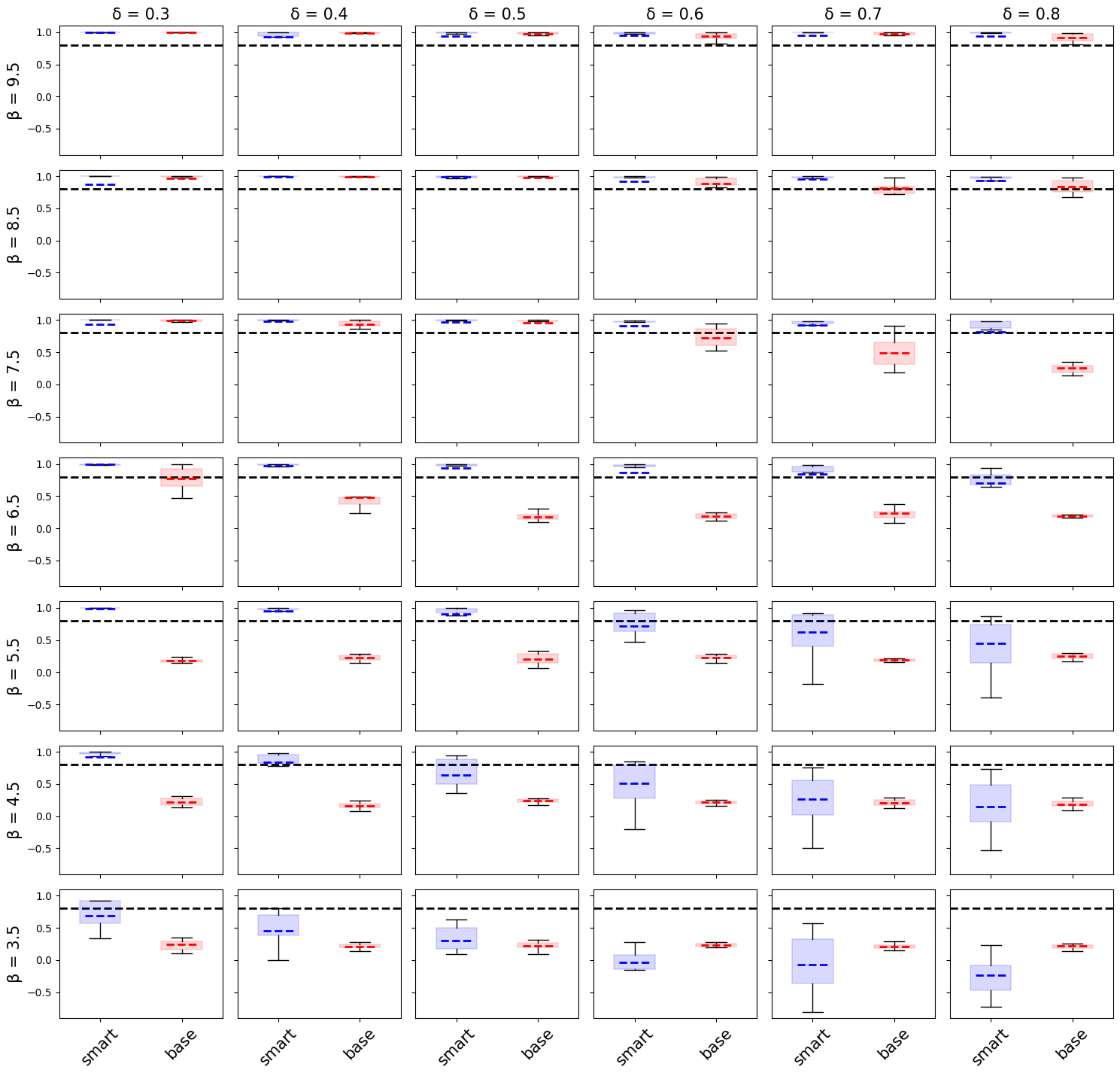}
    \caption{\textbf{Distribution of $M(T)$ per each $\left[ \delta, \, \beta \right]$-configuration}. Simulation settings included: (1) $N$ ABM-ants (base), (2) baseline population plus $N'$ smart-agents trained across the full phase space (smart).}
    \label{fig:boxplot-delta-beta-MT}
\end{figure}

\paragraph{Order parameters across the phase diagram.}
Figures \ref{fig:boxplot-delta-beta-mT}-\ref{fig:boxplot-delta-beta-MT} show the distribution of the order parameters $m(T)$ and $M(T)$, disaggregated with respect to the considered $\left[ \delta, \, \beta \right]$-configurations. 

The distributions of $m(T)$ and $M(T)$ across the evaluation episodes further corroborate the trends observed from the average values reported in Table \ref{tab:order_par_tab1}. In particular, the boxplots show that, in the smart setup, the configurations characterized by average order parameter values above the empirical threshold of $0.8$ generally exhibit distributions concentrated in the high-value region, with median values above the threshold and limited overlap with lower-order regimes. This indicates that the observed increase in both $m(T)$ and $M(T)$ is not driven by a small number of favorable realizations, but rather reflects a systematic and robust effect across the evaluation testbed. Moreover, several $\left[ \delta, \, \beta \right]$-configurations located below the phase transition line in the baseline case display a clear upward shift of the entire distribution under the smart control strategy, confirming that the emergence of ordered states occurs consistently over repeated trials. Specifically, of $m(T)$ are particularly concentrated near their upper bound in the smart setup, supporting the conclusion that the control action reliably promotes the formation of highly structured pheromone fields. In contrast, while the $M(T)$ distributions confirm increases in ant recruitment, a larger variability persists in the bottom-right region of the phase diagram. In this regime, low pheromone sensitivity and high movement randomness inherently reduce control effectiveness.
The smart setup presents in those cases worse values, due to the limited controllability of systems with low pheromone sensitivity and high movement randomness. This, as previously commented, is explained by observing that, in the baseline setup, lower values of $m(T)$ correspond to a more uniform distribution of pheromone across the environment. When combined with the high randomness in agent movement typical of these configurations, this leads to an increased probability that ABM-ants traverse pheromone-containing locations, thereby artificially inflating $M(T)$.

Overall, the distributional analysis confirms that the differences highlighted by the average order parameters are statistically meaningful and reflect genuine shifts in the collective dynamics induced by the control strategies, rather than fluctuations associated with individual evaluation episodes.

\clearpage

\subsection{Limited training scenario: testing}
\label{sec:additional_res_spec}

\begin{figure}[H]
    \centering
    \begin{subfigure}[b]{0.3\textwidth}
        \includegraphics[width=\textwidth]{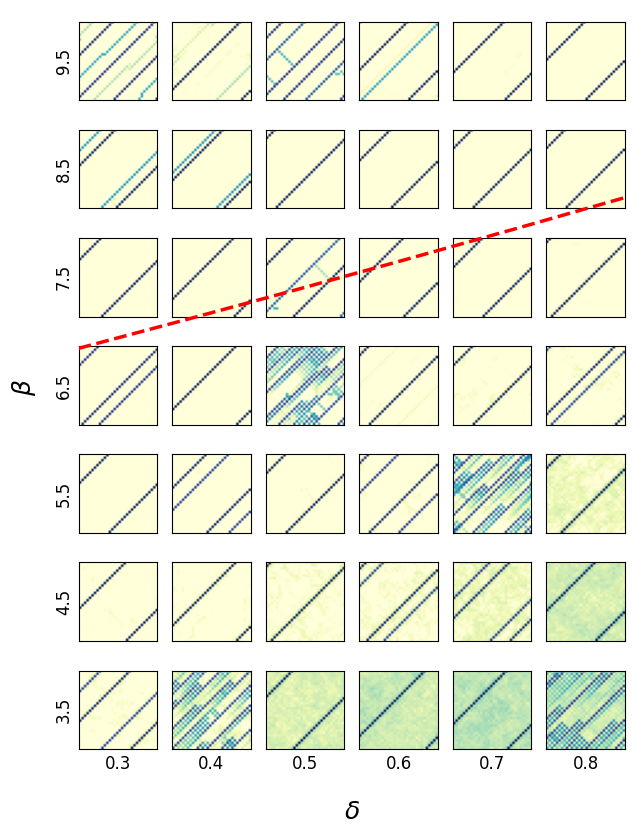}
        \caption{}
        \label{sfig:l3_pher_fix}
    \end{subfigure}
    \hspace{10pt}
    \begin{subfigure}[b]{0.3\textwidth}
        \includegraphics[width=\textwidth]{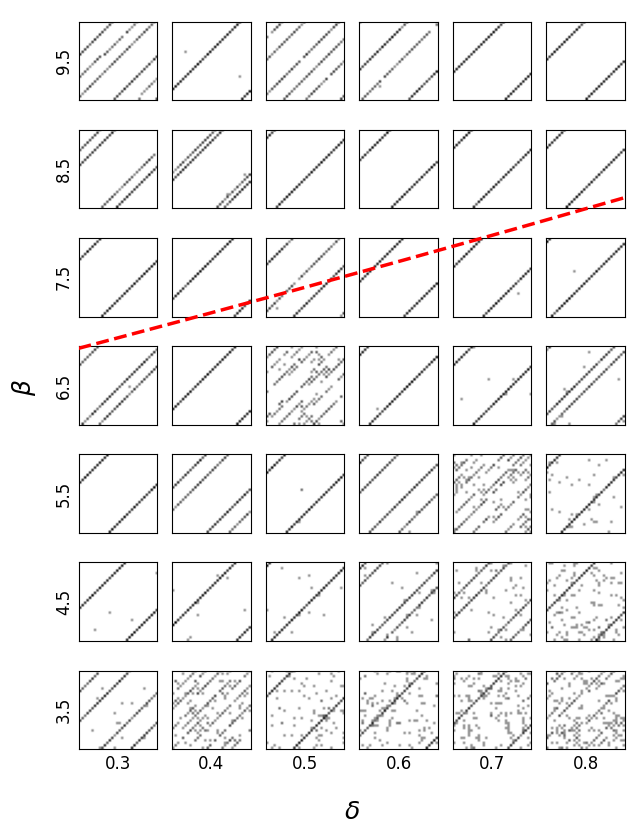}
        \caption{}
        \label{sfig:l3_ants_fix}
    \end{subfigure}
    \vspace{0.3cm}

    \begin{subfigure}[b]{0.3\textwidth}
        \includegraphics[width=\textwidth]{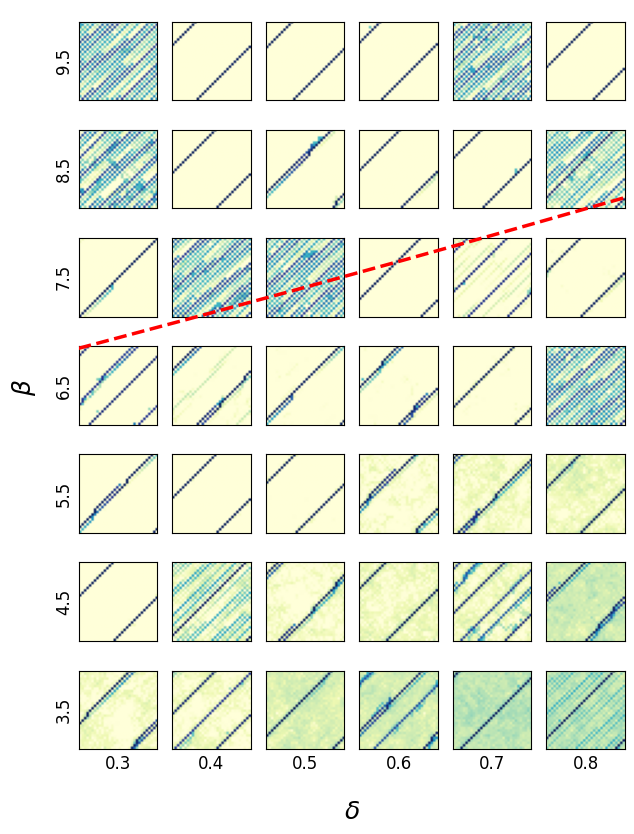}
        \caption{}
        \label{sfig:l4_pher_fix}
    \end{subfigure}
    \hspace{10pt}
    \begin{subfigure}[b]{0.3\textwidth}
        \includegraphics[width=\textwidth]{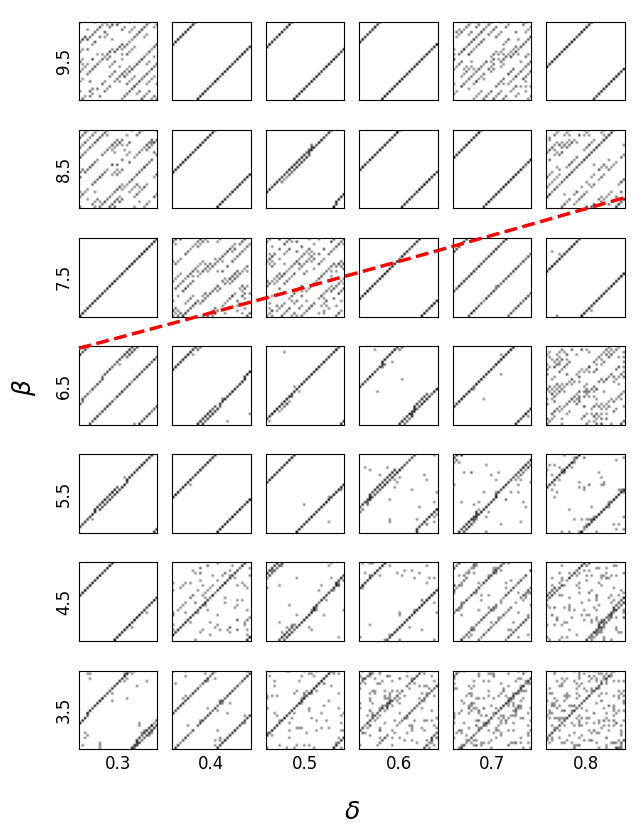}
        \caption{}
        \label{sfig:l4_ants_fix}
    \end{subfigure}
    \vspace{0.3cm}

    \begin{subfigure}[b]{0.3\textwidth}
        \includegraphics[width=\textwidth]{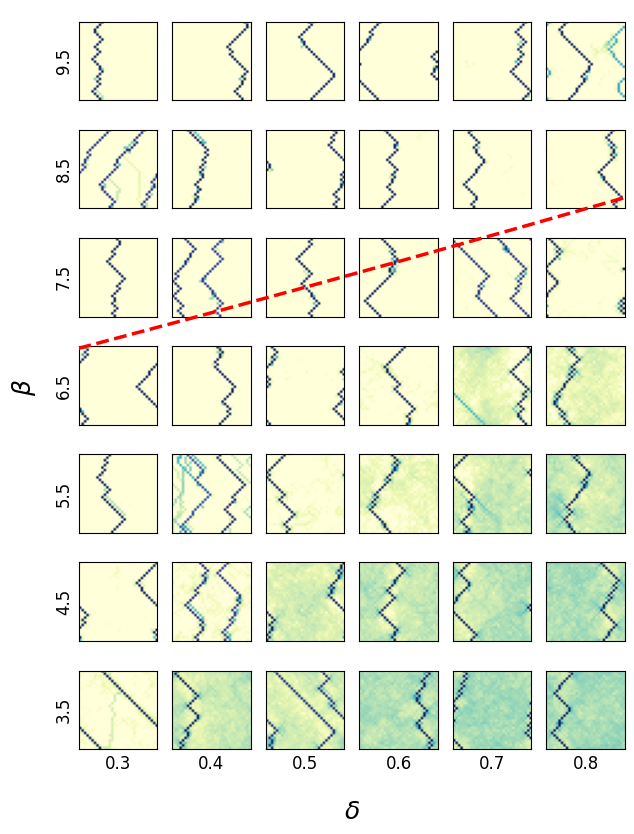}
        \caption{}
        \label{sfig:l5_pher_fix}
    \end{subfigure}
    \hspace{10pt}
    \begin{subfigure}[b]{0.3\textwidth}
        \includegraphics[width=\textwidth]{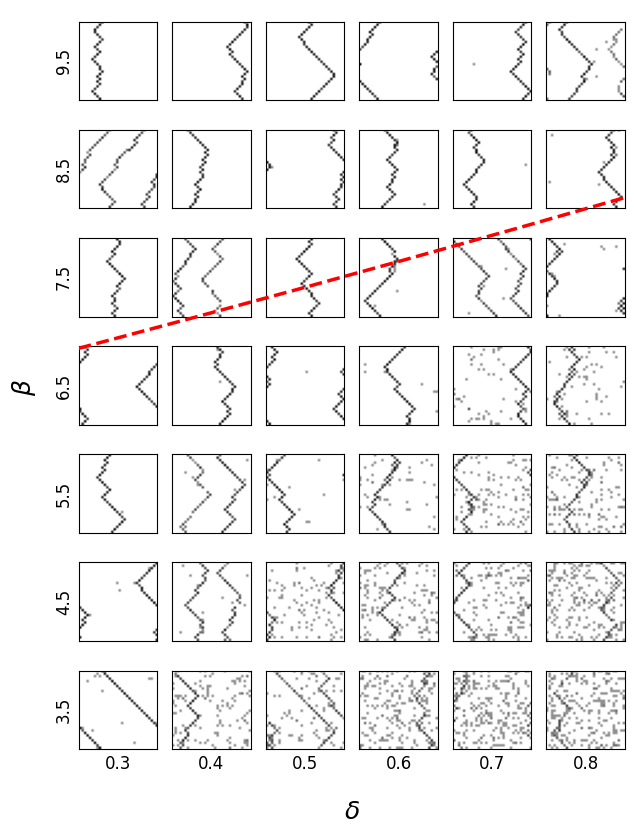}
        \caption{}
        \label{sfig:l5_ants_fix}
    \end{subfigure}

    \caption{\textbf{Limited training scenario: testing}. Phase diagrams of pheromone (left) and ABM-ants density (right) for policies $\pi=3, 4, 5$.}
    \label{fig:grid_part2}
\end{figure}

\begin{figure}[H]
    \centering
    \begin{subfigure}[b]{0.3\textwidth}
        \includegraphics[width=\textwidth]{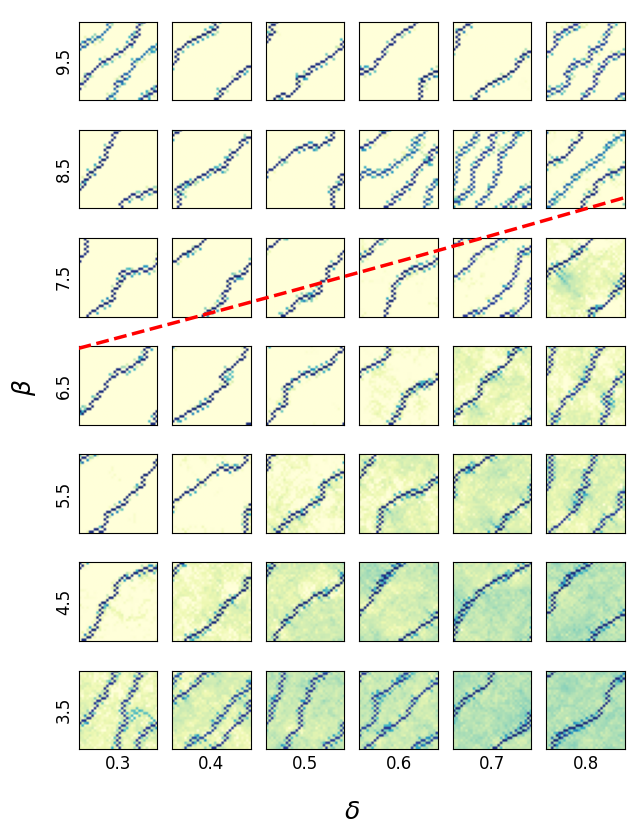}
        \caption{}
        \label{sfig:l6_pher_fix}
    \end{subfigure}
    \hspace{10pt}
    \begin{subfigure}[b]{0.3\textwidth}
        \includegraphics[width=\textwidth]{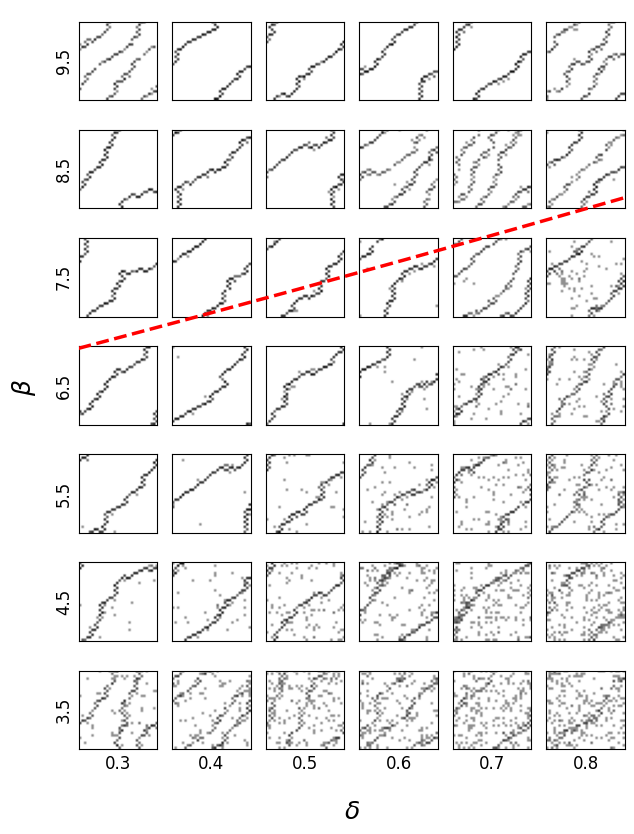}
        \caption{}
        \label{sfig:l6_ants_fix}
    \end{subfigure}
    \vspace{0.3cm}

    \begin{subfigure}[b]{0.3\textwidth}
        \includegraphics[width=\textwidth]{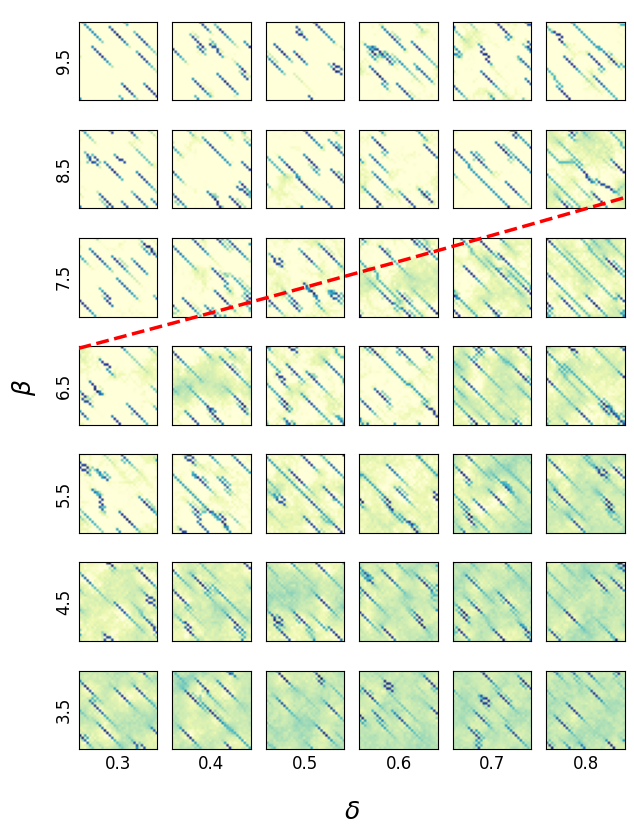}
        \caption{}
        \label{sfig:l7_pher_fix}
    \end{subfigure}
    \hspace{10pt}
    \begin{subfigure}[b]{0.3\textwidth}
        \includegraphics[width=\textwidth]{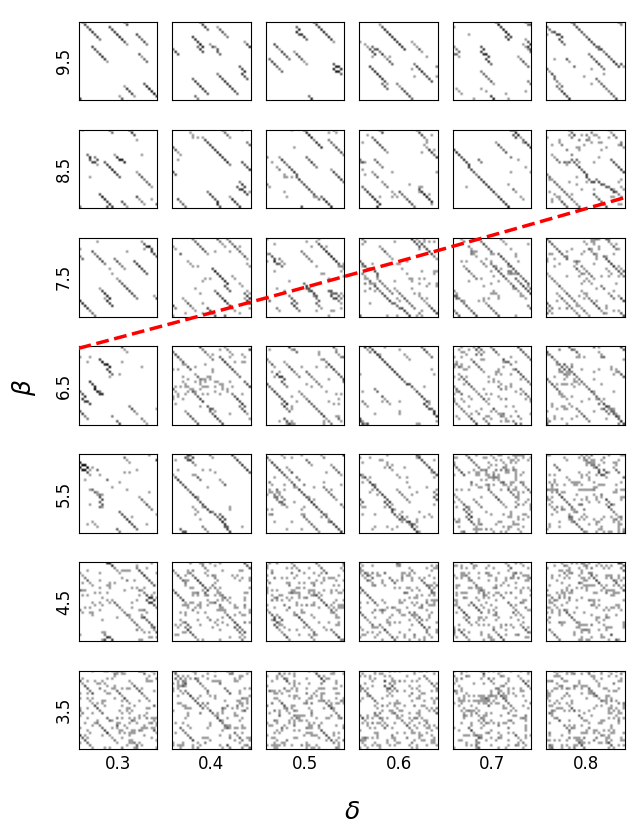}
        \caption{}
        \label{sfig:l7_ants_fix}
    \end{subfigure}
    \vspace{0.3cm}

    \begin{subfigure}[b]{0.3\textwidth}
        \includegraphics[width=\textwidth]{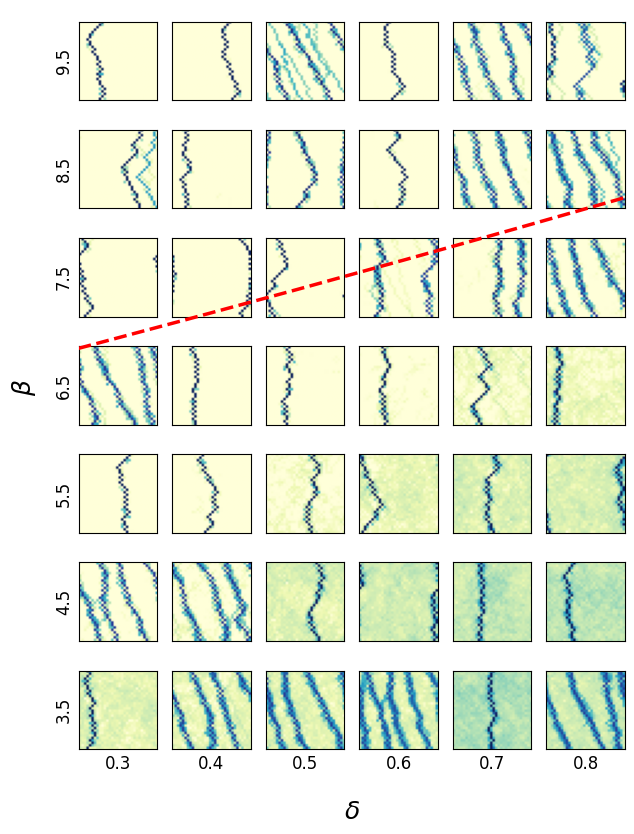}
        \caption{}
        \label{sfig:l8_pher_fix}
    \end{subfigure}
    \hspace{10pt}
    \begin{subfigure}[b]{0.3\textwidth}
        \includegraphics[width=\textwidth]{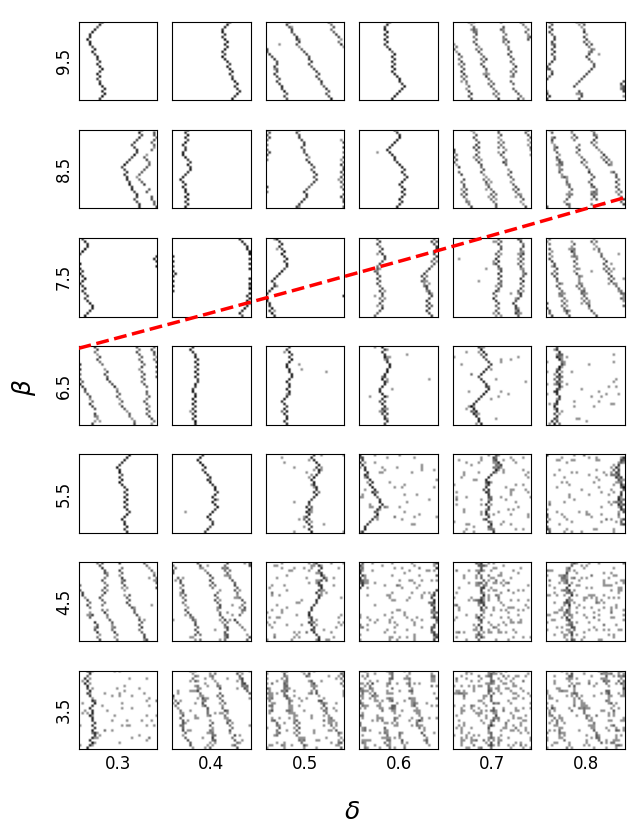}
        \caption{}
        \label{sfig:l8_ants_fix}
    \end{subfigure}

    \caption{\textbf{Limited training scenario: testing}. Phase diagrams of pheromone (left) and ABM-ants density (right) for policies $\pi=6, 7, 8$.}
    \label{fig:grid_part3}
\end{figure}

\paragraph{Final pheromone and individual concentrations.}
The results closely match those obtained in the basic training scenario. Notably, policy $\pi=7$, displayed in Figures \subref{sfig:l7_ants_fix} and \subref{sfig:l7_pher_fix}, failed to converge during training, exhibiting behaviour distinct from the other cases. Similarly, policy $\pi=4$ (Figures \subref{sfig:l4_ants_fix} and \subref{sfig:l4_pher_fix}) occasionally fails to form stable trails. All remaining policies consistently replicate the behaviour observed in Section \ref{sec:Results} for policy $\pi=9$, which is characterized by a shift in the phase line.

\begin{figure}[H]
    \centering
    \begin{subfigure}[b]{0.3\textwidth}
        \includegraphics[width=\textwidth]{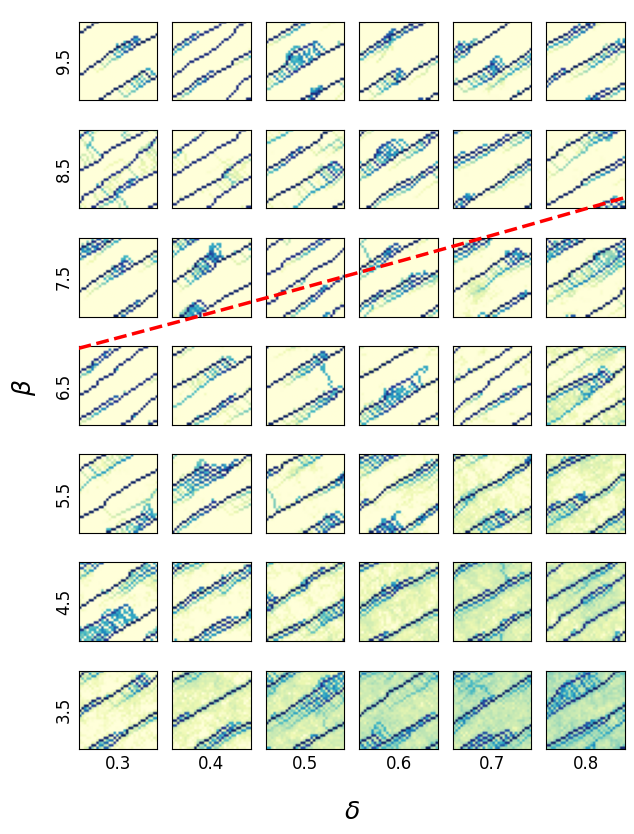}
        \caption{}
        \label{sfig:l0_pher_fix}
    \end{subfigure}
    \hspace{10pt}
    \begin{subfigure}[b]{0.3\textwidth}
        \includegraphics[width=\textwidth]{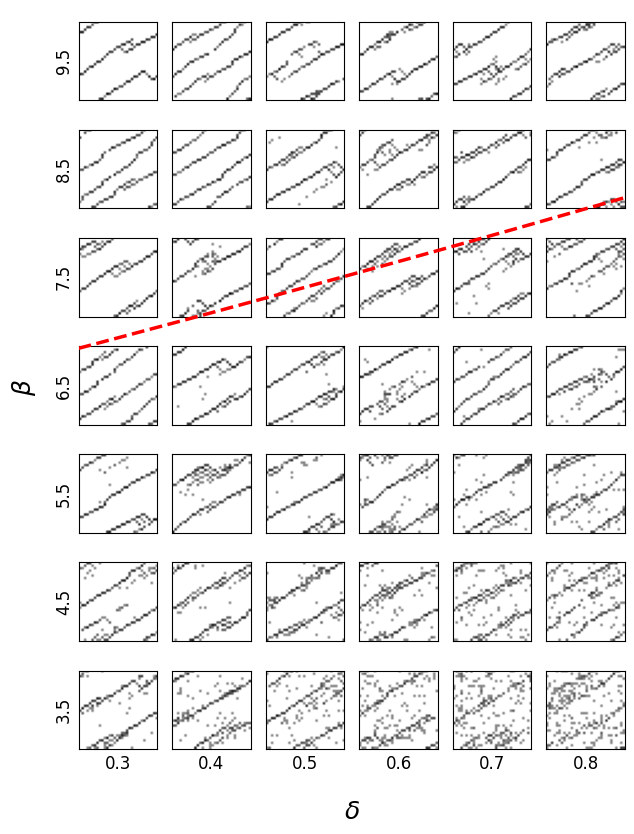}
        \caption{}
        \label{sfig:l0_ants_fix}
    \end{subfigure}
    \vspace{0.3cm}

    \begin{subfigure}[b]{0.3\textwidth}
        \includegraphics[width=\textwidth]{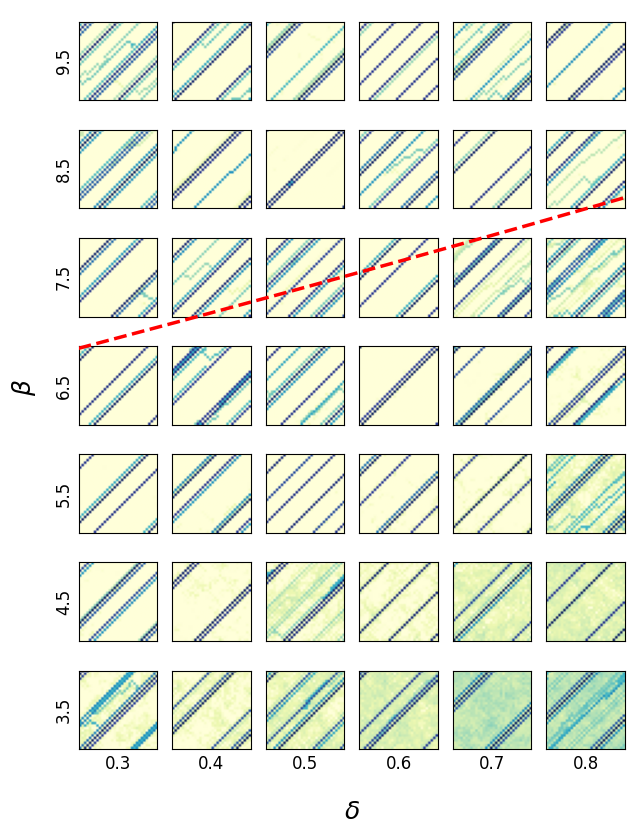}
        \caption{}
        \label{sfig:l1_pher_fix}
    \end{subfigure}
    \hspace{10pt}
    \begin{subfigure}[b]{0.3\textwidth}
        \includegraphics[width=\textwidth]{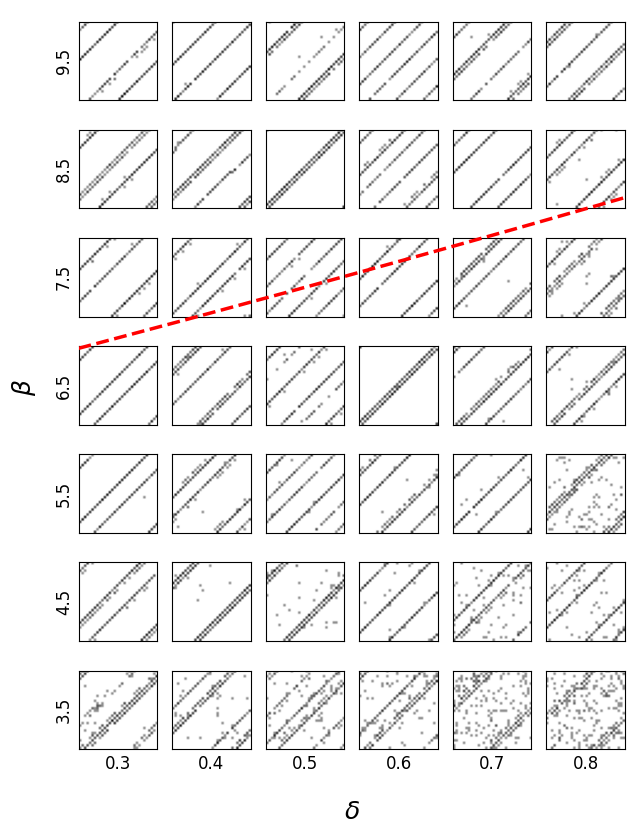}
        \caption{}
        \label{sfig:l1_ants_fix}
    \end{subfigure}
    \vspace{0.3cm}

    \begin{subfigure}[b]{0.3\textwidth}
        \includegraphics[width=\textwidth]{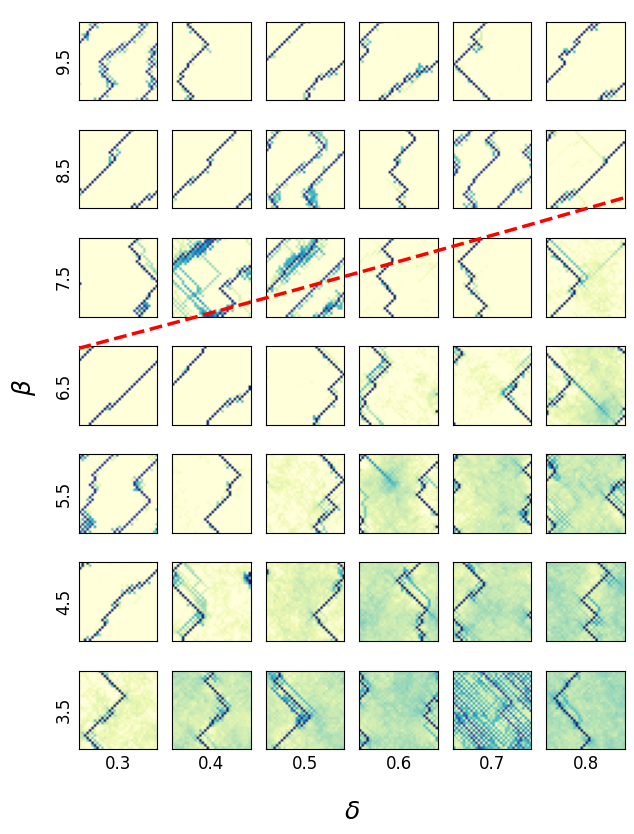}
        \caption{}
        \label{sfig:l2_pher_fix}
    \end{subfigure}
    \hspace{10pt}
    \begin{subfigure}[b]{0.3\textwidth}
        \includegraphics[width=\textwidth]{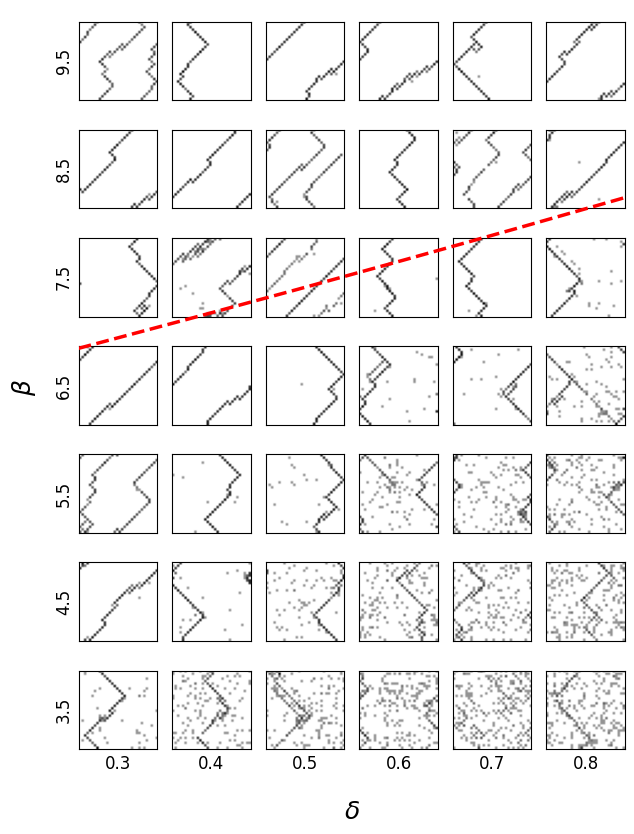}
        \caption{}
        \label{sfig:l2_ants_fix}
    \end{subfigure}

    \caption{\textbf{Limited training scenario: testing}. Phase diagrams of pheromone (left) and ABM-ants density (right) for policies $\pi=0, 1, 2$.}
    \label{fig:grid_part1}
\end{figure}


\end{document}